# Ice-free geomorphometry of Queen Maud Land, East Antarctica: 1. Sôya Coast

I.V. Florinsky[1*], S.O. Zharnova[2]

[1] Institute of Mathematical Problems of Biology, Keldysh Institute of Applied Mathematics, Russian Academy of Sciences, Pushchino, Moscow Region, 142290, Russia

[2] National Research Tomsk State University, 36 Lenin Ave., Tomsk, 634050, Russia

## Abstract

Geomorphometric modeling and mapping of ice-free Antarctic areas is promising for obtaining new quantitative knowledge about the topography of these unique landscapes and for the further use of morphometric information in Antarctic research. Within the framework of a project of creating a physical geographical thematic scientific reference geomorphometric atlas of ice-free terrains of Antarctica, we performed geomorphometric modeling and mapping of key ice-free areas of the Sôya Coast (the east coast of Lützow-Holm Bay, Queen Maud Land, East Antarctica). These include the Flatvaer (Ongul) Islands, Langhovde Hills, Breidvågnipa, Skarvsnes Foreland, Skallen Hills, and Skallevikhalsen Hills. As input data for geomorphometric modeling and mapping, we used five fragments of the Reference Elevation Model of Antarctica. For the six ice-free areas and adjacent glaciers, we derived models and maps of eleven most scientifically important morphometric variables (i.e., slope, aspect, horizontal curvature, vertical curvature, minimal curvature, maximal curvature, catchment area, topographic wetness index, stream power index, total insolation, and wind exposition index). The obtained models and maps describe the ice-free topography of the Sôya Coast in a rigorous, quantitative, and reproducible manner. New morphometric data can be useful for further geological, geomorphological, glaciological, ecological, and hydrological studies of this region.

## 1 Introduction

Topography is the most important component of the environment. Being a result of the interaction of endo- and exogenous processes of various scales and reflecting the geological structure, topography determines prerequisites for the gravity-driven migration and accumulation of moisture and other substances along the land surface and in the near-surface layer, controls thermal, hydrological, and wind regimes of the terrain, etc.

The rigor and reproducibility of topographic studies is ensured by geomorphometry, a science with a developed physical and mathematical theory and a powerful set of computational methods. The subject of geomorphometry is the modeling and analysis of topography as well as relationships between topography and other components of geosystems (Evans, 1972; Moore et al., 1991; Wilson and Gallant, 2000; Shary et al., 2002; Hengl and Reuter, 2009; Minár et al., 2016; Florinsky, 2017, 2025a).

There are three main types of ice-free areas in Antarctica: (1) Antarctic oases, i.e. coastal, shelf, and mountainous ice-free areas of Antarctica; (2) ice-free islands (or areas thereof) situated outside the ice shelves; and (3) ice-free mountain chains (or their portions) and nunataks (Markov et al., 1970; Simonov, 1971; Korotkevich, 1972; Alexandrov, 1985; Pickard, 1986; Beyer and Bölter, 2002; Sokratova, 2010).

Geomorphometric modeling and mapping of Antarctic ice-free areas (Florinsky, 2023a, 2023b; Florinsky and Zharnova, 2025) can be used for obtaining new quantitative knowledge about the topography of these terrains and for the further use of morphometric information in solving problems of geomorphology, geology, glaciology, soil science, ecology, and other sciences. A project has recently been launched to create a physical geographical thematic scientific reference geomorphometric atlas of ice-free areas of Antarctica (Florinsky, 2024, 2025b). As part of this project, we carry out a geomorphometric modeling and mapping of ice-free areas of Queen Maud Land, East Antarctica. In this paper, we present results for key ice-free areas of the Sôya Coast.

* Correspondence to: iflor@mail.ru

## 2 Study areas

The Sôya Coast is the east coast of Lützow-Holm Bay (Queen Maud Land, East Antarctica) (Fig. 1). It is located between 70°S and 68.7°S, and 38.5°E and 40°E. The Sôya Coast is bordered on the west by the Prince Harald Coast and on the east by the Prince Olav Coast. It was first mapped by Norwegian cartographers from air photos of the Lars Christensen Expedition, 1936–37. The Japanese Antarctic Research Expedition have studied this territory since 1957 and named it after a research vessel *Sôya*.[1]

We consider the following key ice-free areas of the Sôya Coast (Figs. 1 and 2, Table 1), from north to south: the Flatvaer (Ongul) Islands (Fig. 2a), Langhovde Hills (Fig. 2b), Breidvågnipa (Fig. 2c), Skarvsnes Foreland (Fig. 2d), Skallen Hills, and Skallevikhalsen Hills (Fig. 2e).

The Flatvaer (Ongul) Islands are a group of small, low-hilled islands with typical elevations of 25–30 m above sea level (ASL) separated from the mainland by the Ongul Sound. The Langhovde Hills are an extensive rocky hill terrain with elevations of 75–300 and 100–450 m ASL in its northern and southern portions, respectively. Breidvågnipa is a small rocky mountain peninsula with heights of up to 325 m ASL.

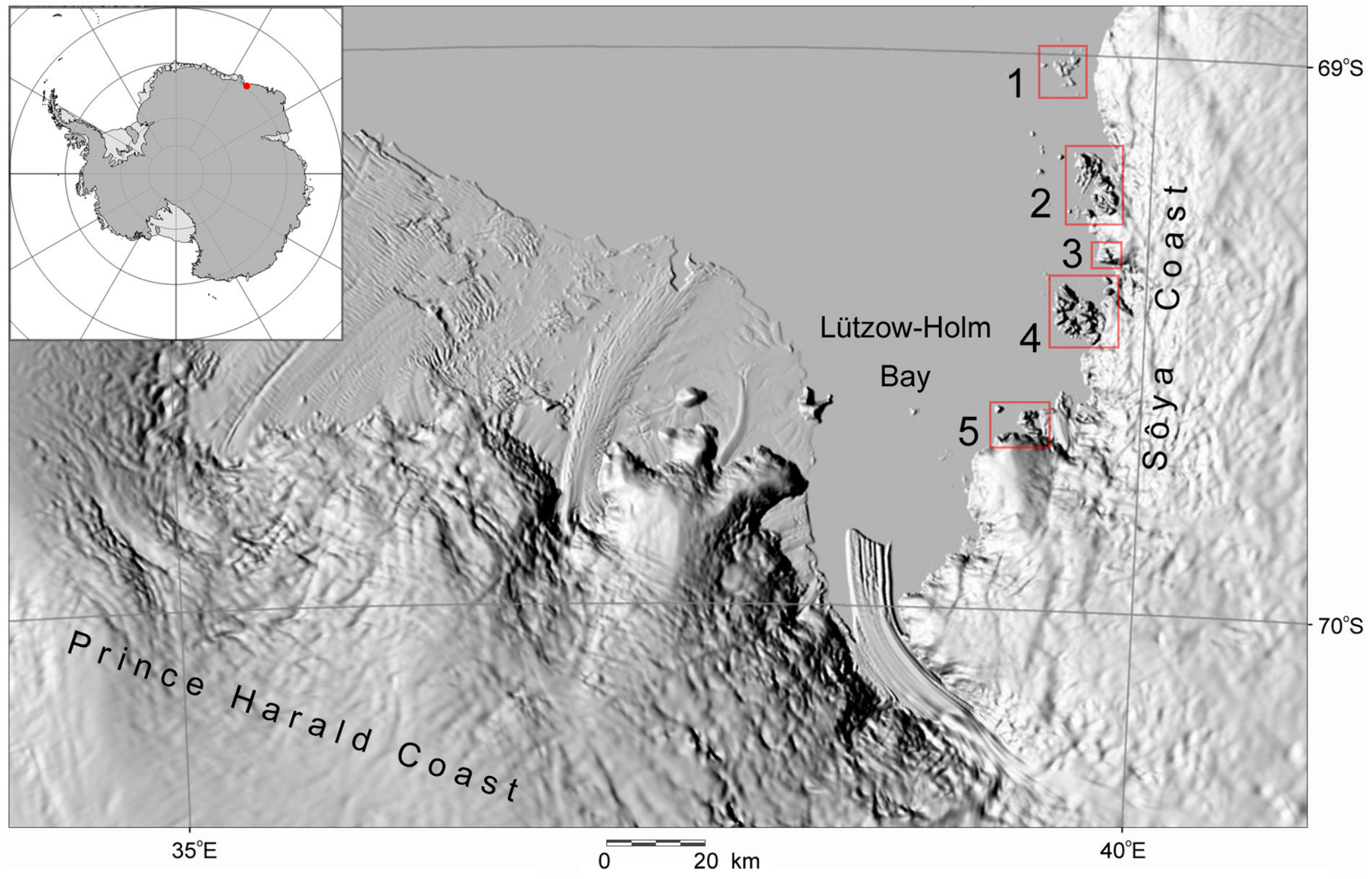

**Fig. 1** Geographical location of the study areas: 1 – Flatvaer (Ongul) Islands, 2 – Langhovde Hills, 3 – Breidvågnipa, 4 – Skarvsnes Foreland, 5 – Skallen Hills and Skallevikhalsen Hills. The hill-shaded map was produced by REMA Exployer (PGC, 2022–2024).

[1] Due to this activity, some geographical features of the Sôya Coast (including study areas) have double names assigned by both the Norwegian and Japanese cartographic authorities (Ishikawa T. et al., 1976, 1977; GSI, 1978; Ishikawa M. et al., 1994; Osanai et al., 2004; GIAJ, 2012; SCAR, 2000–2014). In such cases, we use names of the both origin, with the second name in brackets.

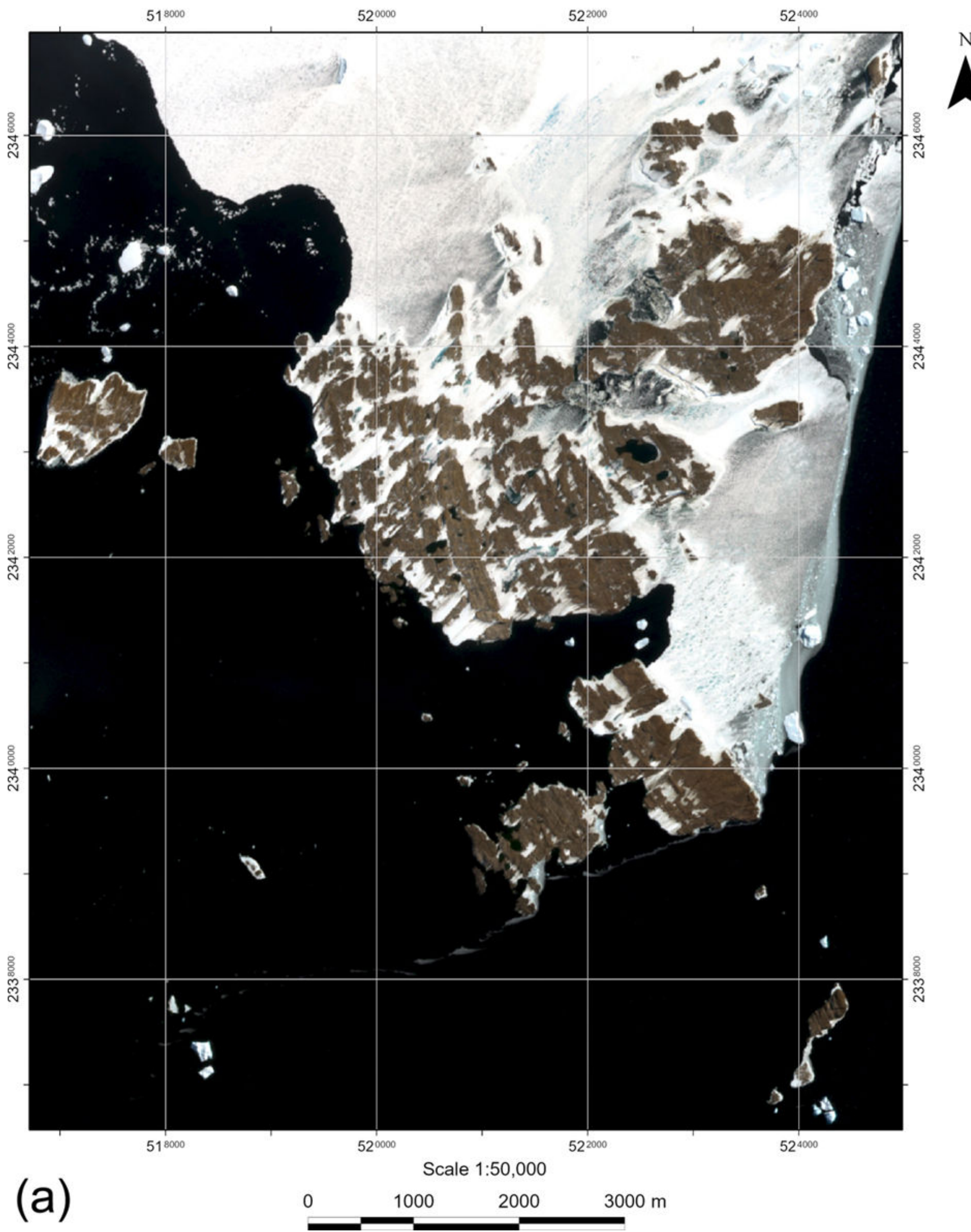

**Fig. 2** Satellite images of the study areas: (a) Flatvaer (Ongul) Islands.

*(Continued)*

The Skarvsnes Foreland is an extensive rocky hill terrain indented by several coves with typical elevations of 100–350 and 150–300 m ASL in its northern and southern parts, respectively. The Skallen Hills are rocky coastal hills with elevations of 50–150 m ASL. The Skallevikhalsen Hills is a line of coastal rocky hills with elevations of 100–200 m ASL. These five areas are separated from each other by glaciers, and are bounded to the east by the ice sheet and to the west by Lützow-Holm Bay.

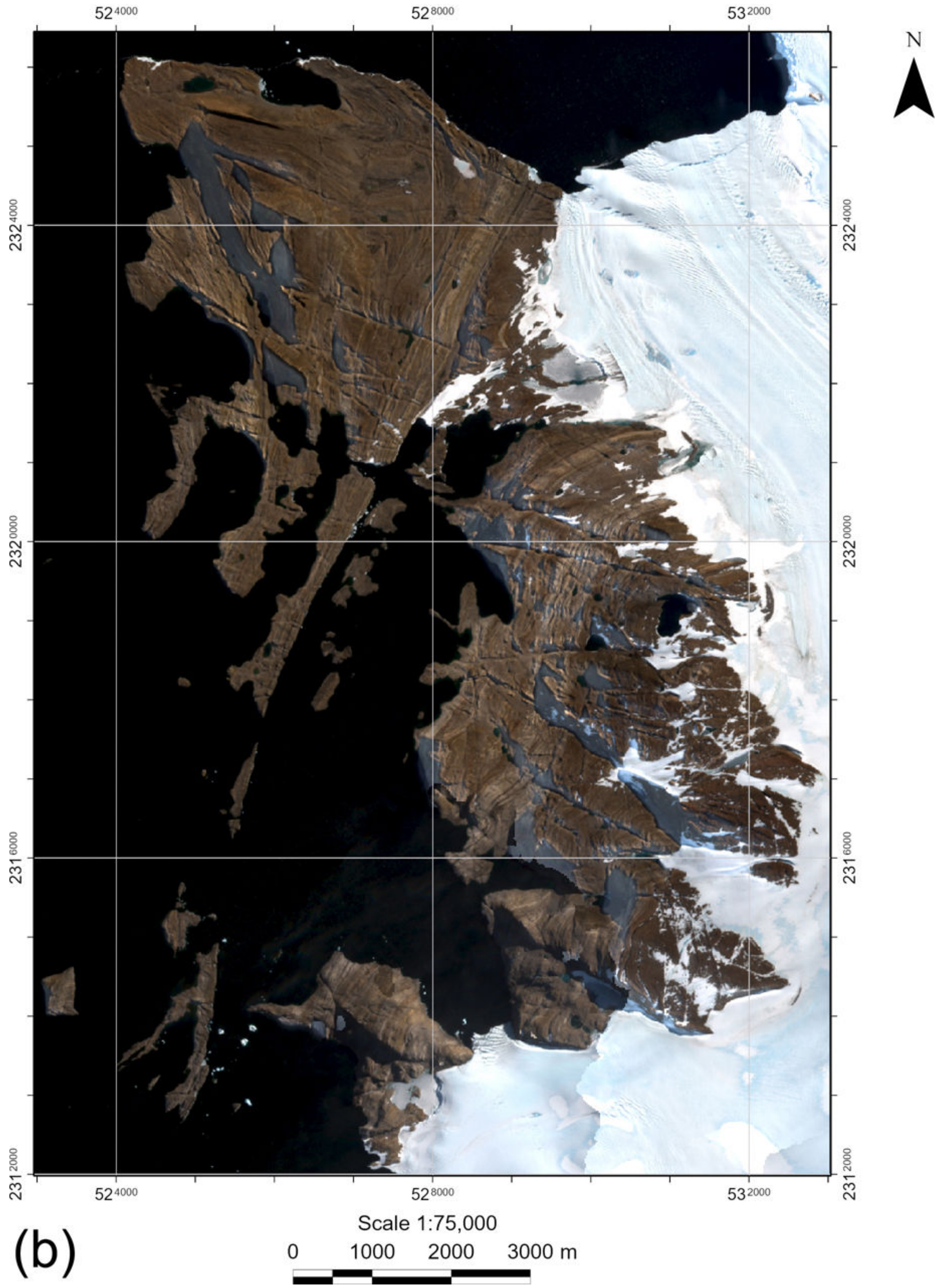

**Fig. 2, cont'd** Satellite images of the study areas: (b) Langhovde Hills.

*(Continued)*

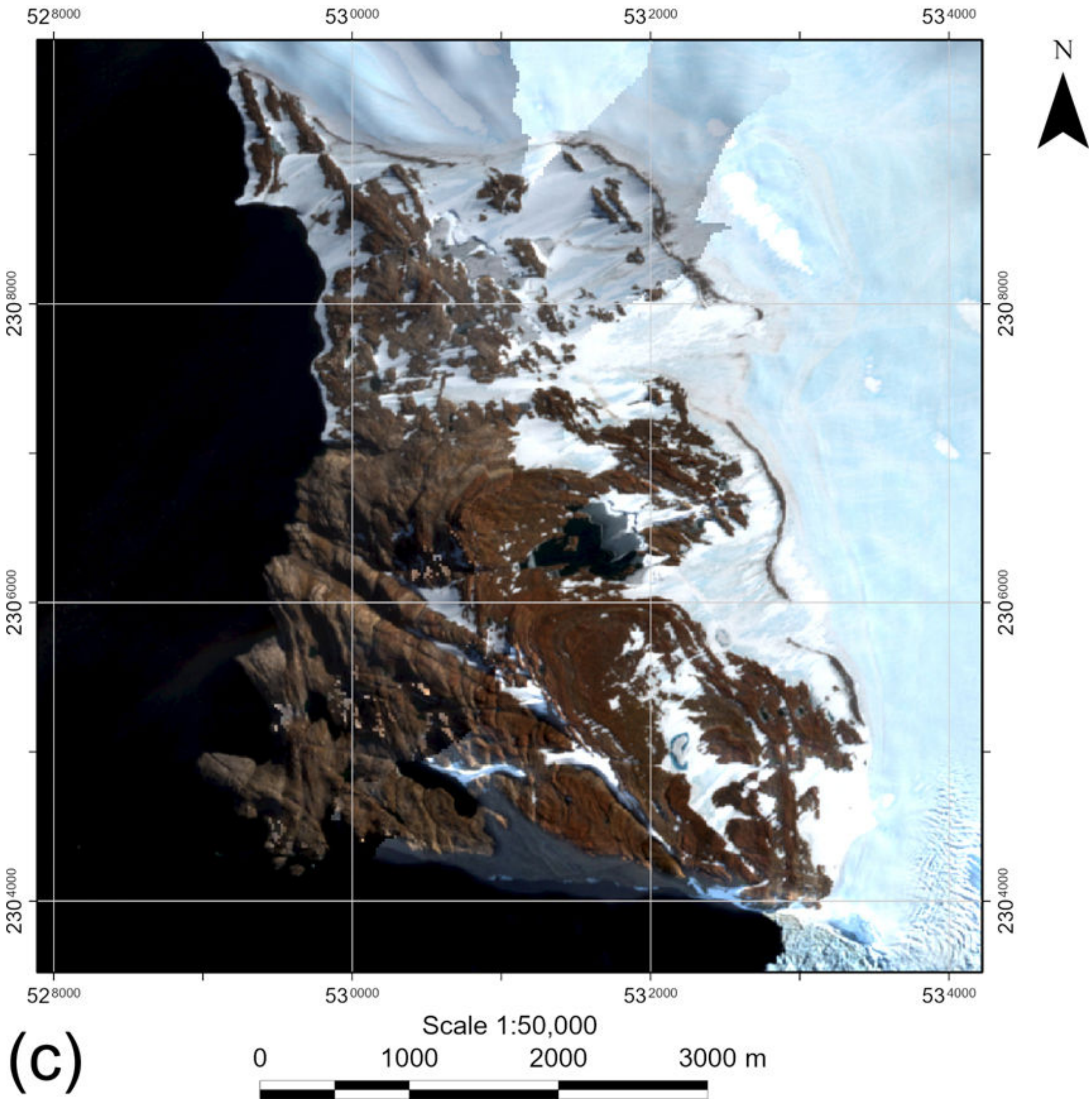

**Fig. 2, cont'd** Satellite images of the study areas: (c) Breidvågnipa.

*(Continued)*

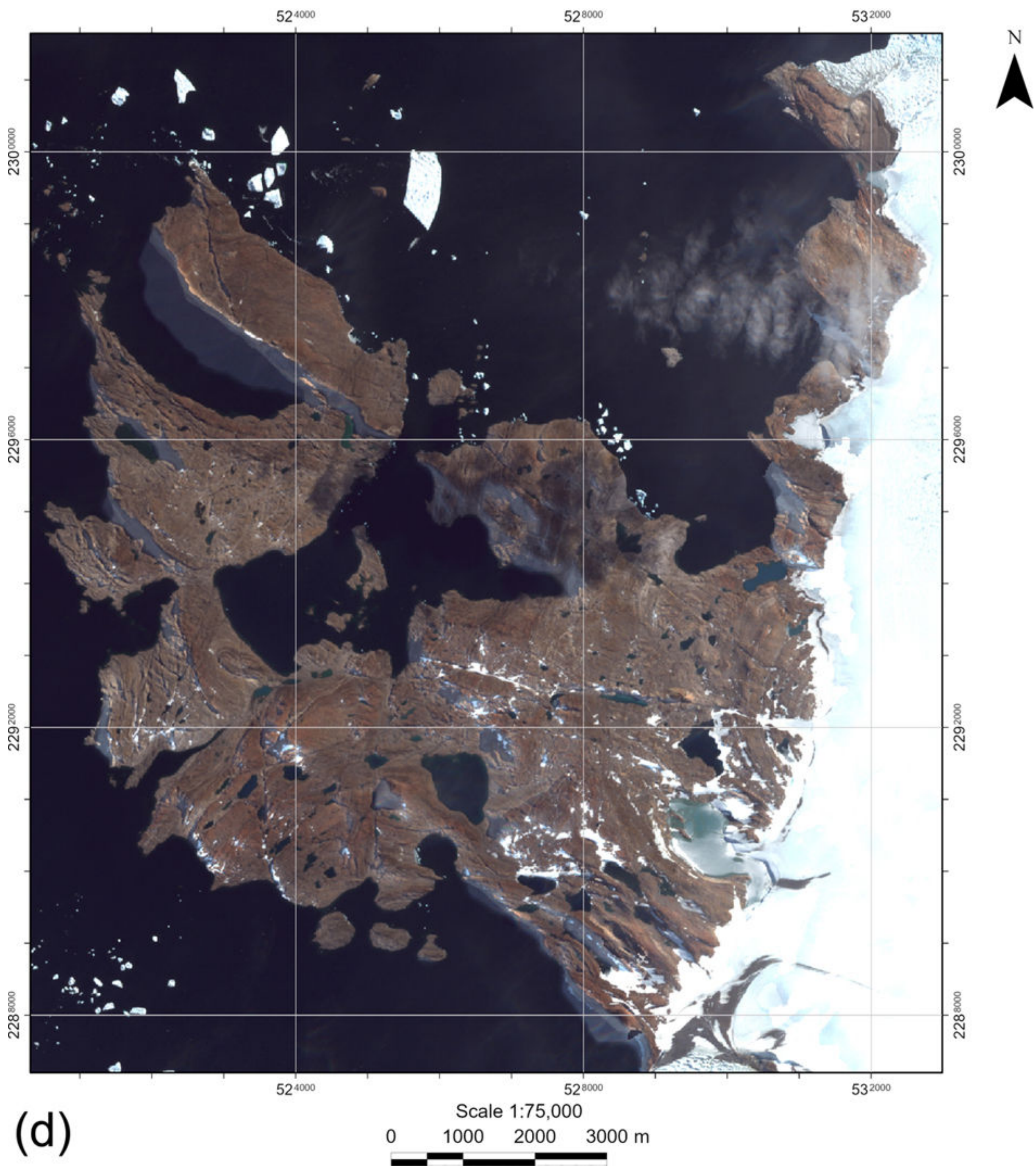

**Fig. 2, cont'd** Satellite images of the study areas: (d) Skarvsnes Foreland.

*(Continued)*

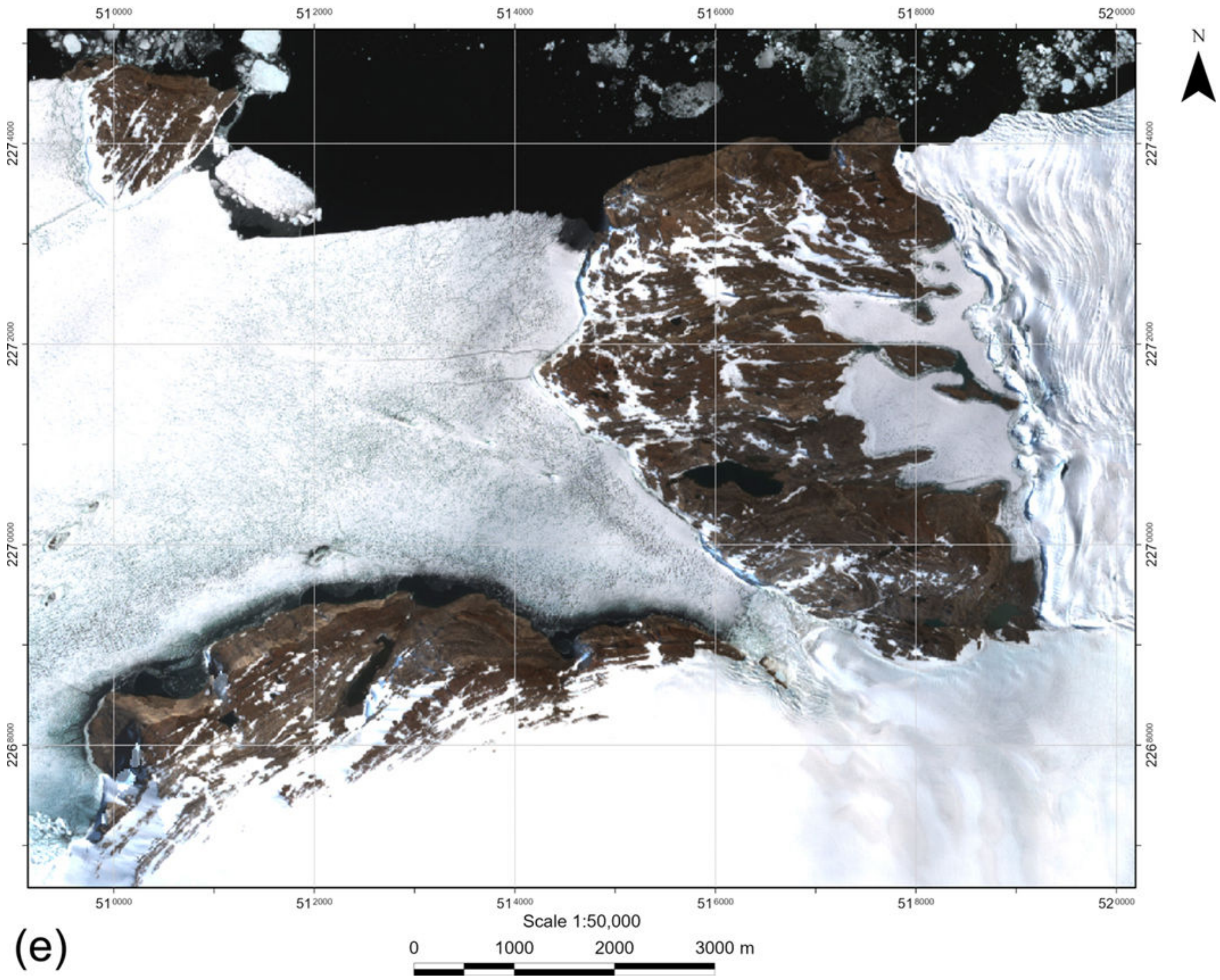

**Fig. 2, cont'd** Satellite images of the study areas: (e) Skallen Hills (left) and Skallevikhalsen Hills (right).
Fragments of a Sentinel-2A MSI scene, 15.02.2025 (ESA, 2025). True color synthesis with brightness and contrast enhancement. UTM projection (zone 37S).

**Table 1** Characteristics of the study areas (Fig. 1) and their DEMs.

| # | Area name | Geographical coordinates | Area, $km^2$ | DEM size, m | DEM size, point | Points with elevation values |
|---|---|---|---|---|---|---|
| 1 | Flatvaer (Ongul) Islands | 69.02852° S, 39.54379° E | 15.5 | 10,488 × 12,264 | 1311 × 1533 | 226,621 |
| 2 | Langhovde Hills | 69.22389° S, 39.69693° E | 68.5 | 20,592 × 19,704 | 2574 × 2463 | 2,175,742 |
| 3 | Breidvågnipa | 69,35772° S, 39,78459° E | 15.5 | 6,504 × 6,392 | 813 × 799 | 501,277 |
| 4 | Skarvsnes Foreland | 69.47392° S, 39.68363° E | 79.0 | 15,240 × 14,664 | 1905 × 1833 | 1,482,642 |
| 5 | Skallen Hills<br>Skallevikhalsen Hills | 69.66762° S, 39.43763° E<br>69.69327° S, 39.31203° E | 17.3<br>10.7 | 11,712 × 9,960 | 1464 × 1245 | 810,695 |

The Langhovde Hills, Breidvågnipa, and Skarvsnes Foreland are characterized by glaciated landforms including giant *roches moutonnee*, deep U-shaped valleys, troughs, and cirques. At the Skallevikhalsen Hills, staircase landforms predominate. The Flatvaer (Ongul) Islands and Skallen Hills have undulating hilly landscapes consisting of *roches moutonnee*, crescentic gouges, and grooves (Omoto, 1977; Yoshida, 1983; Hirakawa et al., 1984).

The study areas are mainly composed of Precambrian metamorphic rocks including various kinds of gneiss interbedded with granite, charnockite, marble, and quartzite. Most of the metamorphic rocks belong to the granulite facies (Tatsumi and Kikuchi, 1959; Banno et al., 1964; Ishikawa T. et al., 1976, 1977; Yoshida, 1977; Matsumoto et al., 1979; Ishikawa M. et al., 1994; Osanai et al., 2004).

The Syowa Station, a Japanese year-round research station is located on East Ongul Island, one of the Flatvaer (Ongul) Islands. At the Langhovde Hills, there is an ecological and ornithological Antarctic Specially Protected Area No. 141 *Yukidori Valley* (ATCM, 2024).

## 3 Materials and methods

For modeling and mapping, we used five digital elevation models (DEMs) with a grid spacing of 8 m. They were extracted from the photogrammetrically derived Reference Elevation Model of Antarctica (REMA) (Howat et al., 2019; PGC, 2018–2024).

### *3.1 Preprocessing*

REMA is presented in the polar stereographic projection with the elevation datum of the WGS84 ellipsoid. There is no bathymetry in REMA; such cells have 'no data' values. The extracted DEMs were reprojected into the UTM projection (zone 37S) preserving the original 8-m grid spacing (Table 1). Ellipsoidal elevations were transformed into orthometric ones.

### *3.2 Calculations*

Digital models of eleven, most scientifically important morphometric variables were derived from the 8-m gridded reprojected DEMs. The list of morphometric variables (Table 2) includes six local attributes: slope ($G$), aspect ($A$), horizontal curvature ($k_h$), vertical curvature ($k_v$), minimal curvature ($k_{min}$), and maximal curvature ($k_{max}$); one nonlocal variable—catchment area ($CA$); two combined variables: topographic wetness index ($TWI$) and stream power index ($SPI$); as well as two two-field-specific attributes—total insolation ($TIns$) and wind exposition index ($WEx$). Formulas and detailed interpretations of these variables can be found elsewhere (Shary et al., 2002; Florinsky 2017, 2025a, chap. 2).

To derive models of local variables (i.e., $G$, $A$, $k_h$, $k_v$, $k_{min}$, and $k_{max}$), we used the finite-difference method by Evans (1980). To compute models of $CA$, we applied the maximum-gradient based multiple flow direction algorithm by Qin et al. (2007) to preprocessed sink-filled DEMs. $CA$ models were logarithmized. To derive models of combined morphometric variables (i.e., $TWI$ and $SPI$), we used previously calculated models of $CA$ and $G$. To compute models of $TIns$ and $WEx$, we applied two related methods by Böhner (2004). $TIns$ were estimated for one mid-summer day (1st January) with a temporal step of 0.5 h.

**Table 2** Definitions and interpretations of key morphometric variables (Florinsky, 2017, 2025a).

| Variable, notation, and unit | Definition and interpretation |
|---|---|
| Slope, *G* (°) | An angle between the tangential and horizontal planes at a given point of the topographic surface. Relates to the velocity of gravity-driven flows. |
| Aspect, *A* (°) | An angle between the north direction and the horizontal projection of the two-dimensional vector of gradient counted clockwise, from 0° to 360°, at a given point of the surface. Relates to the direction of gravity-driven flows. |
| Horizontal (tangential) curvature, $k_h$ ($m^{-1}$) | A curvature of a normal section tangential to a contour line at a given point of the surface. A measure of flow convergence and divergence. Gravity-driven lateral flows converge where $k_h < 0$, and diverge where $k_h > 0$. $k_h$ mapping reveals crest and valley spurs. |
| Vertical (profile) curvature, $k_v$ ($m^{-1}$) | A curvature of a normal section having a common tangent line with a slope line at a given point of the surface. A measure of relative deceleration and acceleration of gravity-driven flows. They are decelerated where $k_v < 0$, and are accelerated where $k_v > 0$. $k_v$ mapping reveals terraces and scarps. |
| Minimal curvature, $k_{min}$ ($m^{-1}$) | A curvature of a principal section with the lowest value of curvature at a given point of the surface. $k_{min} > 0$ corresponds to local convex landforms (e.g., hills), while $k_{min} < 0$ relates to elongated concave landforms (e.g., troughs). |
| Maximal curvature, $k_{max}$ ($m^{-1}$) | A curvature of a principal section with the highest value of curvature at a given point of the surface. $k_{max} > 0$ corresponds to elongated convex landforms (e.g., crests), while $k_{max} < 0$ relate to local concave landforms (e.g., holes). |
| Catchment area, *CA* ($m^2$) | An area of a closed figure formed by a contour segment at a given point of the surface and two flow lines coming from upslope to the contour segment ends. A measure of the contributing area. |
| Topographic wetness index, *TWI* | A ratio of catchment area to slope gradient at a given point of the topographic surface. A measure of the extent of flow accumulation. |
| Stream power index, *SPI* | A product of catchment area and slope gradient at a given point of the surface. A measure of potential flow erosion and related landscape processes. |
| Total insolation, *TIns* (kWh/$m^2$) | A measure of the topographic surface illumination by solar light flux. Total potential incoming solar radiation, a sum of direct and diffuse insolations. |
| Wind exposition index, *WEx* | A measure of an average exposition of slopes to wind flows of all possible directions at a given point of the topographic surface. |

### *3.3 Mapping*

First, we created hypsometric maps of the key ice-free areas of the Sôya Coast from the reprojected DEMs. Two gradient hypsometric tint scales were used to depict the topography. To display the elevations of the ice-free topography, we applied the green-yellow part of the standard spectral hypsometric scale of color plasticity (Kovaleva, 2014). To display the elevations of the glacier topography, we utilized a modified hypsometric tint scale for polar regions (Patterson and Jenny, 2011). Then, hypsometric tintings were combined with achromatic hill shading derived from the DEM by a standard procedure (Figs. 3a, 4a, 5a, 6a, and 7a).

Finally, we placed labels of geographical names on the hypsometric maps. This information was derived from available maps of the study areas (Ishikawa T. et al., 1976, 1977; GSI, 1978; Ishikawa M. et al., 1994; Osanai et al., 2004; GIAJ, 2012) as well as the SCAR Composite Gazetteer of Antarctica (2000–2014). We utilized the following principles:

- Some features of the Sôya Coast have double names assigned independently by the Norwegian and Japanese researchers and cartographic authorities. We used names of all the origin, with a second name in brackets.
- Japanese maps utilize two romanization systems for transcribing Japanese words into the Latin alphabet (i.e., either the Kunrei-shiki system or the Hepburn one). The Kunrei-shiki system is, as a rule, used in the Composite Gazetteer of Antarctica (SCAR, 2000–2014).

We used transcriptions of the Japanese geographic names from the Composite Gazetteer.

- Specific proper names were used without modification. However, we replaced generic names of Norwegian and Japanese origin (i.e., breen, bukta, fjell, fjella, fjellet, høgda, isen, kammen, knausane, holme, øy, øya, øyane, sund, ryggen, toppen; dai, daira, dake, hama, hyoga, ike, irie, iwa, kubo, minato, misaki, nagaone, san, sawa, seto, sima, soto, to, ura, wan, yama) with English ones (i.e., bay, beach, cape, cirque, cove, glacier, heights, icefield, hills, island, lake, mount, mountains, peak, point, peninsula, ridge, rock, sound, strait, terrace, valley) for consistency and avoiding tautologies.

Second, from the calculated digital morphometric models, we produced five series of morphometric maps for the key ice-free areas of the Sôya Coast (Figs. 3b–l, 4b–l, 5b–l, 6b–l, and 7b–l). For optimal visual perception of morphometry, we used the following principles for applying gradient tint scales:

1. *G* and *CA* can take only positive values. To map *G* and *CA*, we applied a standard gray tint scale; the minimum and maximum values of *G* or *CA* correspond to white and black, respectively (Figs. 3b and 3h, 4b and 4h, 5b and 5h, 6b and 6h, 7b and 7h).

2. $k_h$, $k_v$, $k_{min}$, and $k_{max}$ can take both negative and positive values, having opposite physical mathematical sense and interpretation. To map curvatures, we used a two-color tint scale consisting of two contrasting parts, blue and orange (negative and positive values, respectively). The most and least saturated shades of blue or orange colors correspond to the absolute maximum and absolute minimum values, respectively, of $k_h$, $k_v$, $k_{min}$, and $k_{max}$ (Figs. 3d–g, 4d–g, 5d–g, 6d–g, and 7d–g).

3. *TWI* and *SPI* can take only positive values. To map these indices, we applied a standard spectral tint scale; the minimum and maximum values of *TWI* or *SPI* correspond to violet and red, respectively (Figs. 3i and 3j, 4i and 4j, 5i and 5j, 6i and 6j, 7i and 7j).

4. *TIns* is a nonnegative variable. To map it, we used an orange tint scale: the minimum and maximum *TIns* values correspond to the darkest and lightest orange shades depicting the least and most illuminated areas, respectively (Figs. 3k, 4k, 5k, 6k, and 7k).

5. *WEx* is a positive dimensionless variable, which values below and above 1 relate to wind-shadowed and wind-exposed areas, respectively. To map this index, we applied a two-color tint scale consisting of two contrasting parts, orange and violet (values below and above 1, respectively). The darkest shades of orange and violet colors correspond to the minimum and maximum *WEx* values, respectively, while the lightest shades of the colors correspond to 1 (Figs. 3l, 4l, 5l, 6l, and 7l).

Maps of the Flatvaer (Ongul) Islands (Figs. 2a and 3), Breidvågnipa (Figs. 2c and 4), Skallen Hills and Skallevikhalsen Hills (Figs. 2e and 7) are presented in 1:50,000 scale, while maps of Langhovde Hills (Figs. 2b and 4) and Skarvsnes Foreland (Figs. 2d and 6) are produced in 1:75,000 scale.

There is no lake bathymetry data in REMA: lake cells contain interpolated values of lake coastal elevations, that is, artifacts. Also, REMA has multiple island-like artifacts related to icebergs. On the produced maps, the lakes and island-like artifacts were masked.

To create masks, we used maps of three spectral indices, namely, modified normalized difference water index (MNDWI) (Xu, 2006), ferrous minerals ratio (FMR), and iron oxide ratio (IOR) (Segal, 1982). We derived the spectral index maps from a satellite scene captured on 15.02.2025 by the Sentinel-2A multispectral imager (MSI) (ESA, 2025). To map spectral indices, we used composite legends allowing for the classification of objects by moisture level and iron content. For example, MNDWI maps (Fig. A1) display ice and snow in blue, open water bodies and light snow in green, while ice-free rocks in gray. FMR maps (Fig. A2) highlight rocks in red, orange, and yellow. IOR maps (Fig. A3) show open water bodies in gray, ice and snow in green, while rock areas in red, orange, and yellow.

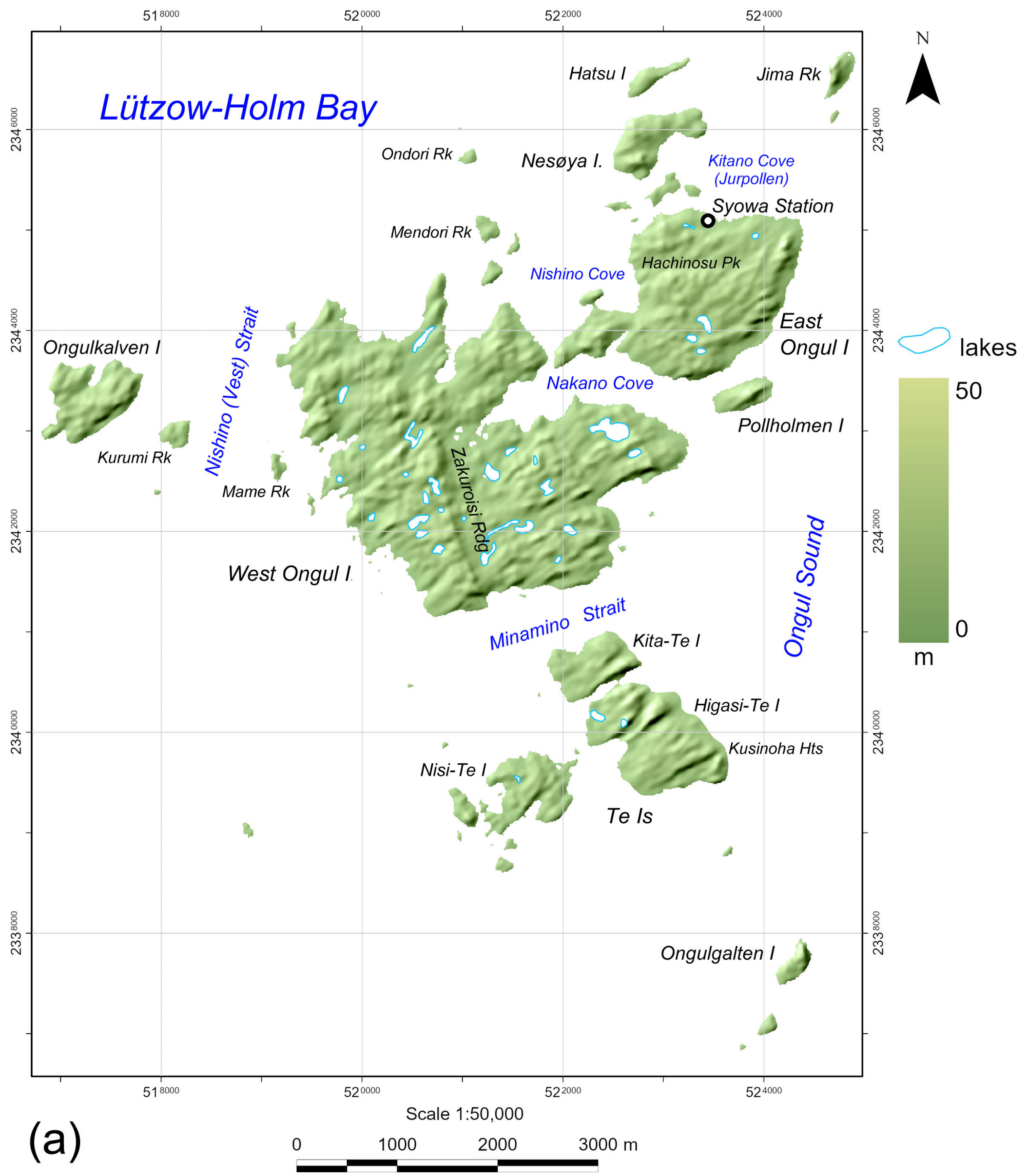

**Fig. 3** Flatvaer (Ongul) Islands: (a) Elevation.

*(Continued)*

For DEM processing and geomorphometric calculations, we used a software SAGA 9.8.1 (Conrad et al., 2015). For morphometric mapping, Sentinel-2A MSI data processing, spectral index mapping, and map masking, we utilized ArcGIS Pro 3.0.1 (ESRI, 2015–2024).

## 4 Results and discussion

Geomorphometric modeling and mapping resulted in five series of large-scale morphometric maps for the key ice-free areas of the Sôya Coast including the Flatvaer (Ongul) Islands (Fig. 3), Langhovde Hills (Fig. 4), Breidvågnipa (Fig. 5), Skarvsnes Foreland (Fig. 6), as well as Skallen Hills and Skallevikhalsen Hills (Fig. 7).

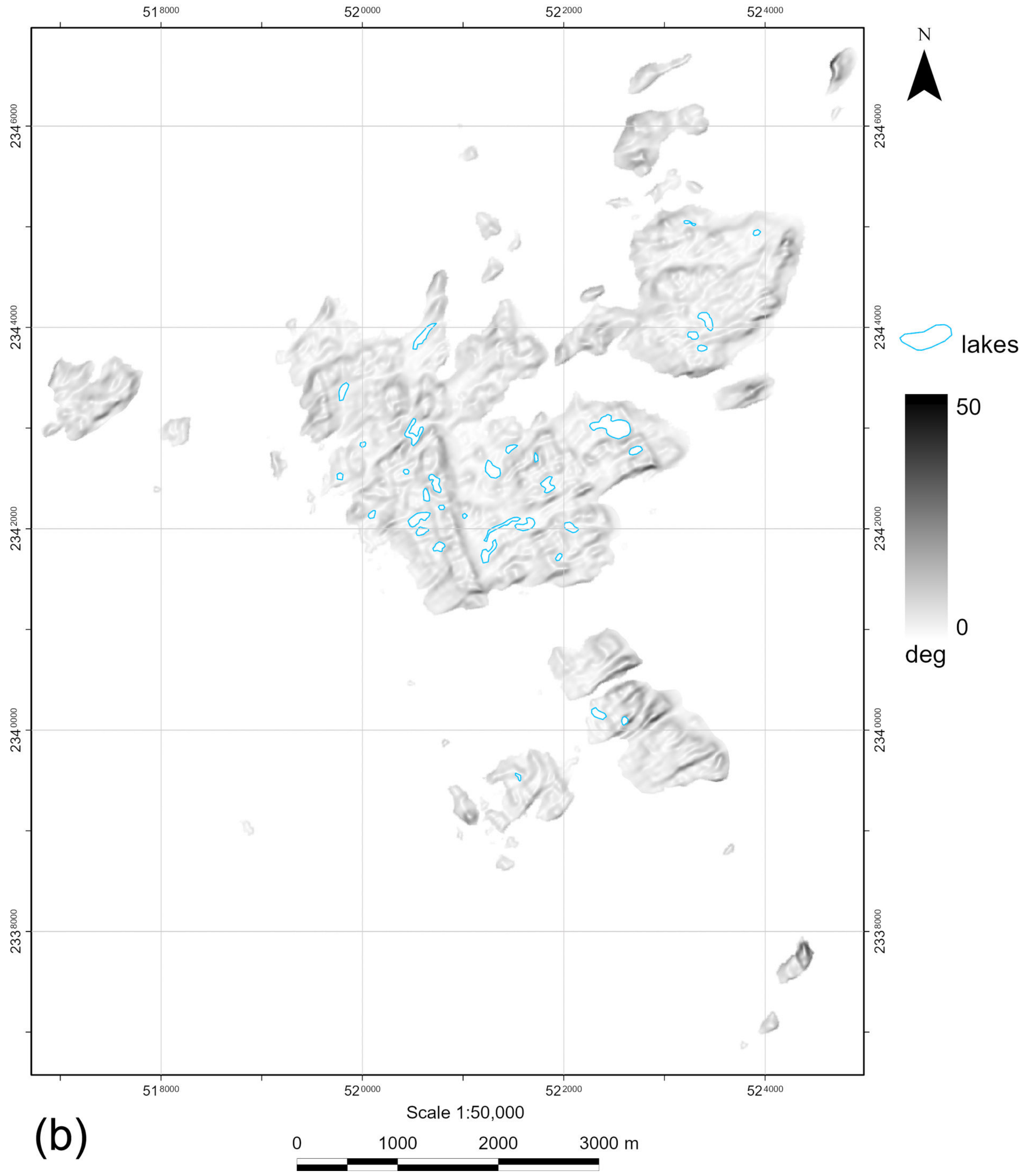

**Fig. 3, cont'd** Flatvaer (Ongul) Islands: (b) Slope.

*(Continued)*

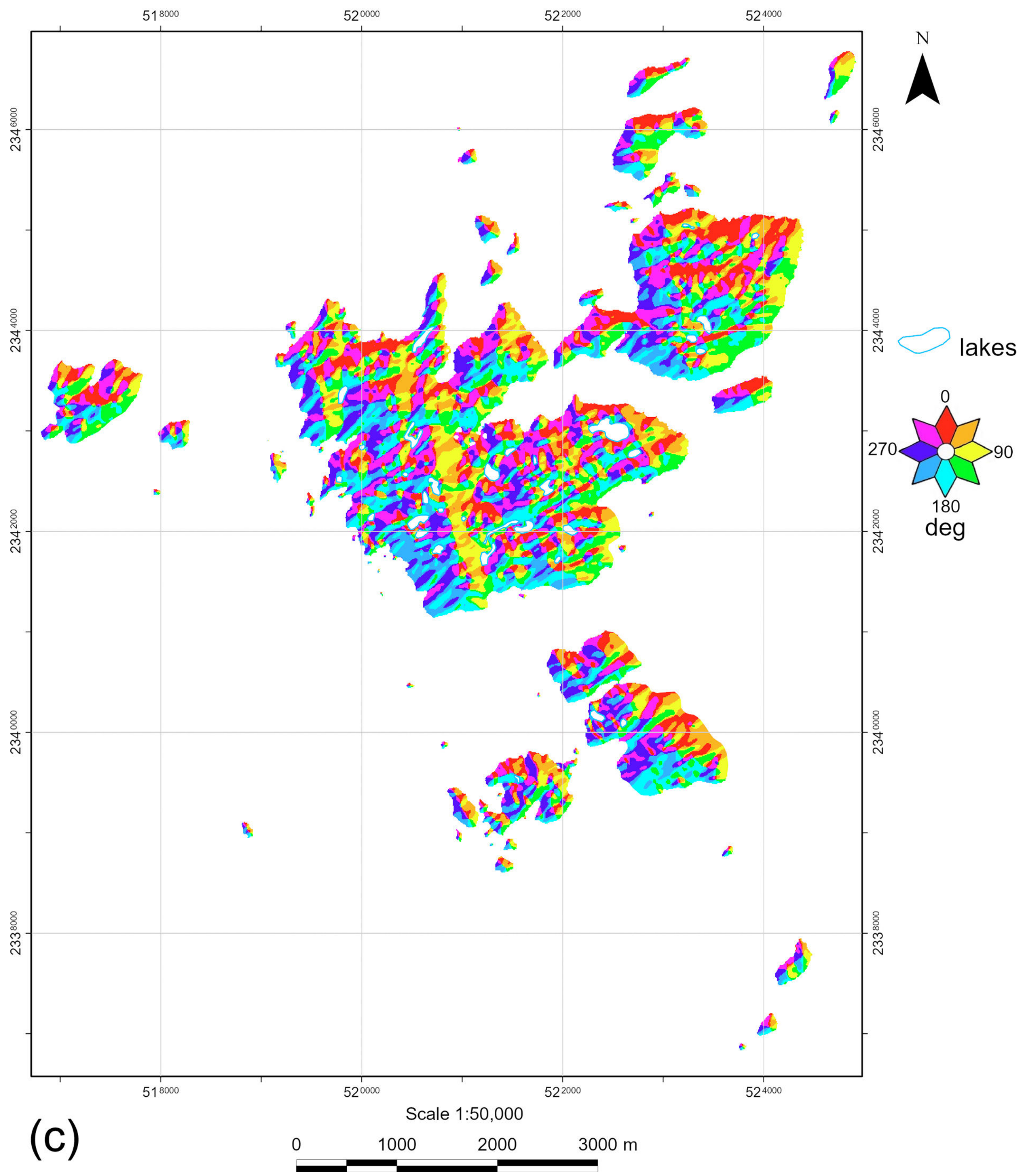

**Fig. 3, cont'd** Flatvaer (Ongul) Islands: (c) Aspect.

*(Continued)*

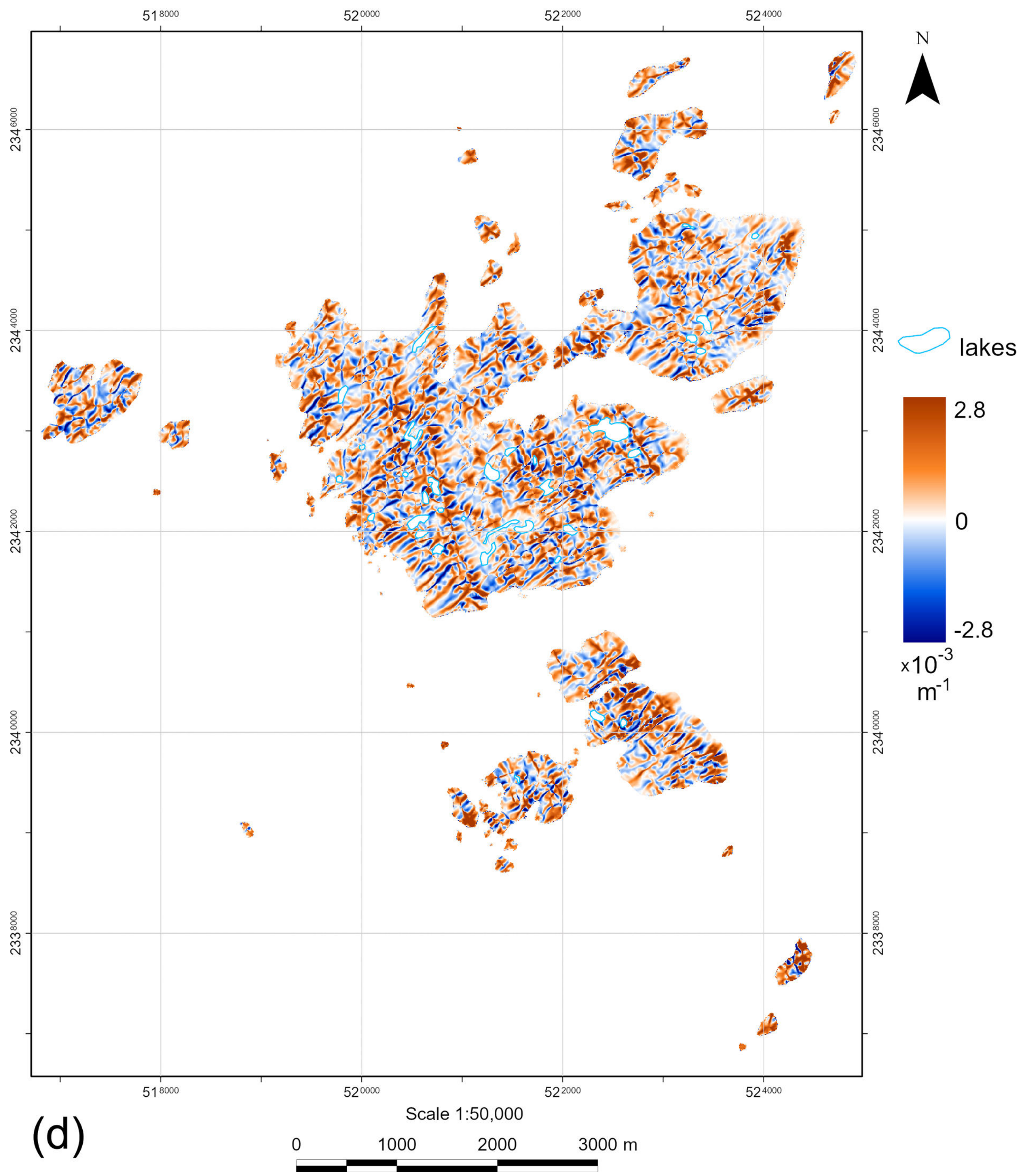

**Fig. 3, cont'd** Flatvaer (Ongul) Islands: (d) Horizontal curvature.

*(Continued)*

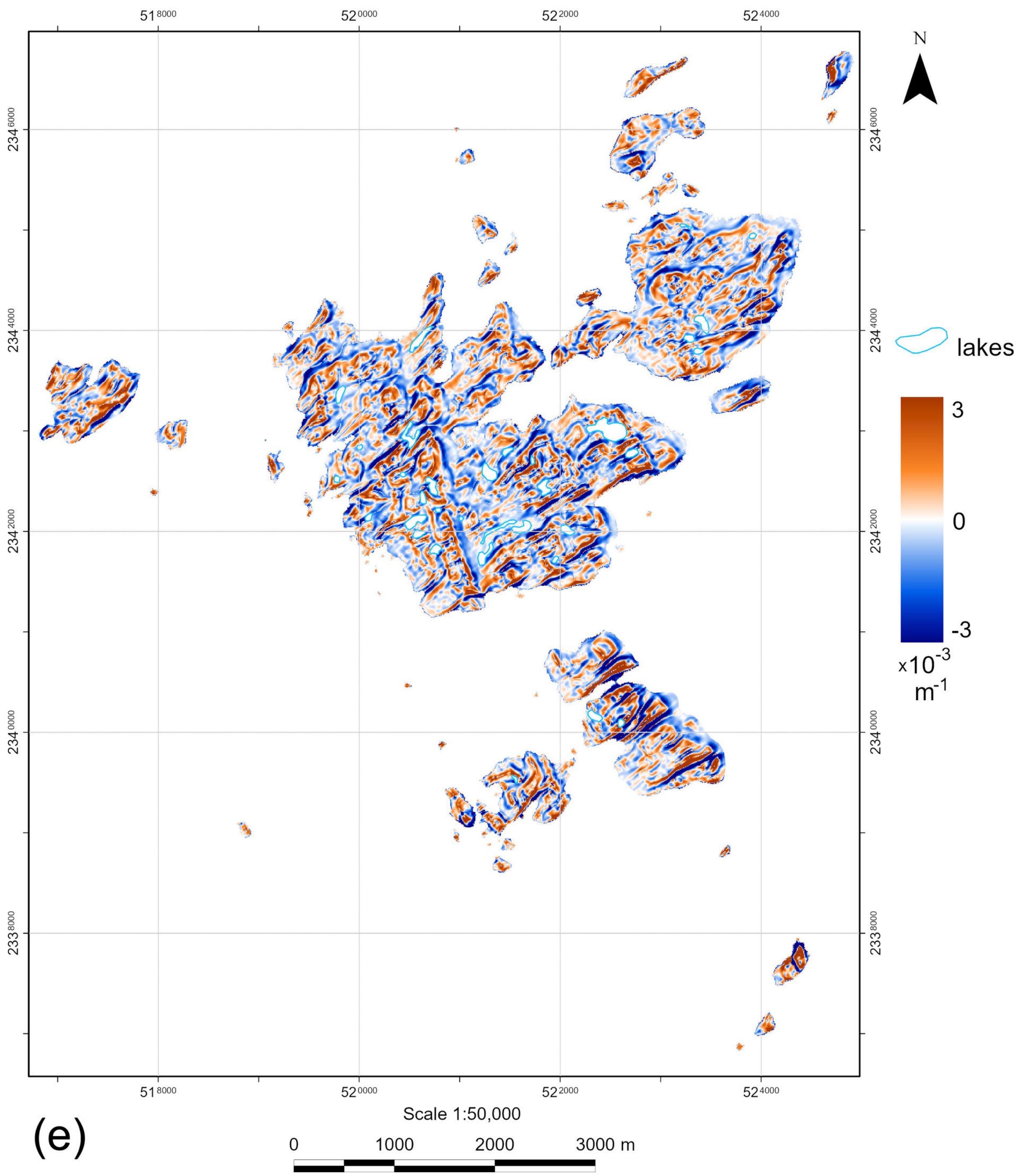

**Fig. 3, cont'd** Flatvaer (Ongul) Islands: (e) Vertical curvature.

*(Continued)*

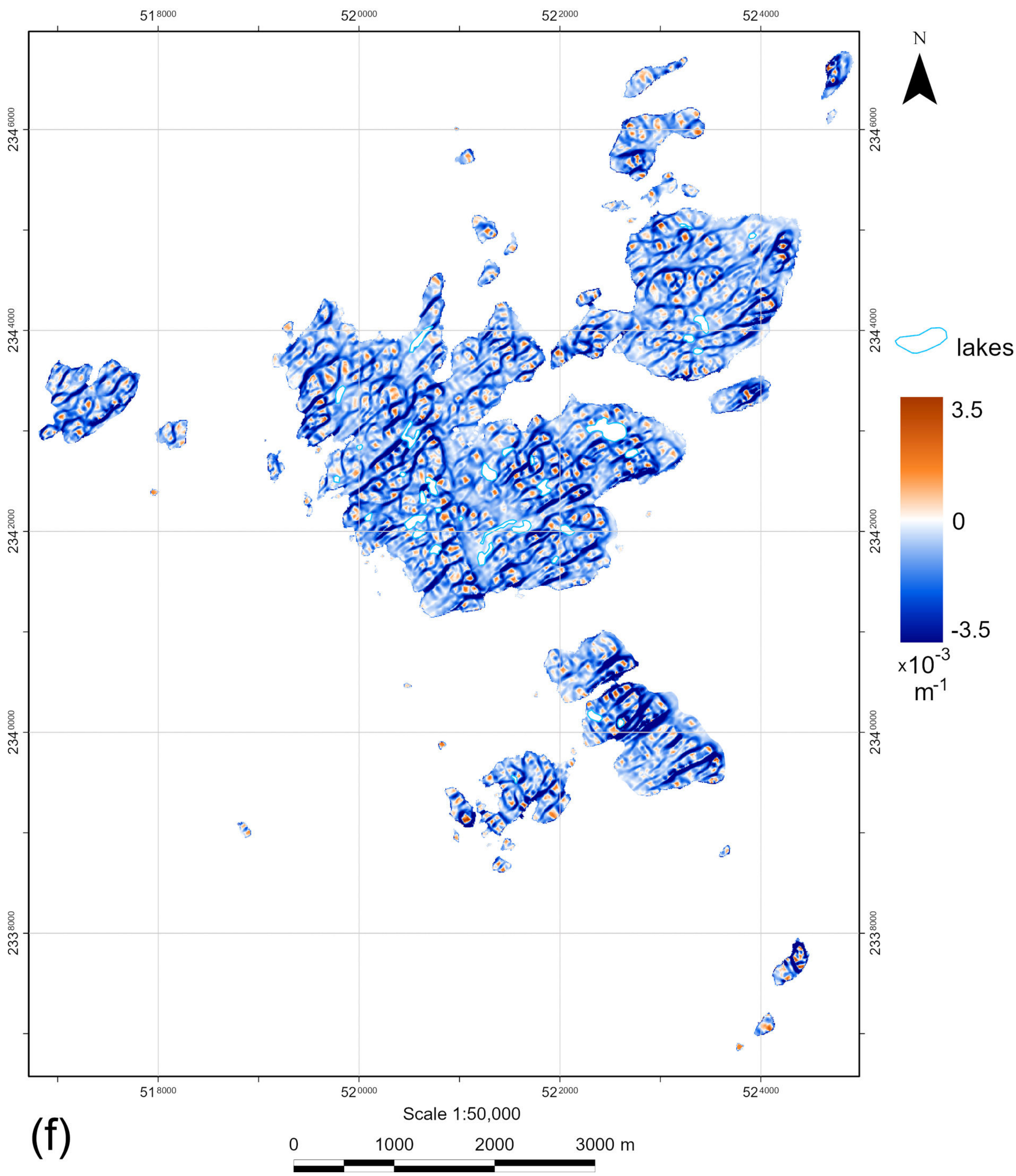

**Fig. 3, cont'd** Flatvaer (Ongul) Islands: (f) Minimal curvature.

*(Continued)*

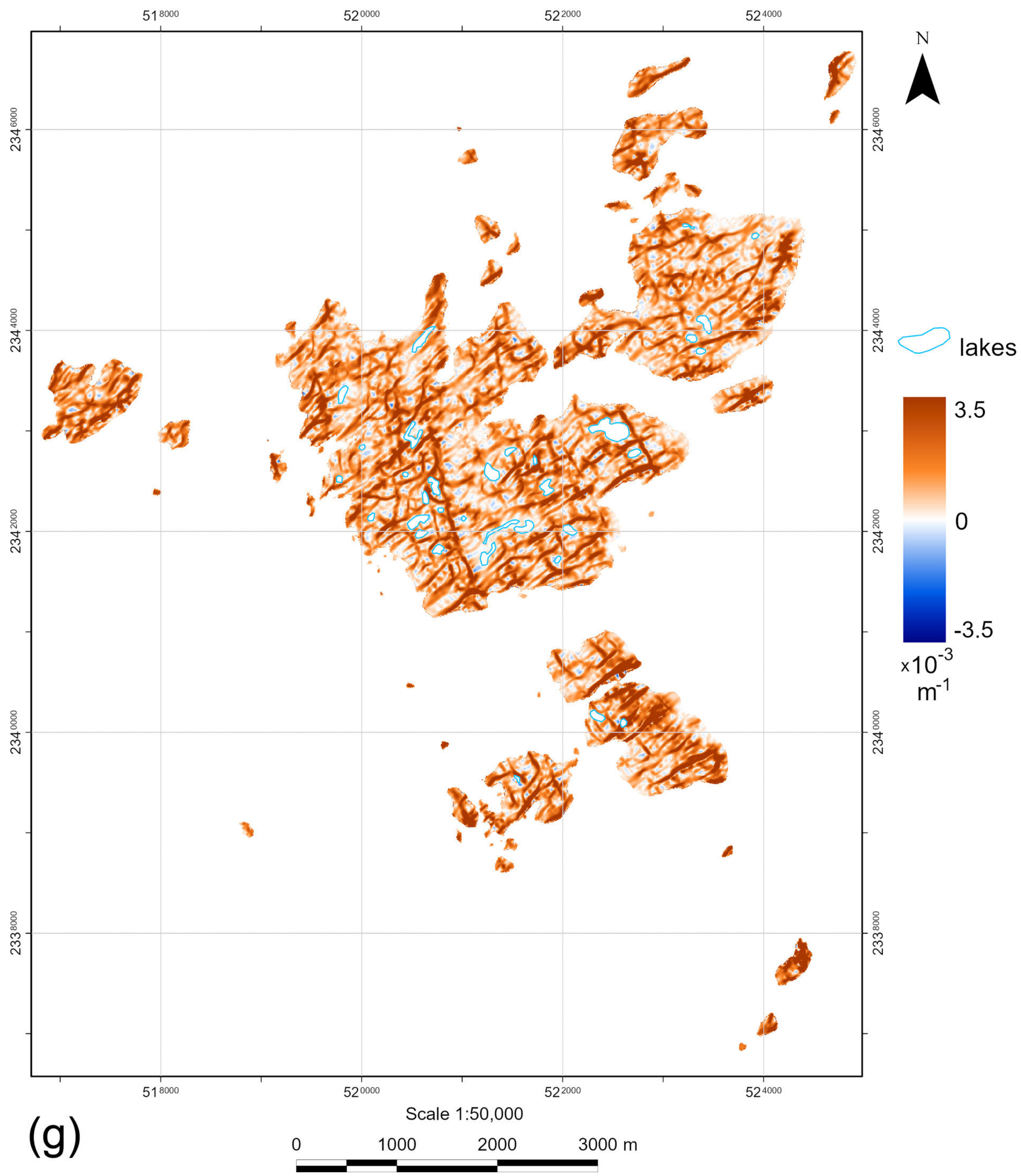

**Fig. 3, cont'd** Flatvaer (Ongul) Islands: (g) Maximal curvature.

*(Continued)*

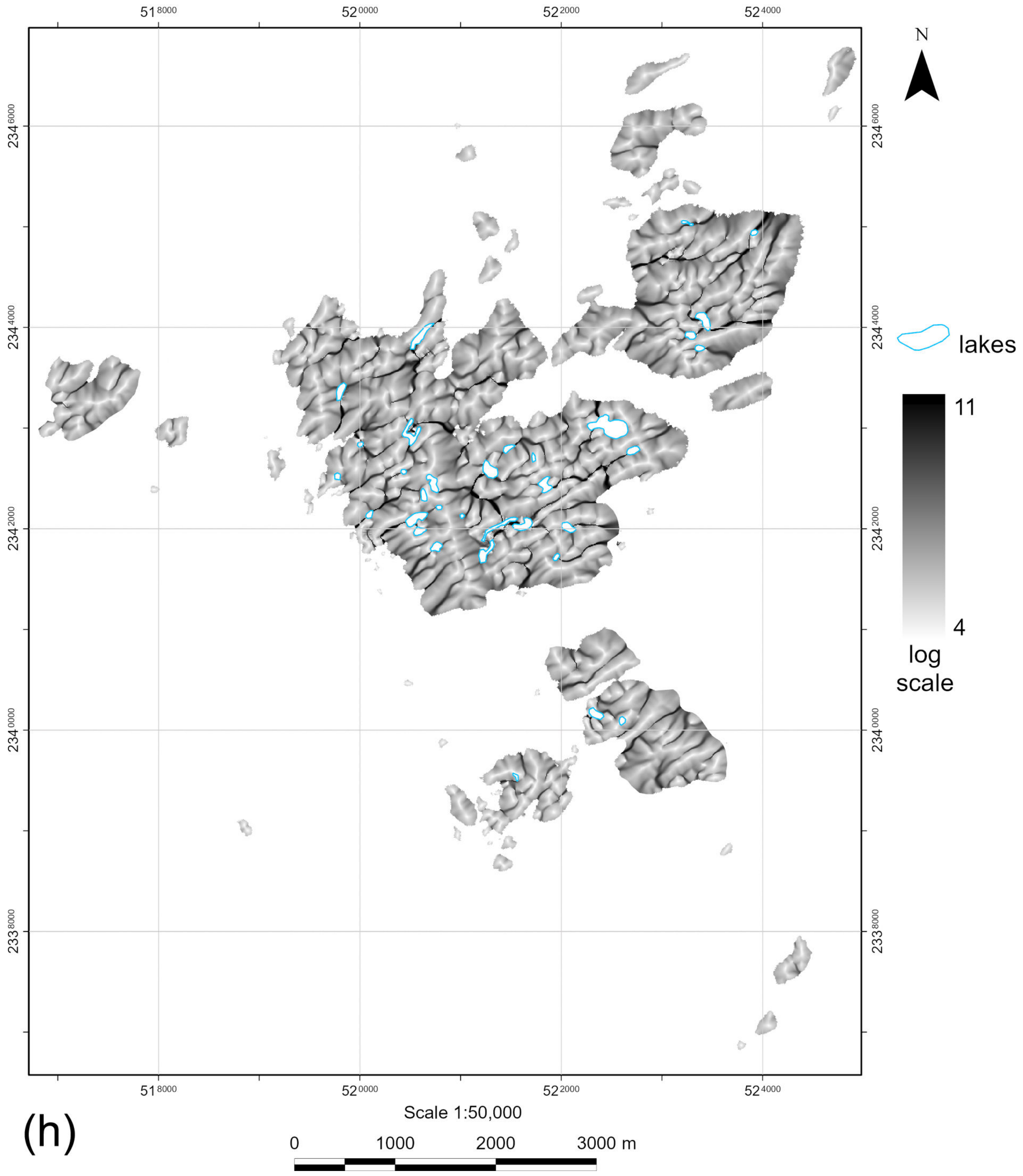

**Fig. 3, cont'd** Flatvaer (Ongul) Islands: (h) Catchment area.

*(Continued)*

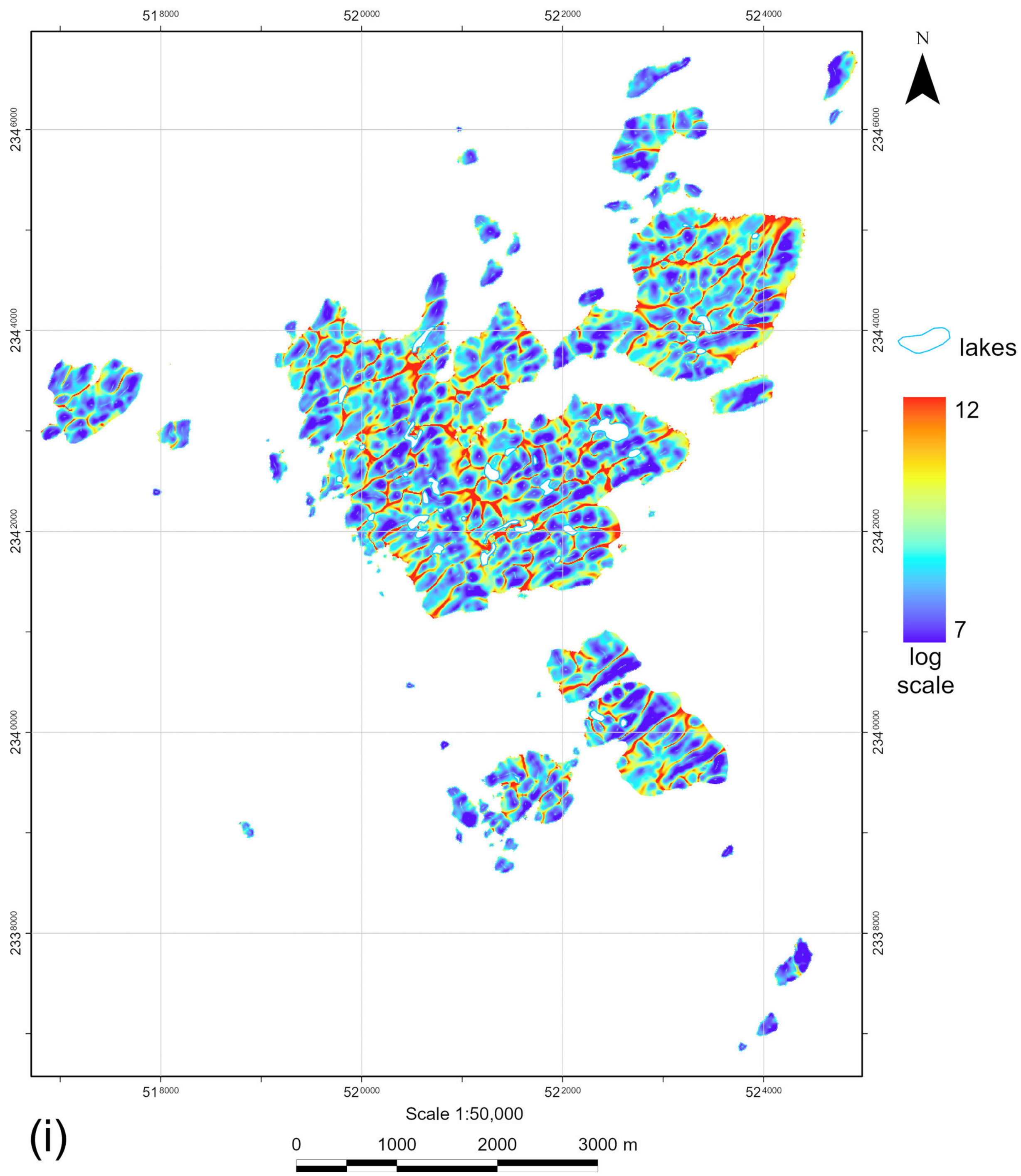

**Fig. 3, cont'd** Flatvaer (Ongul) Islands: (i) Topographic wetness index.

*(Continued)*

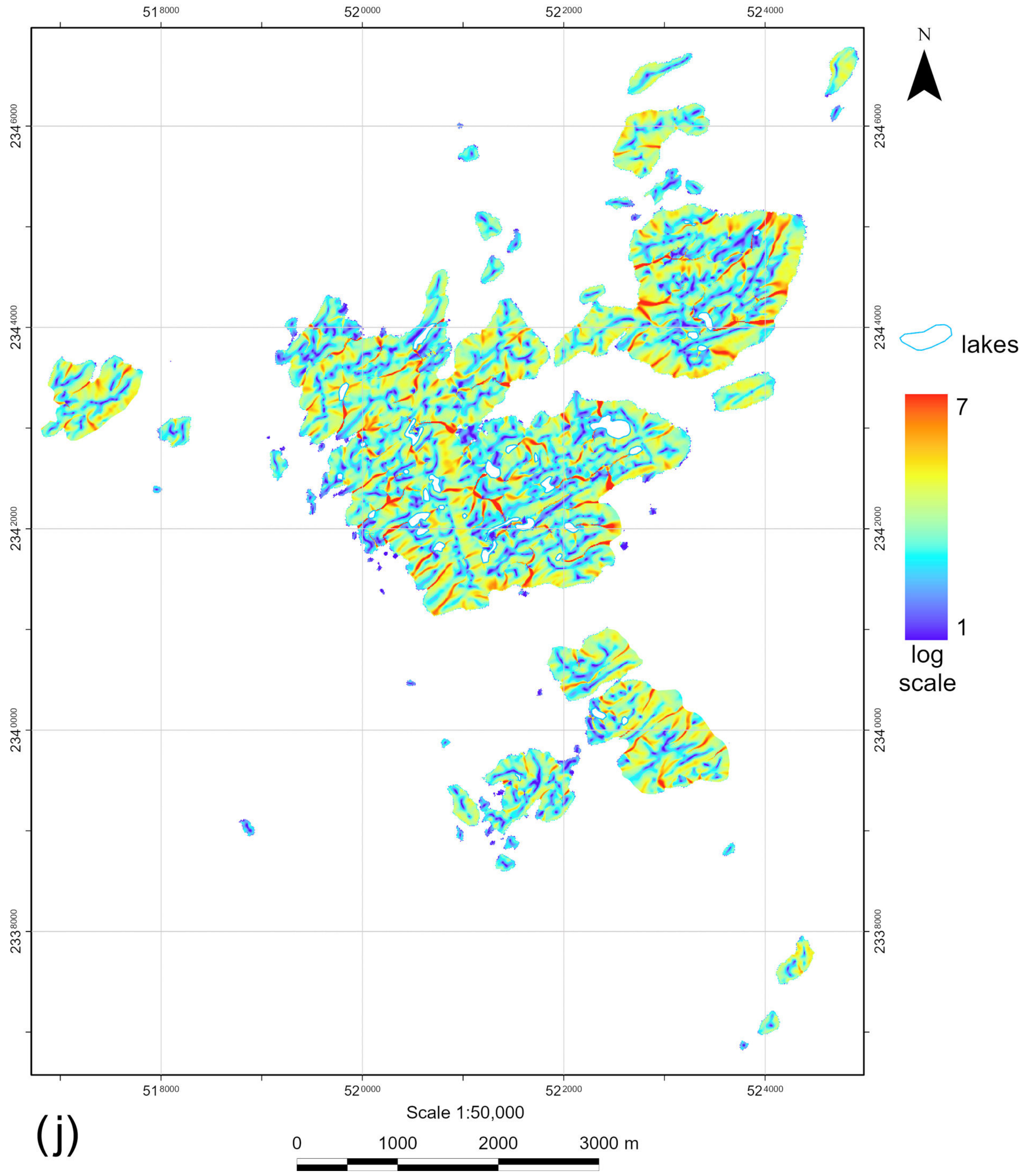

**Fig. 3, cont'd** Flatvaer (Ongul) Islands: ( j) Stream power index.

*(Continued)*

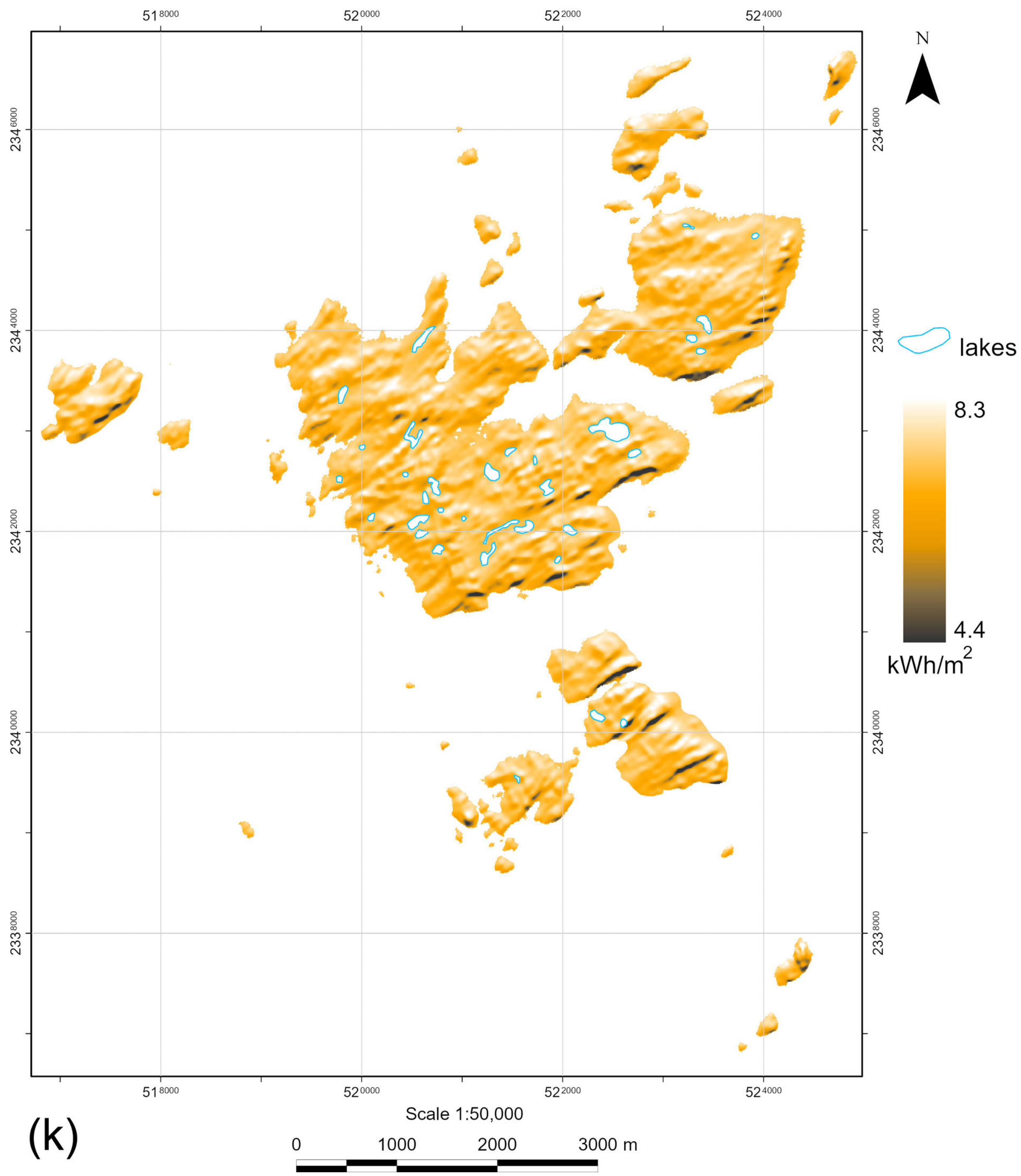

**Fig. 3, cont'd** Flatvaer (Ongul) Islands: (k) Total insolation.

*(Continued)*

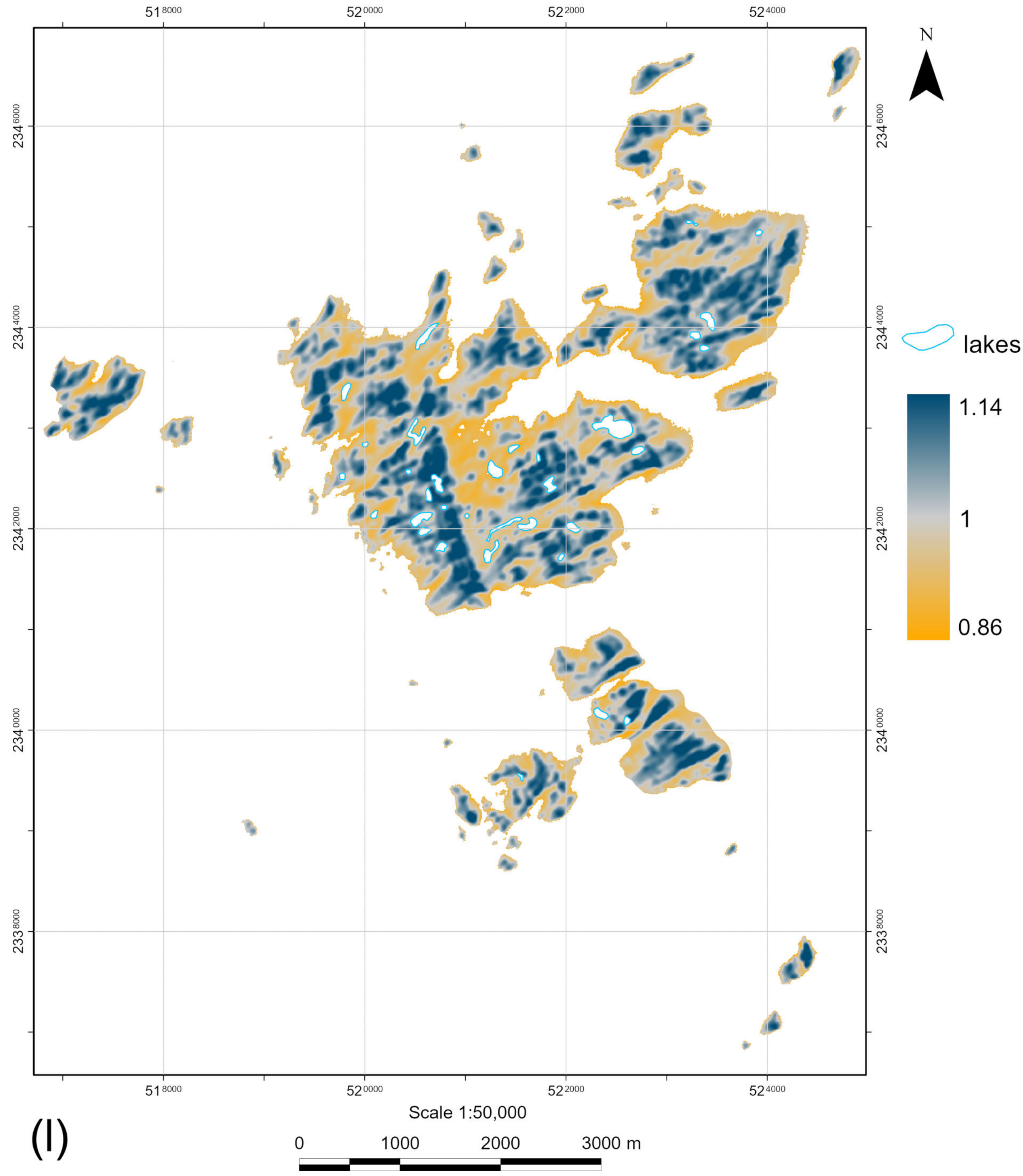

**Fig. 3, cont'd** Flatvaer (Ongul) Islands: (l) Wind exposition index.

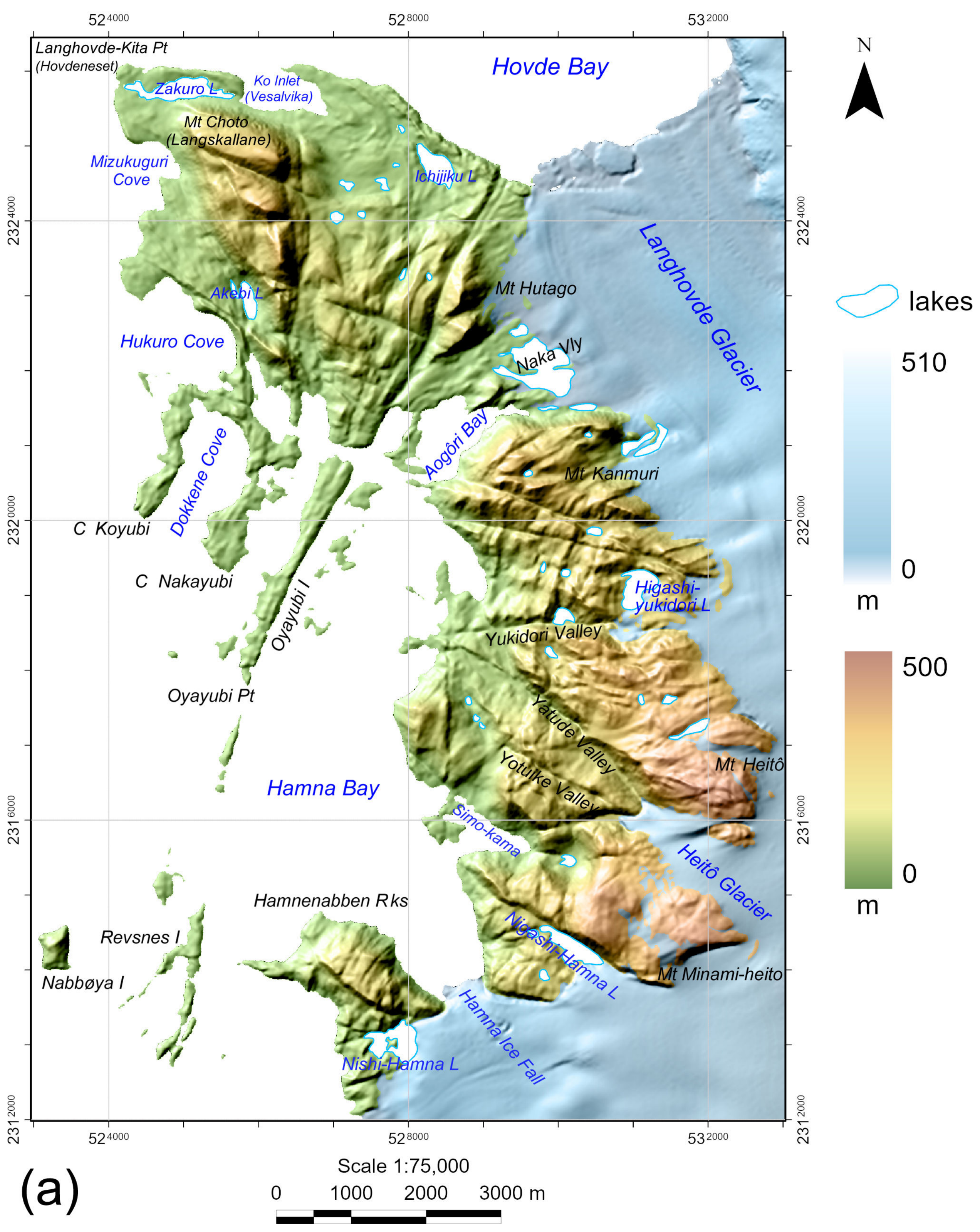

**Fig. 4** Langhovde Hills: (a) Elevation.

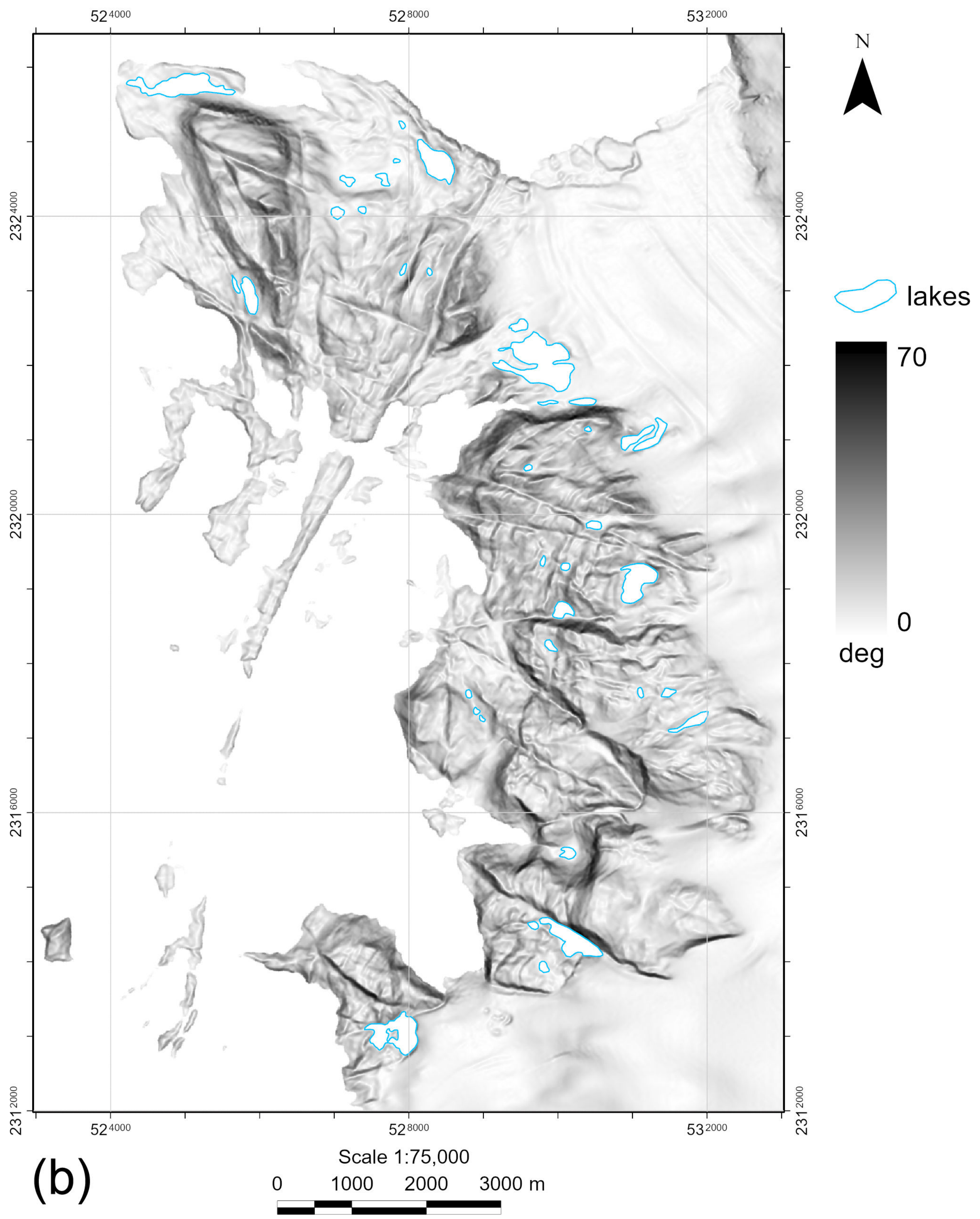

**Fig. 4, cont'd** Langhovde Hills: (b) Slope.

*(Continued)*

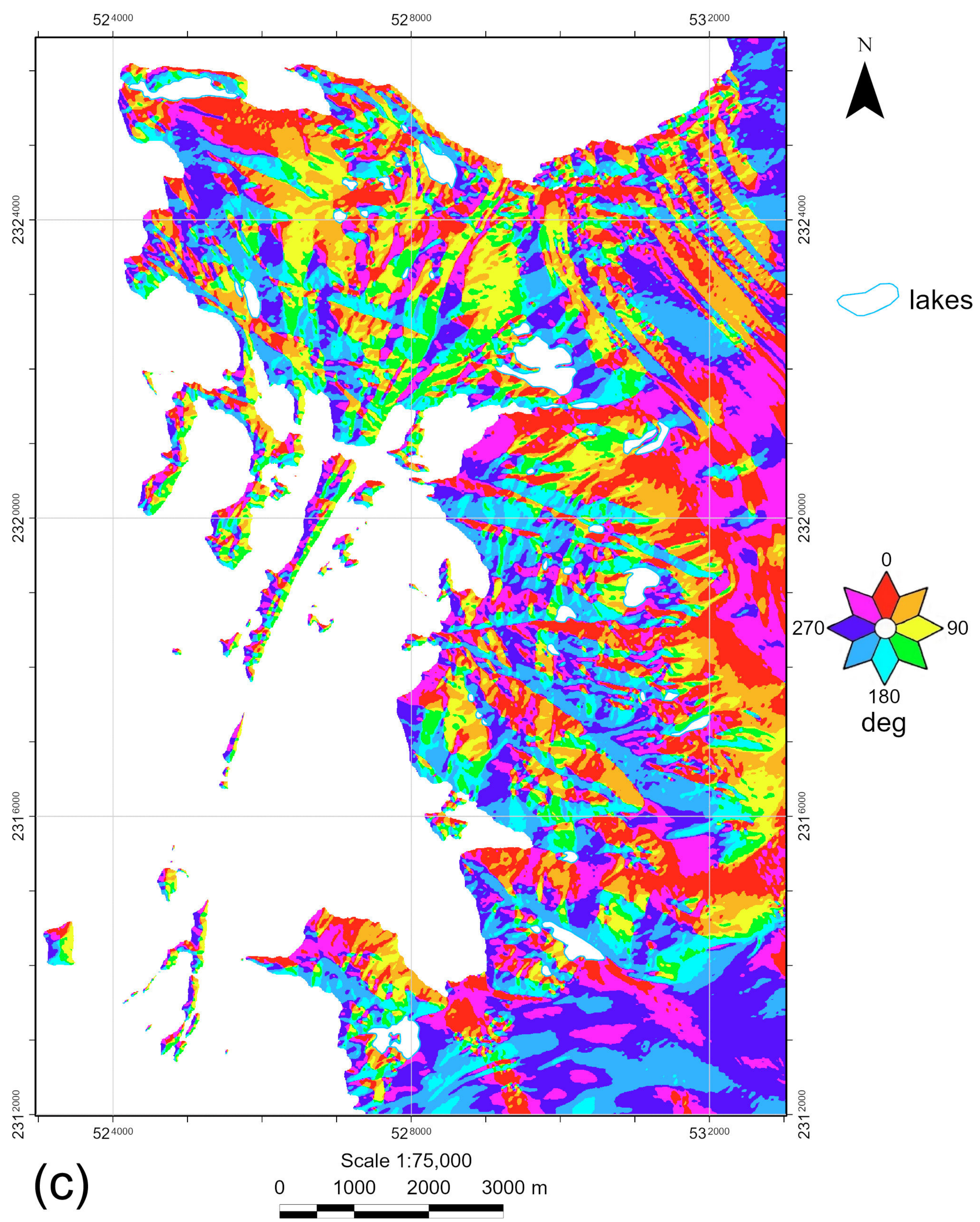

**Fig. 4, cont'd** Langhovde Hills: (c) Aspect.

*(Continued)*

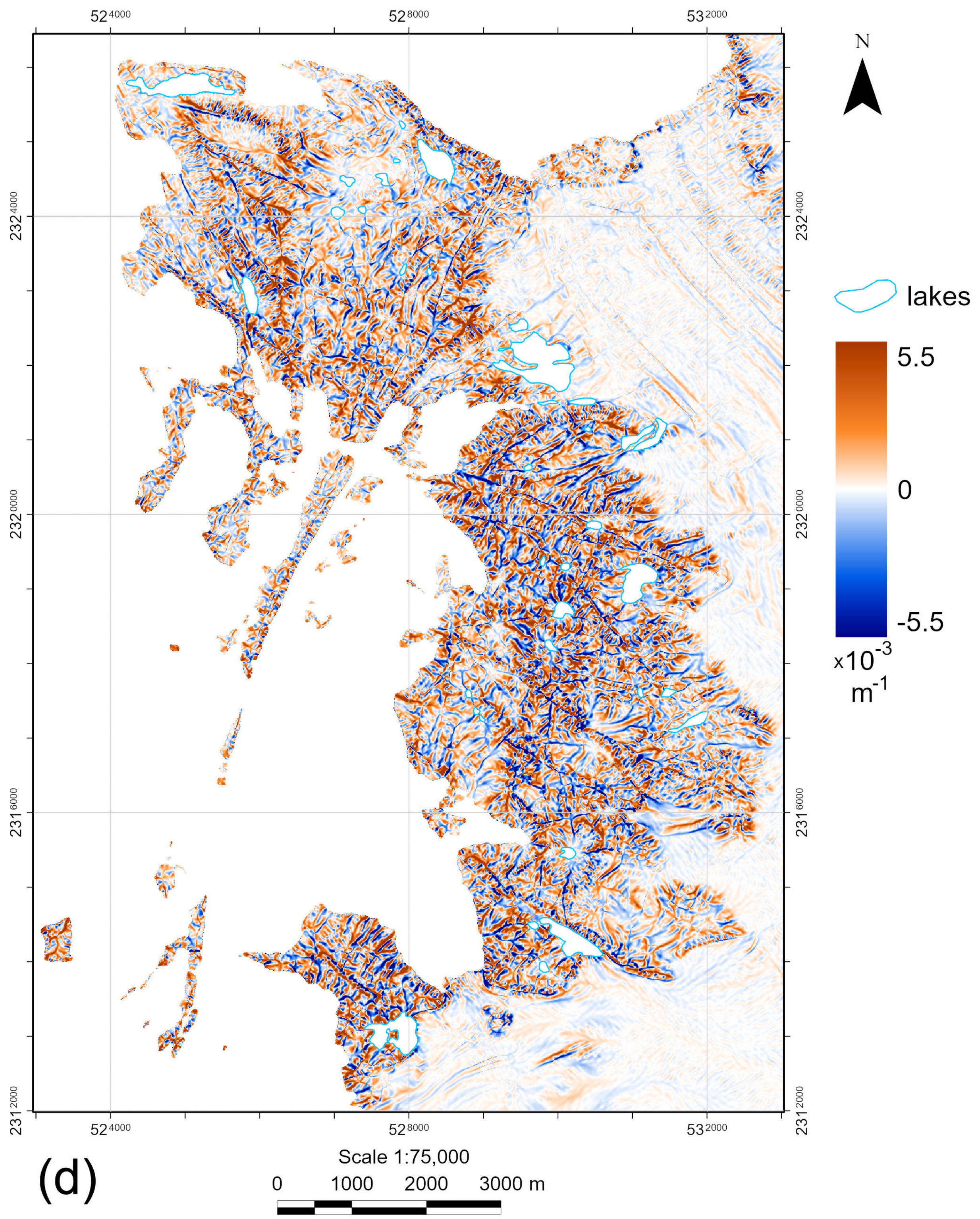

**Fig. 4, cont'd** Langhovde Hills: (d) Horizontal curvature.

*(Continued)*

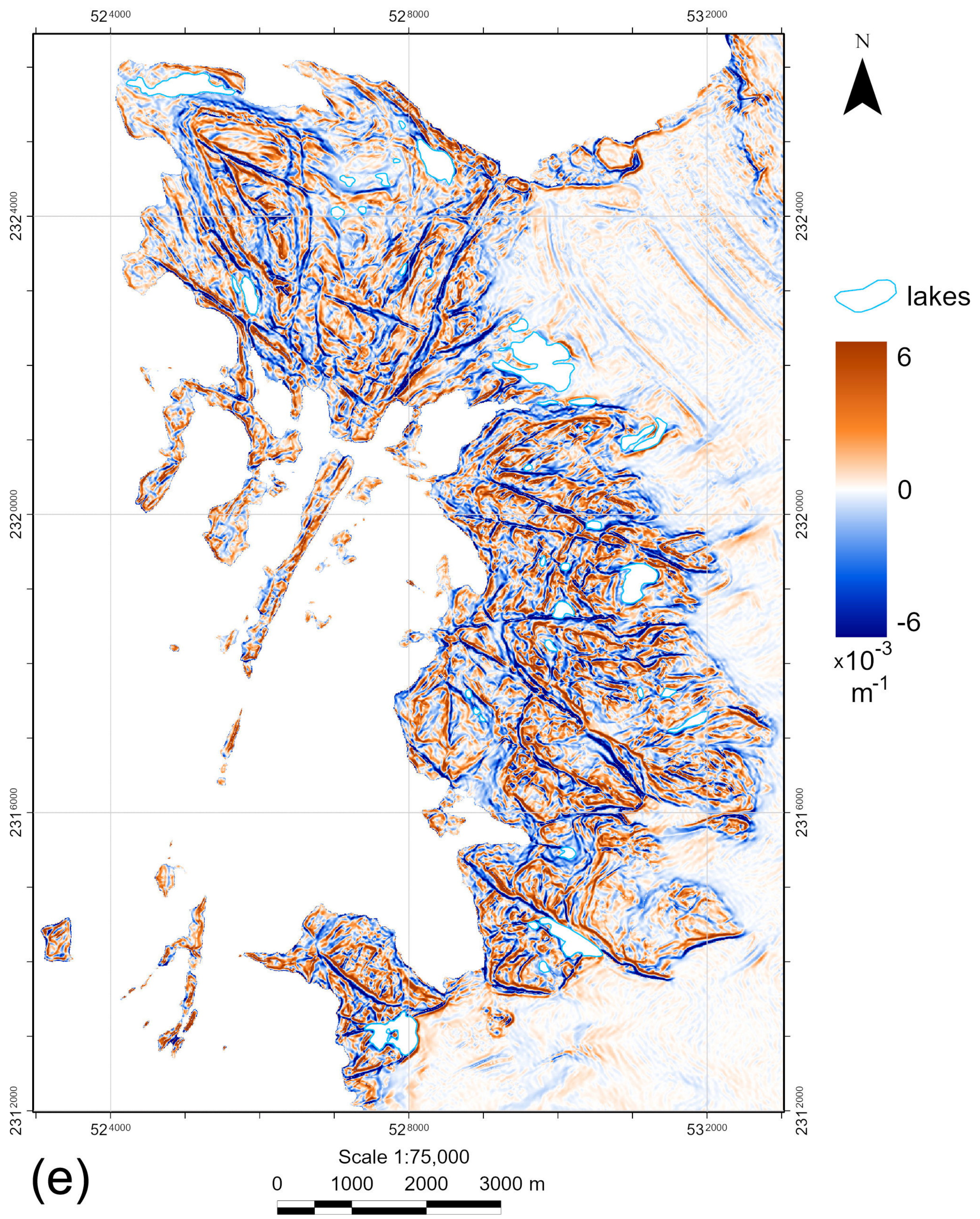

**Fig. 4, cont'd** Langhovde Hills: (e) Vertical curvature.

*(Continued)*

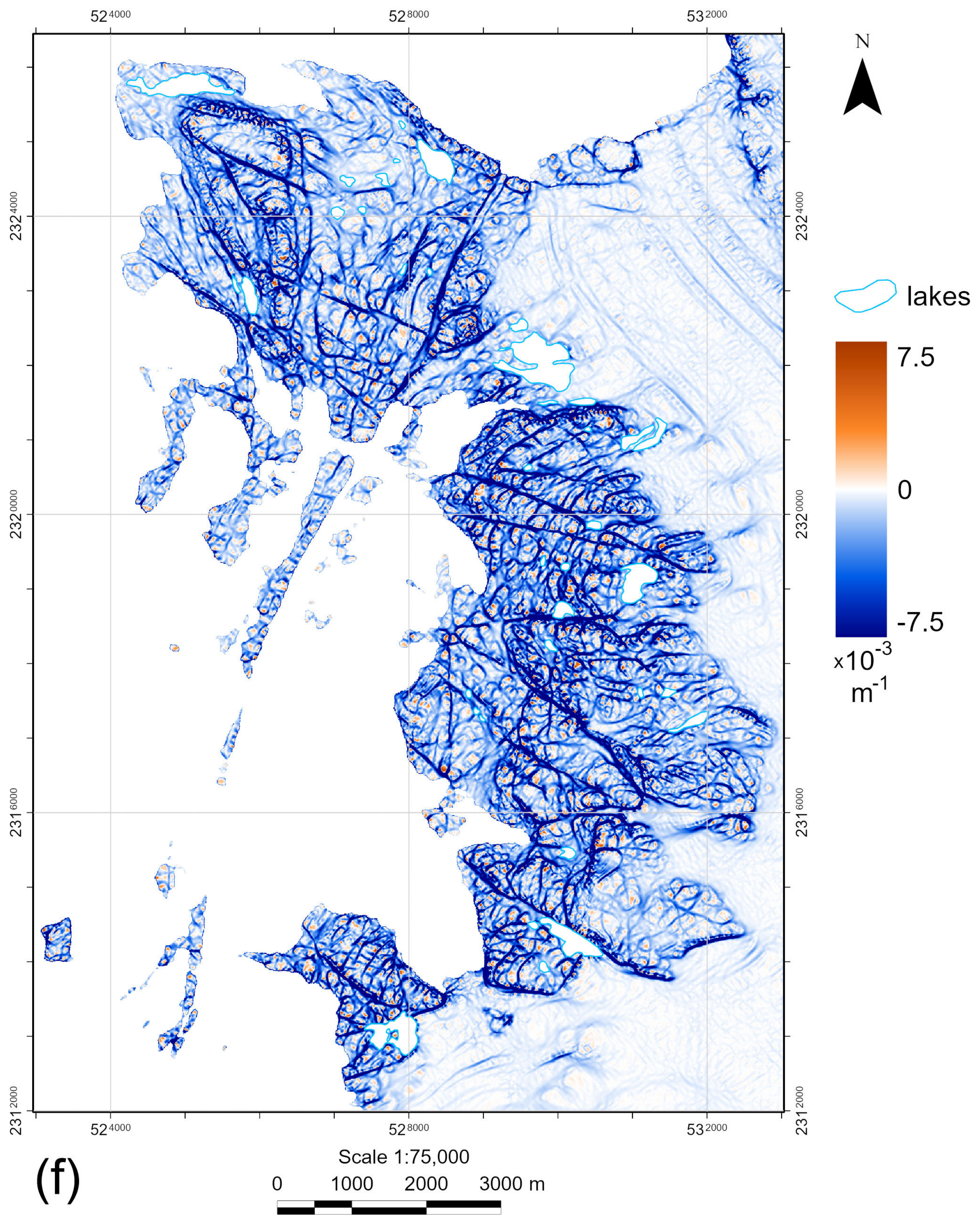

**Fig. 4, cont'd** Langhovde Hills: (f) Minimal curvature.

*(Continued)*

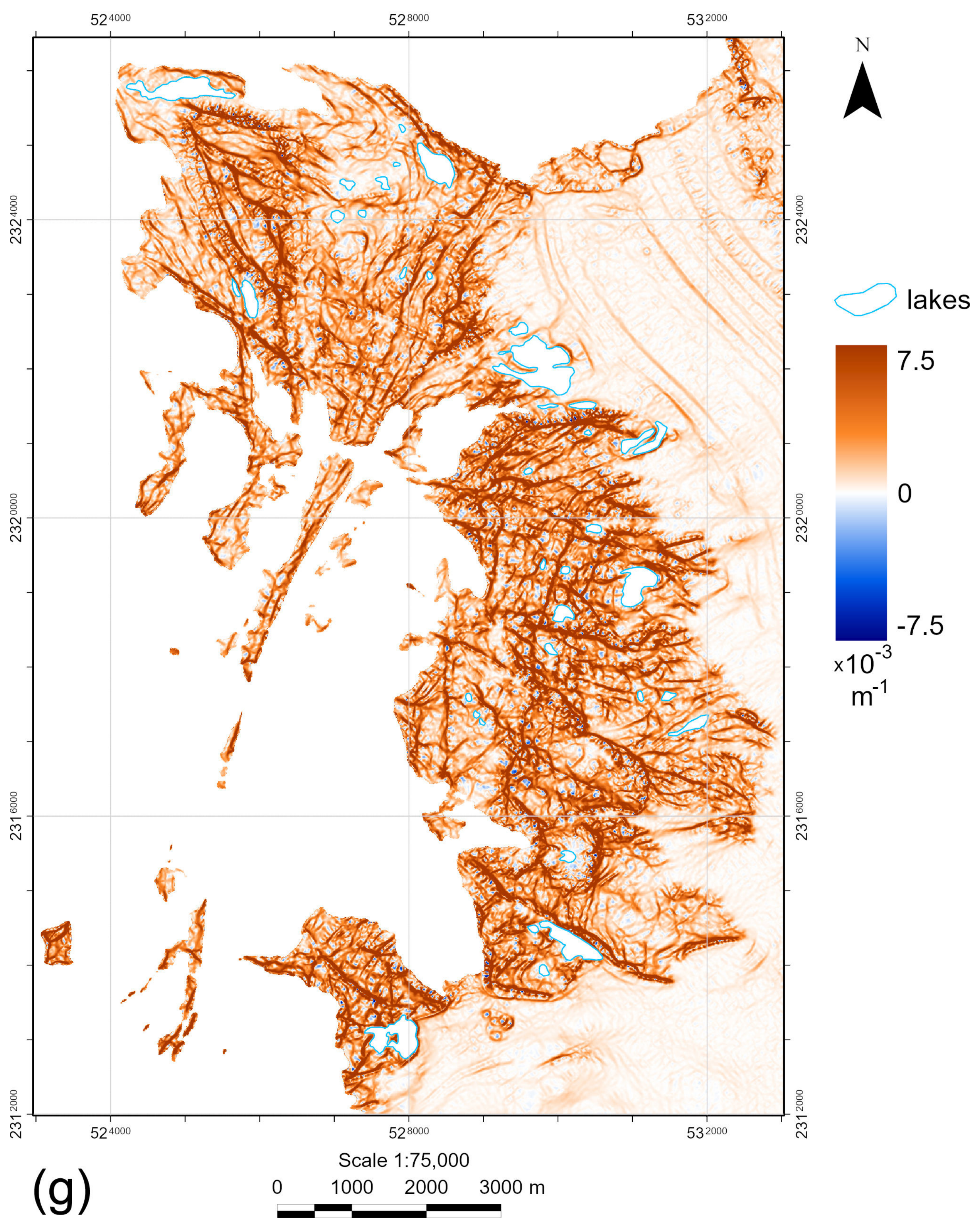

**Fig. 4, cont'd** Langhovde Hills: (g) Maximal curvature.

*(Continued)*

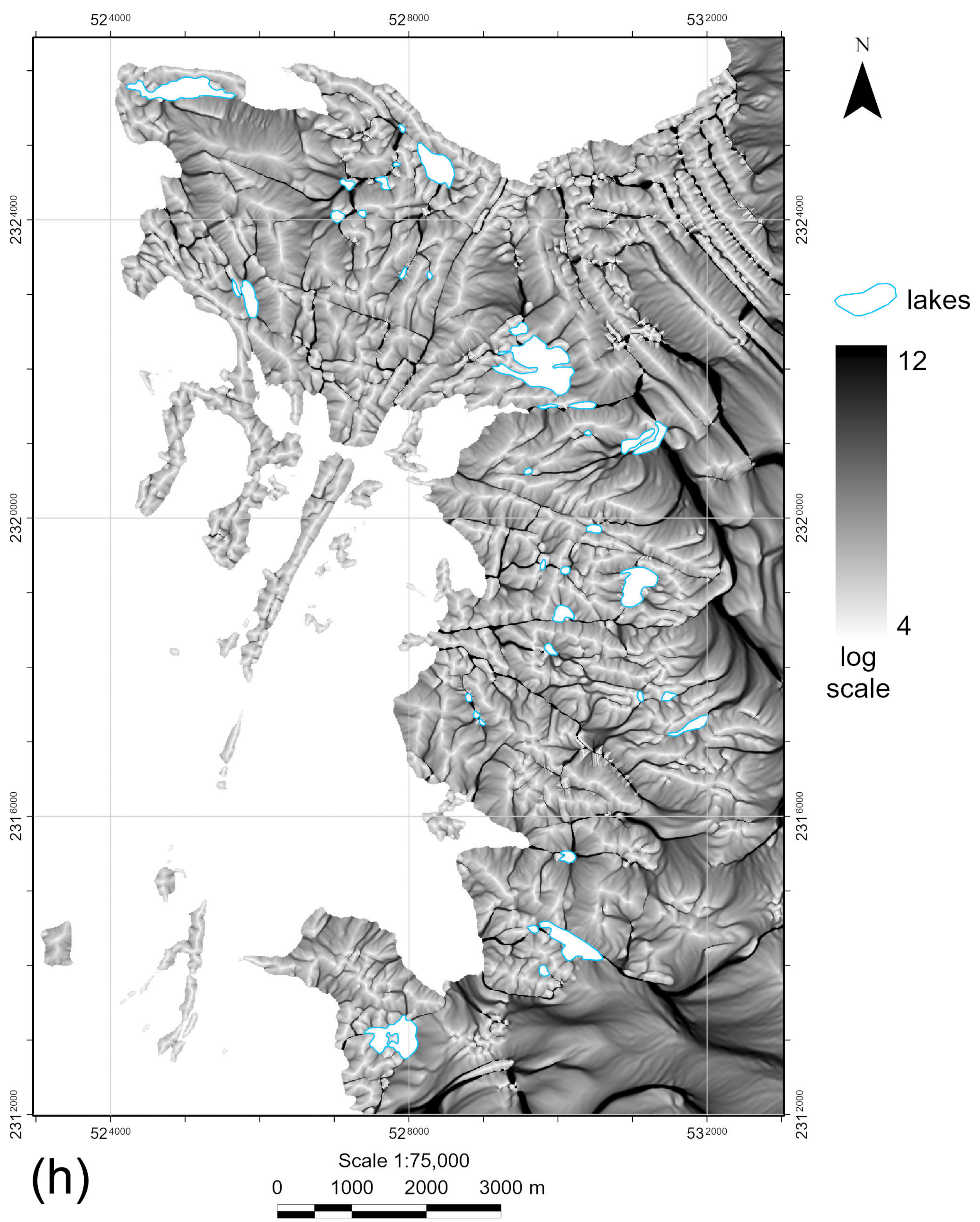

**Fig. 4, cont'd** Langhovde Hills: (h) Catchment area.

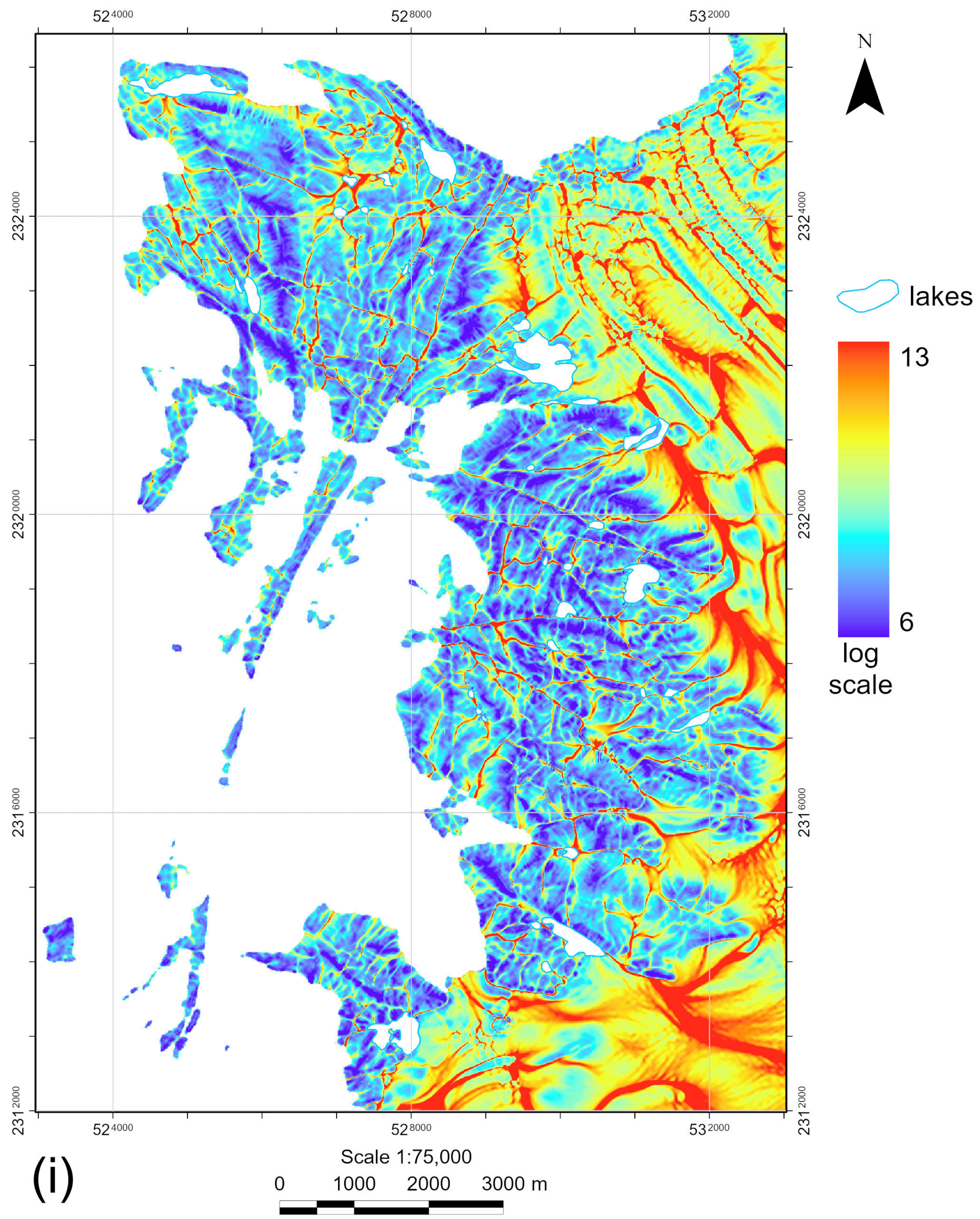

**Fig. 4, cont'd** Langhovde Hills: (i) Topographic wetness index.

*(Continued)*

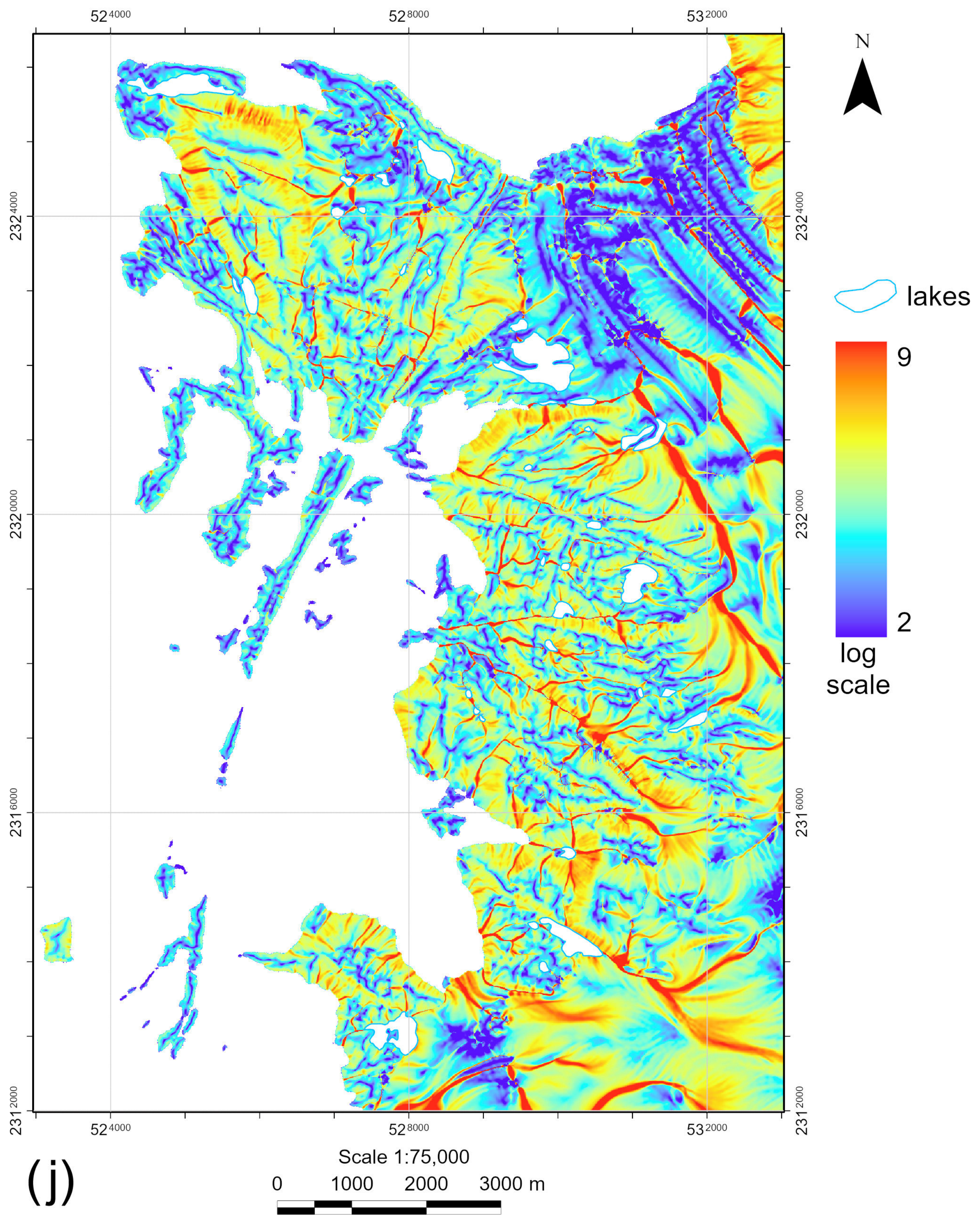

**Fig. 4, cont'd** Langhovde Hills: ( j) Stream power index.

*(Continued)*

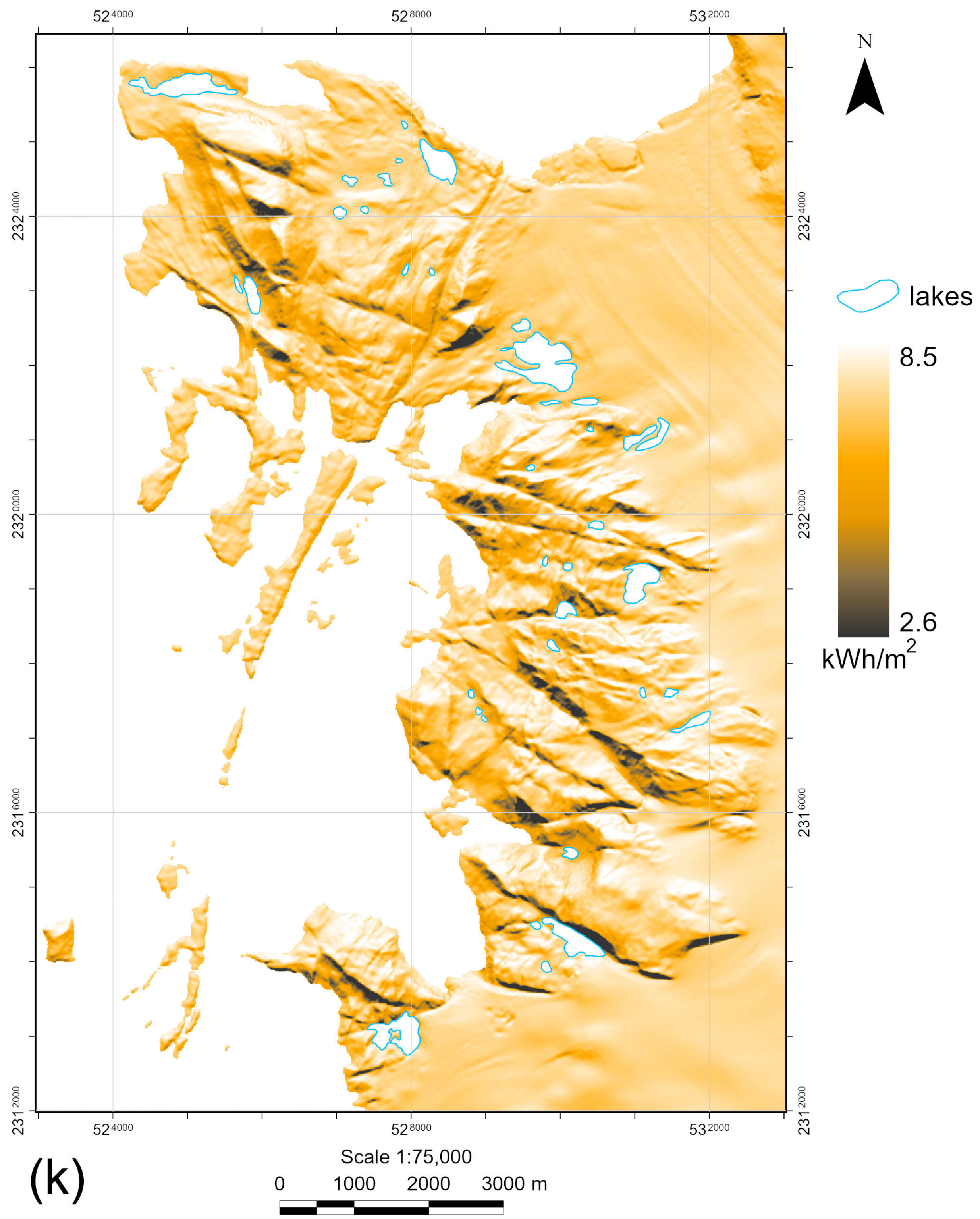

**Fig. 4, cont'd** Langhovde Hills: (k) Total insolation.

*(Continued)*

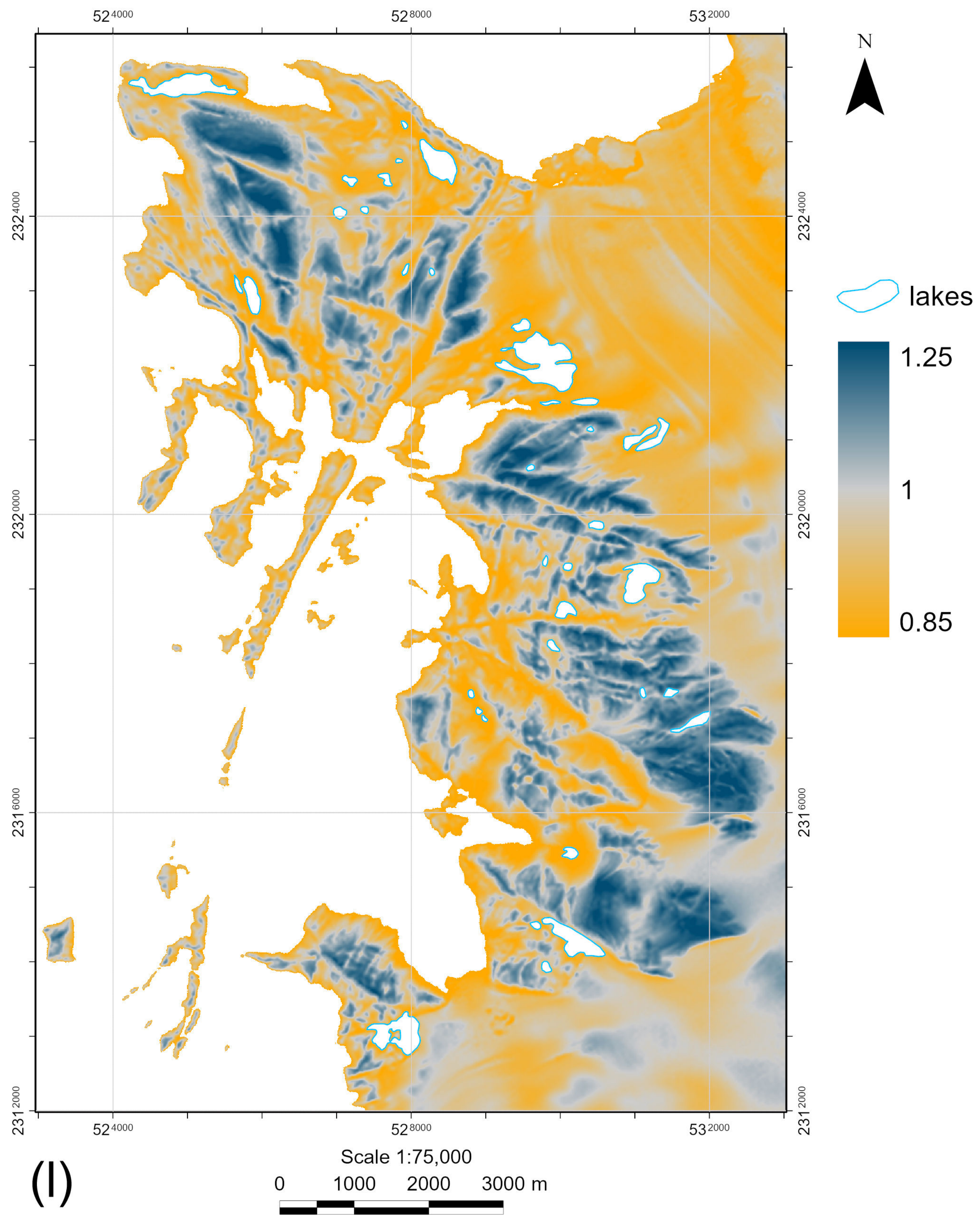

**Fig. 4, cont'd** Langhovde Hills: (l) Wind exposition index.

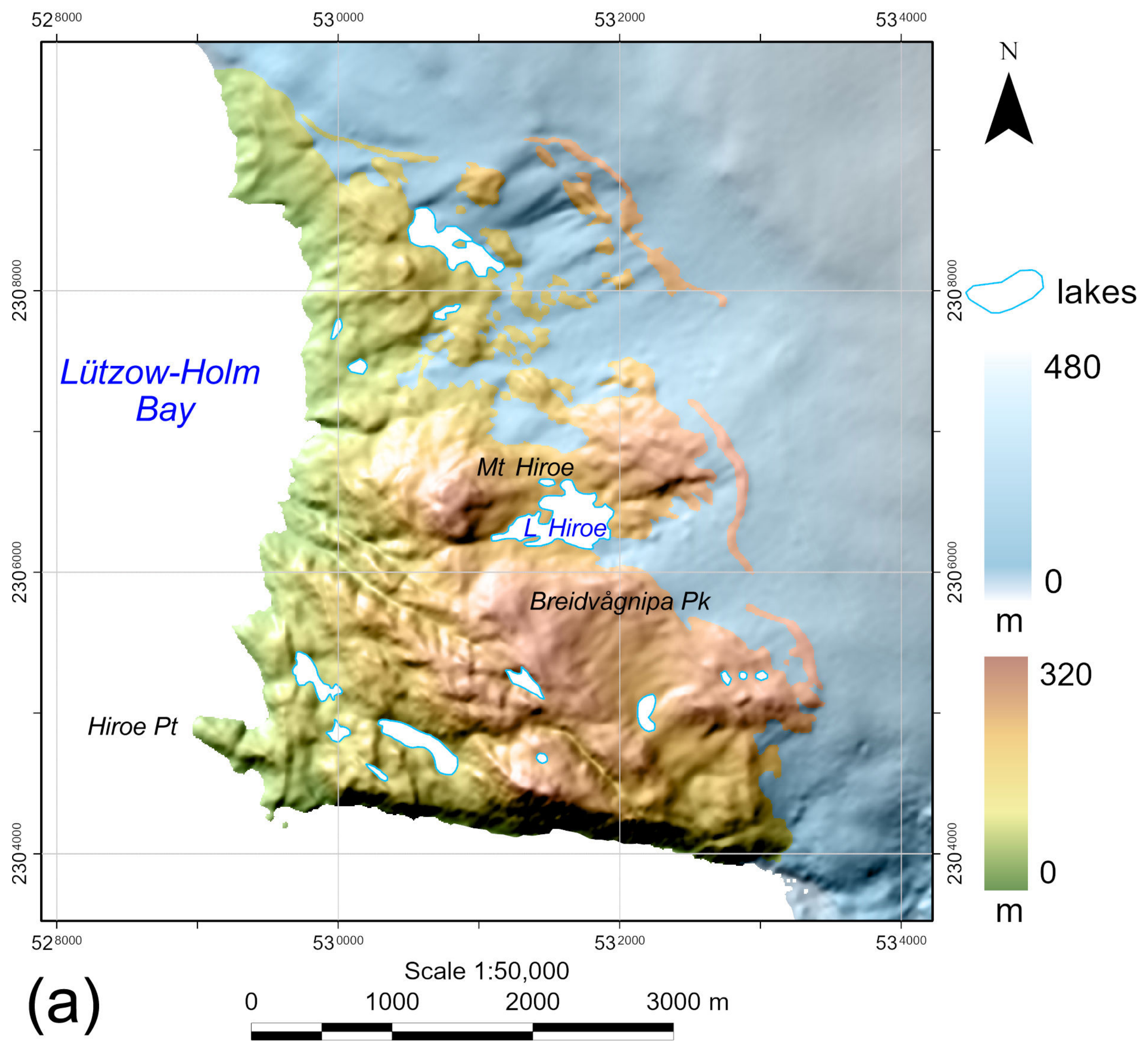

**Fig. 5** Breidvågnipa: (a) Elevation.

*(Continued)*

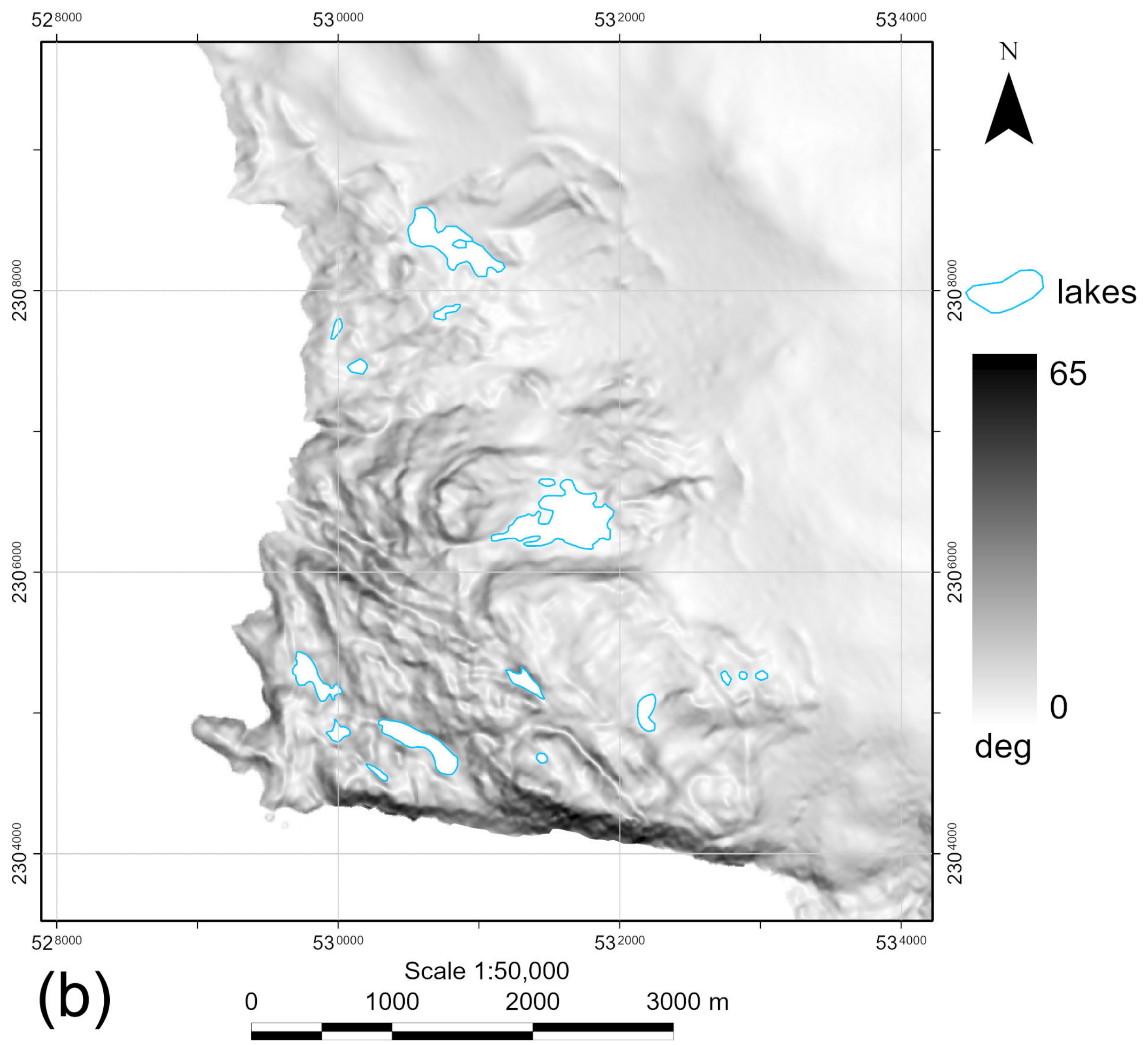

**Fig. 5, cont'd** Breidvågnipa: (b) Slope.

*(Continued)*

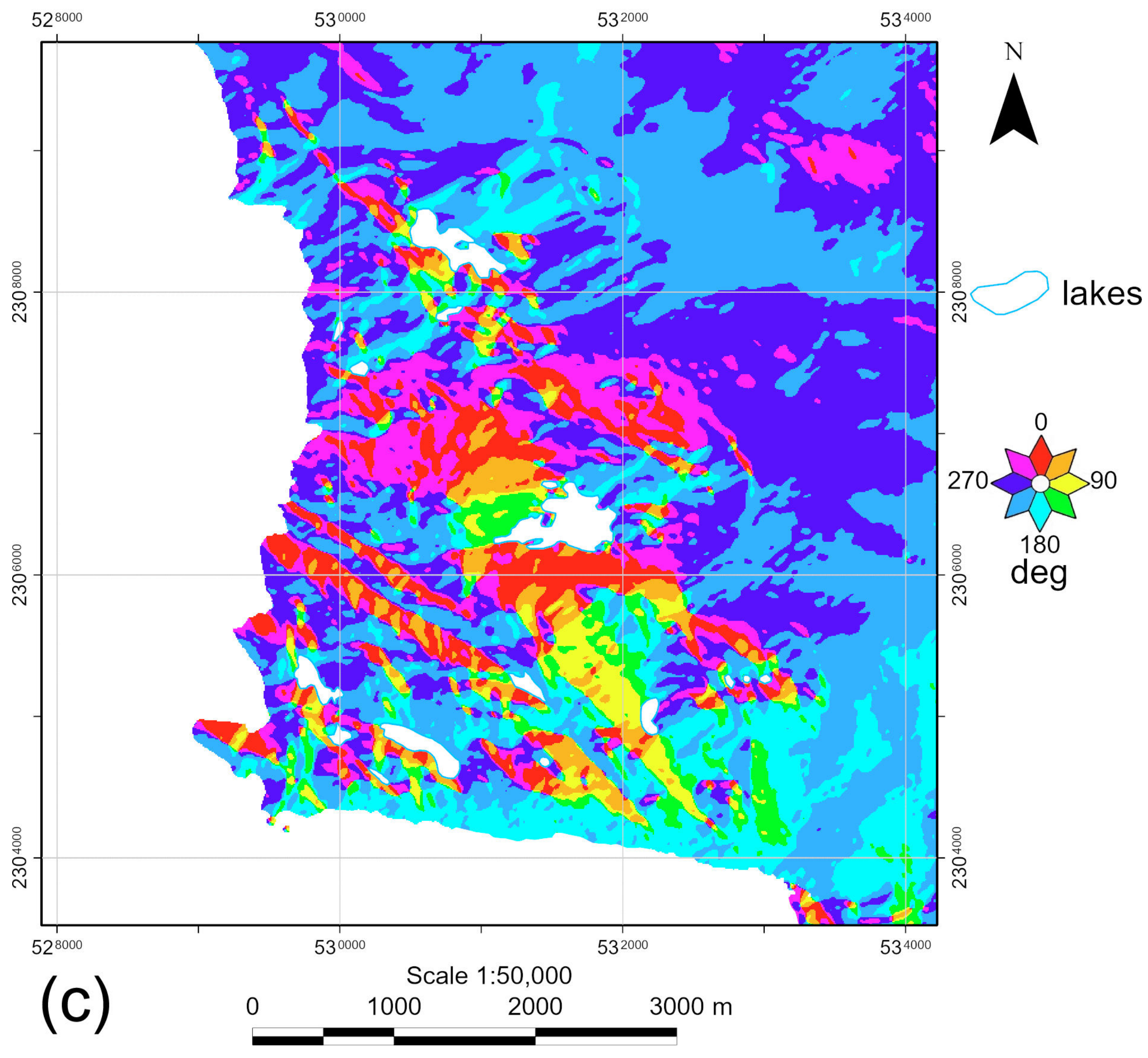

**Fig. 5, cont'd** Breidvågnipa: (c) Aspect.

*(Continued)*

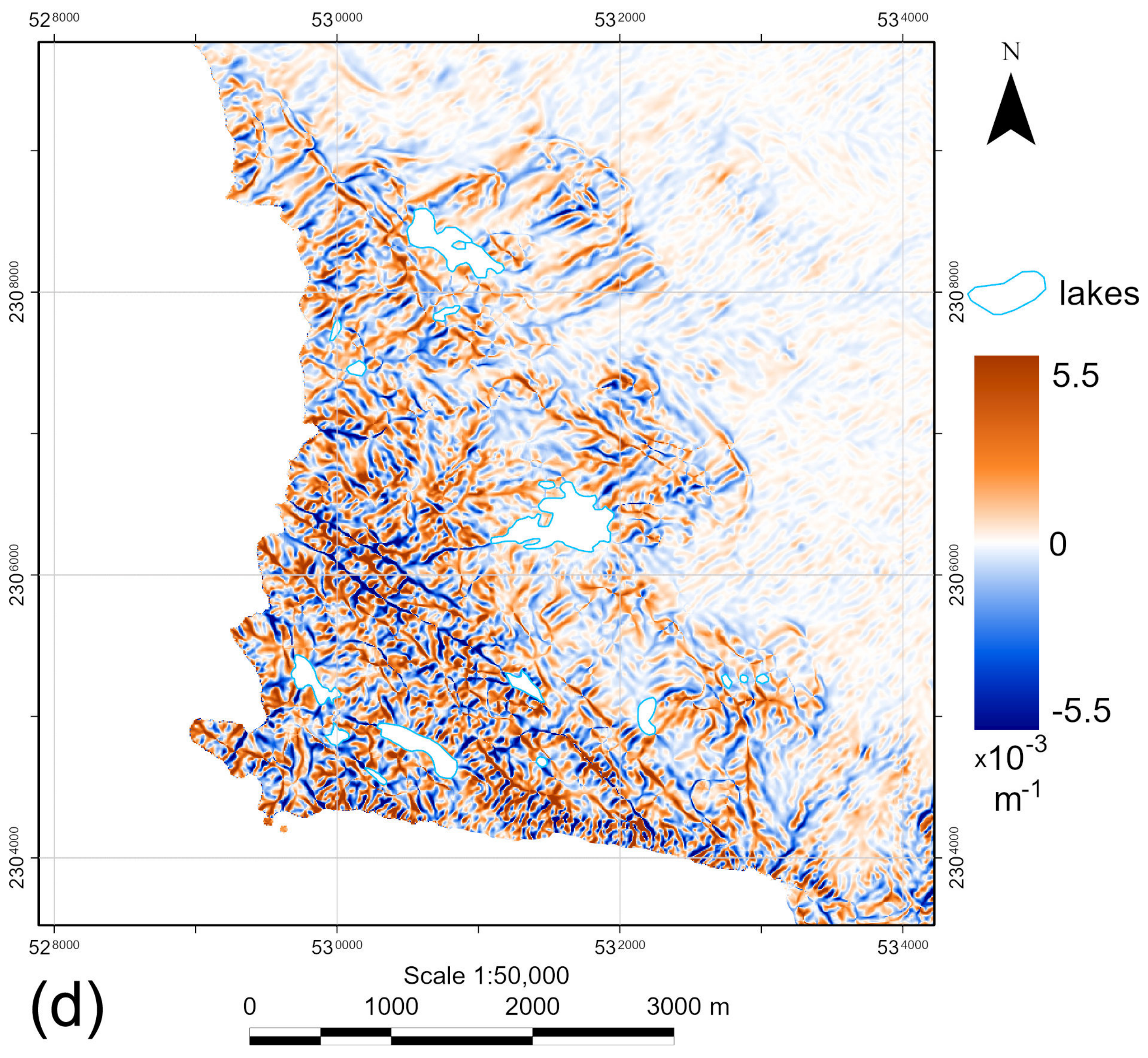

**Fig. 5, cont'd** Breidvågnipa: (d) Horizontal curvature.

*(Continued)*

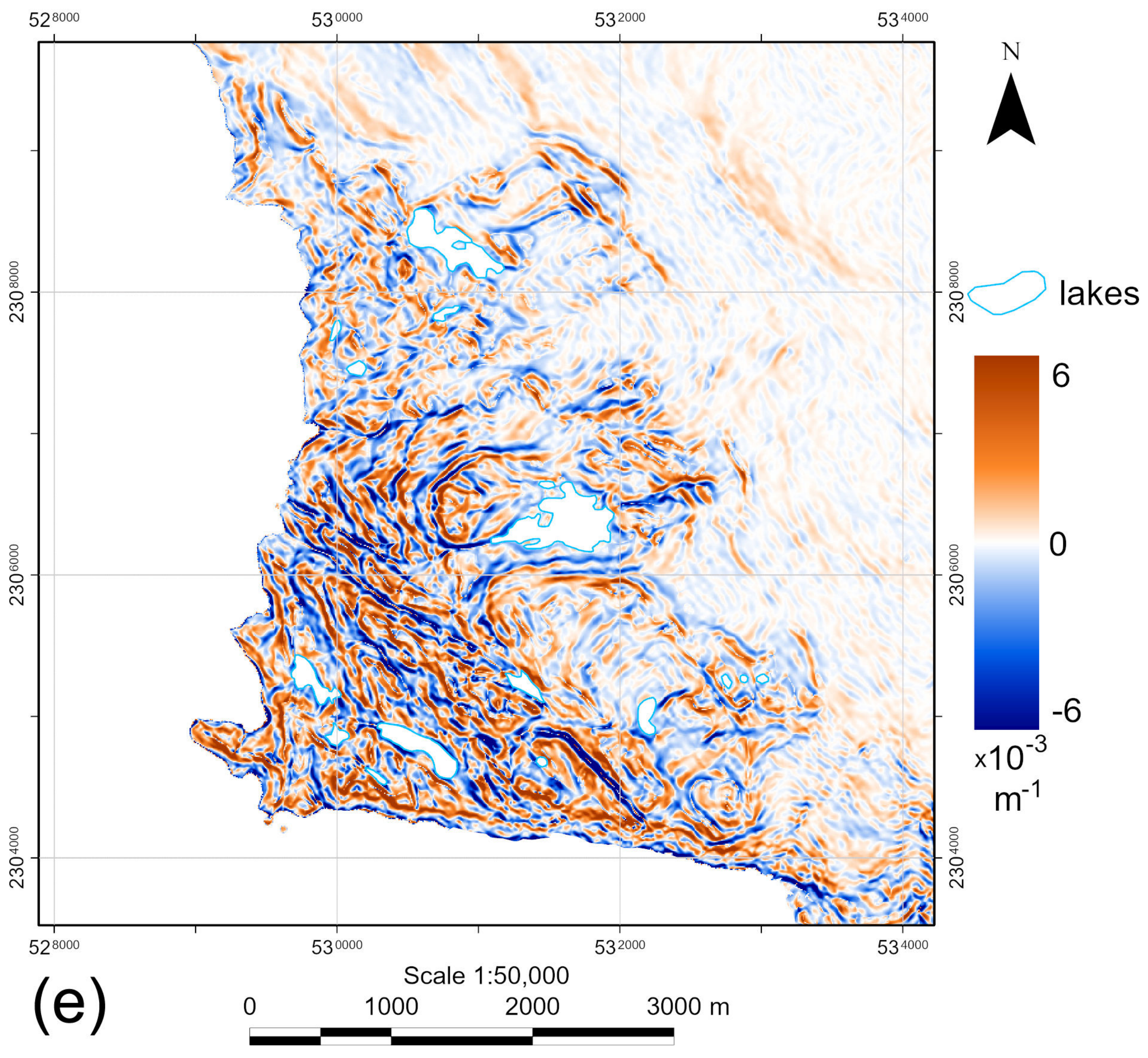

**Fig. 5, cont'd** Breidvågnipa: (e) Vertical curvature.

*(Continued)*

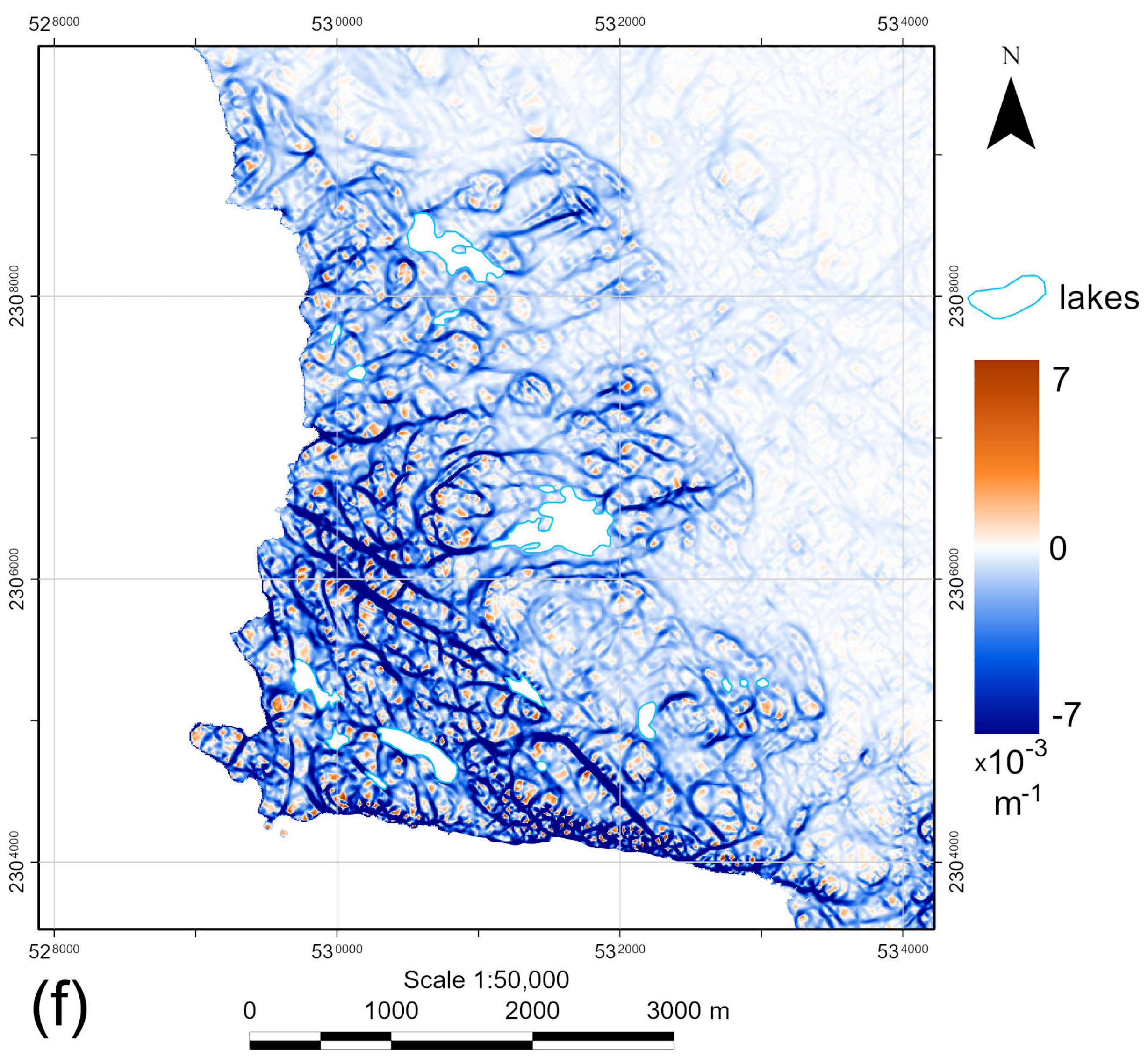

**Fig. 5, cont'd** Breidvågnipa: (f) Minimal curvature.

*(Continued)*

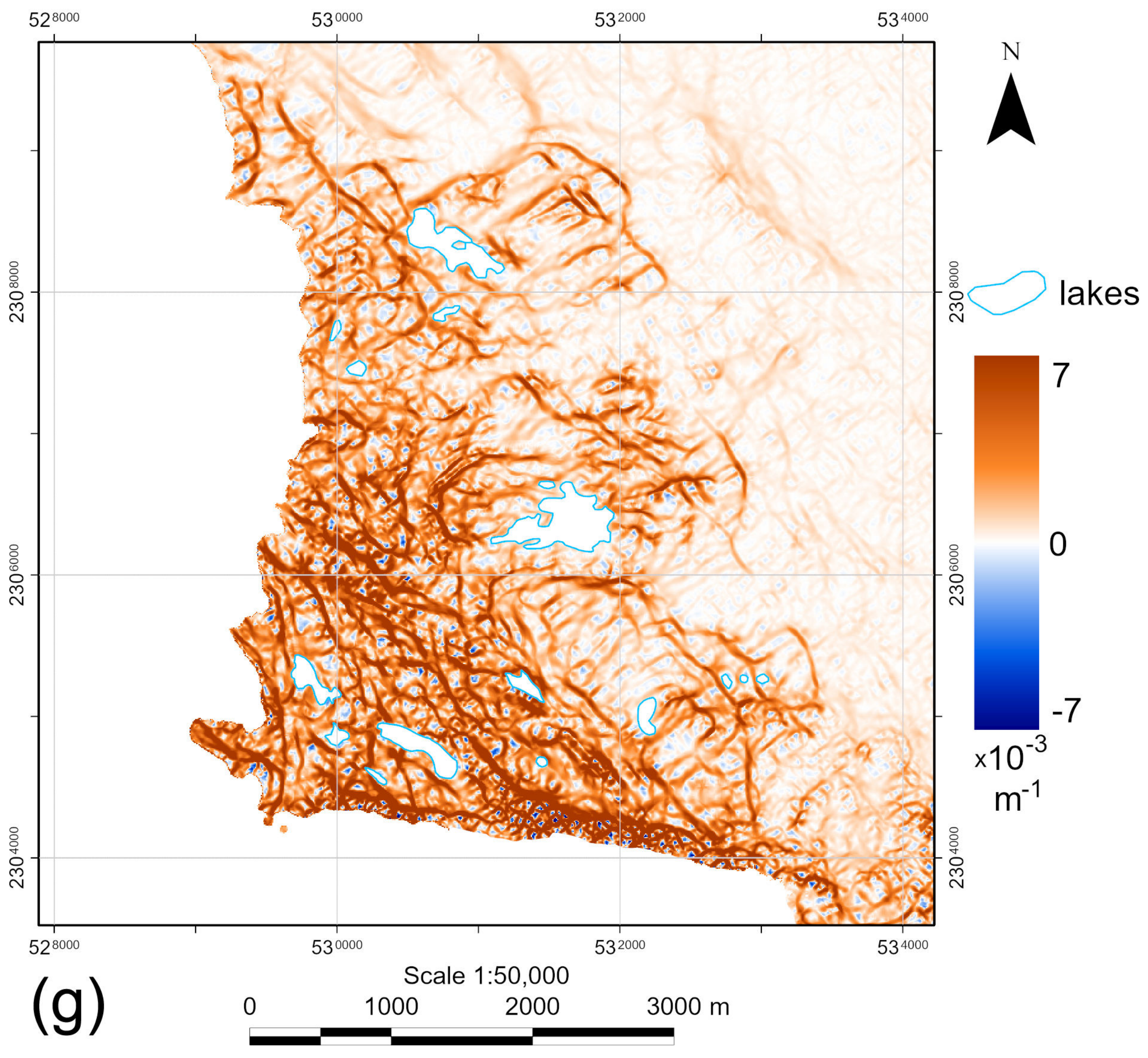

**Fig. 5, cont'd** Breidvågnipa: (g) Maximal curvature.

*(Continued)*

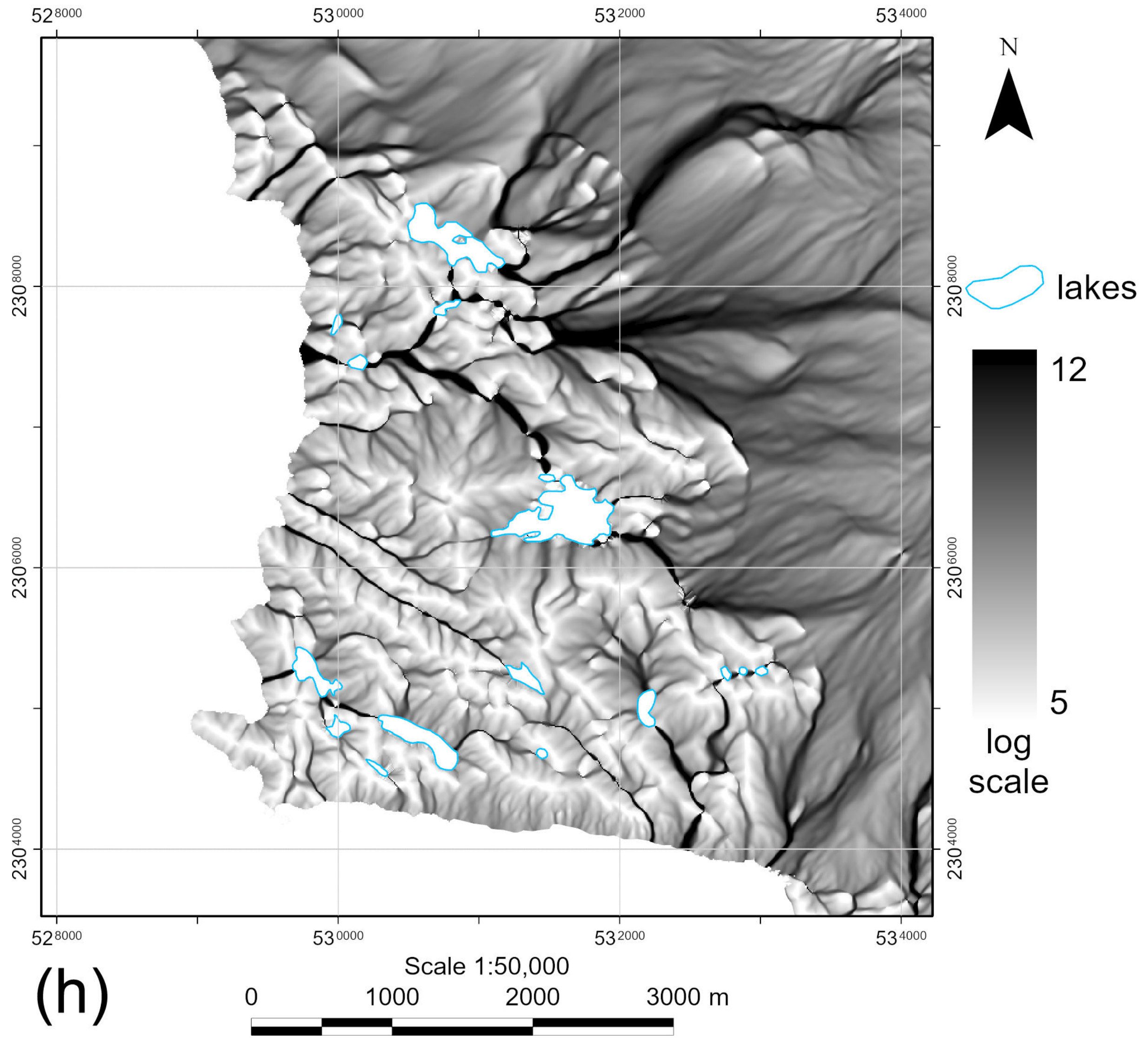

**Fig. 5, cont'd** Breidvågnipa: (h) Catchment area.

*(Continued)*

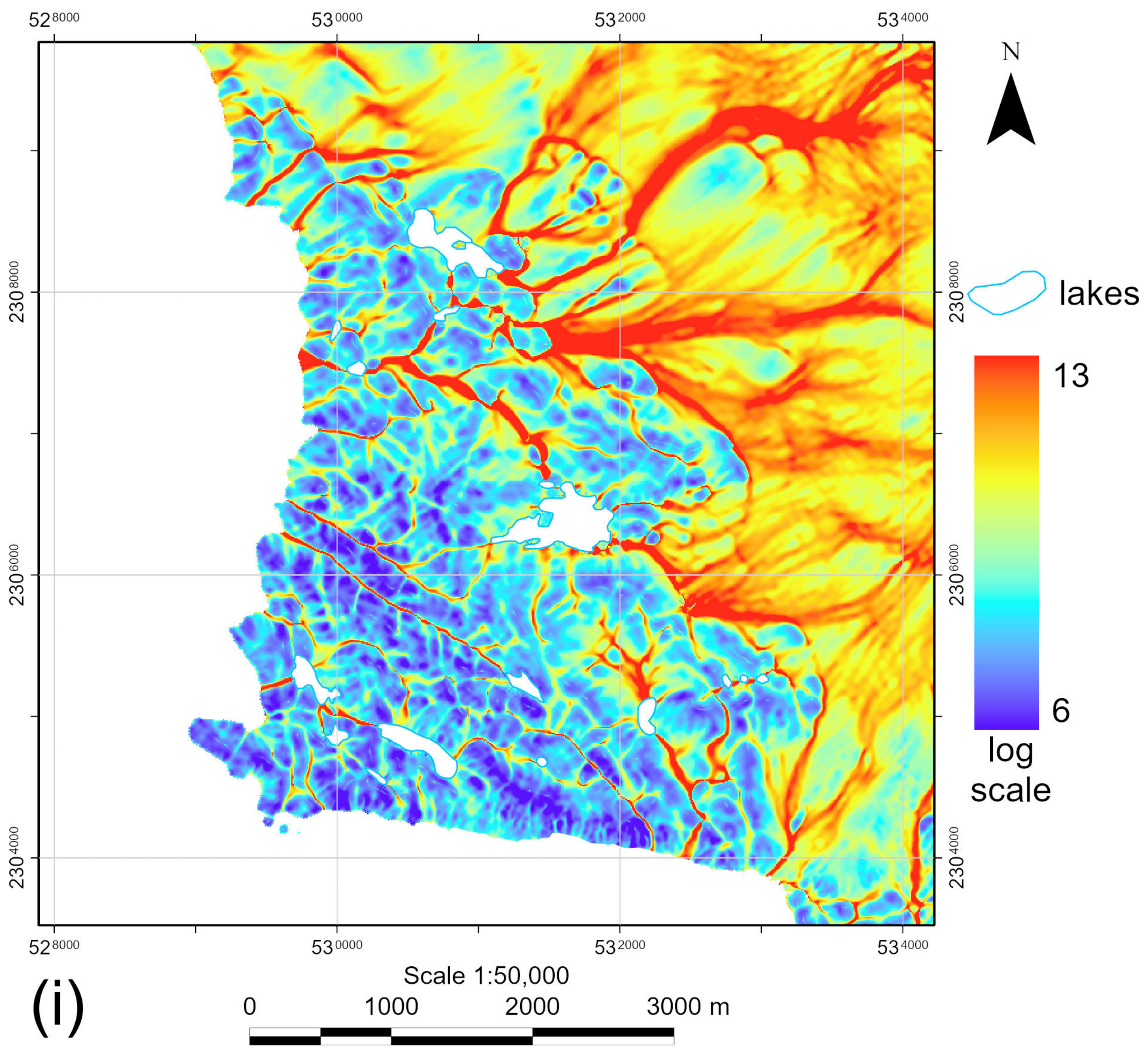

**Fig. 5, cont'd** Breidvågnipa: (i) Topographic wetness index.

*(Continued)*

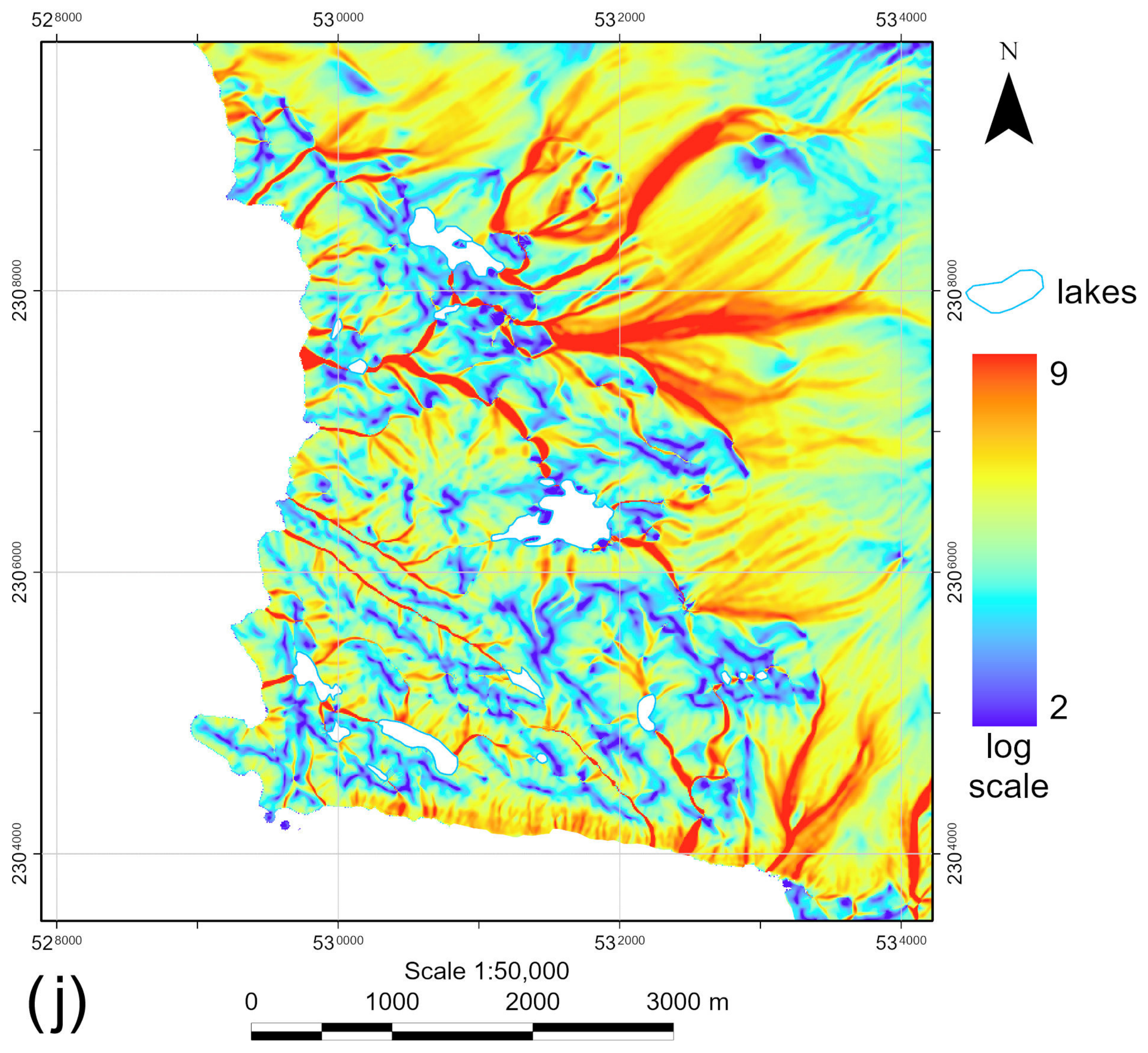

**Fig. 5, cont'd** Breidvågnipa: ( j) Stream power index.

*(Continued)*

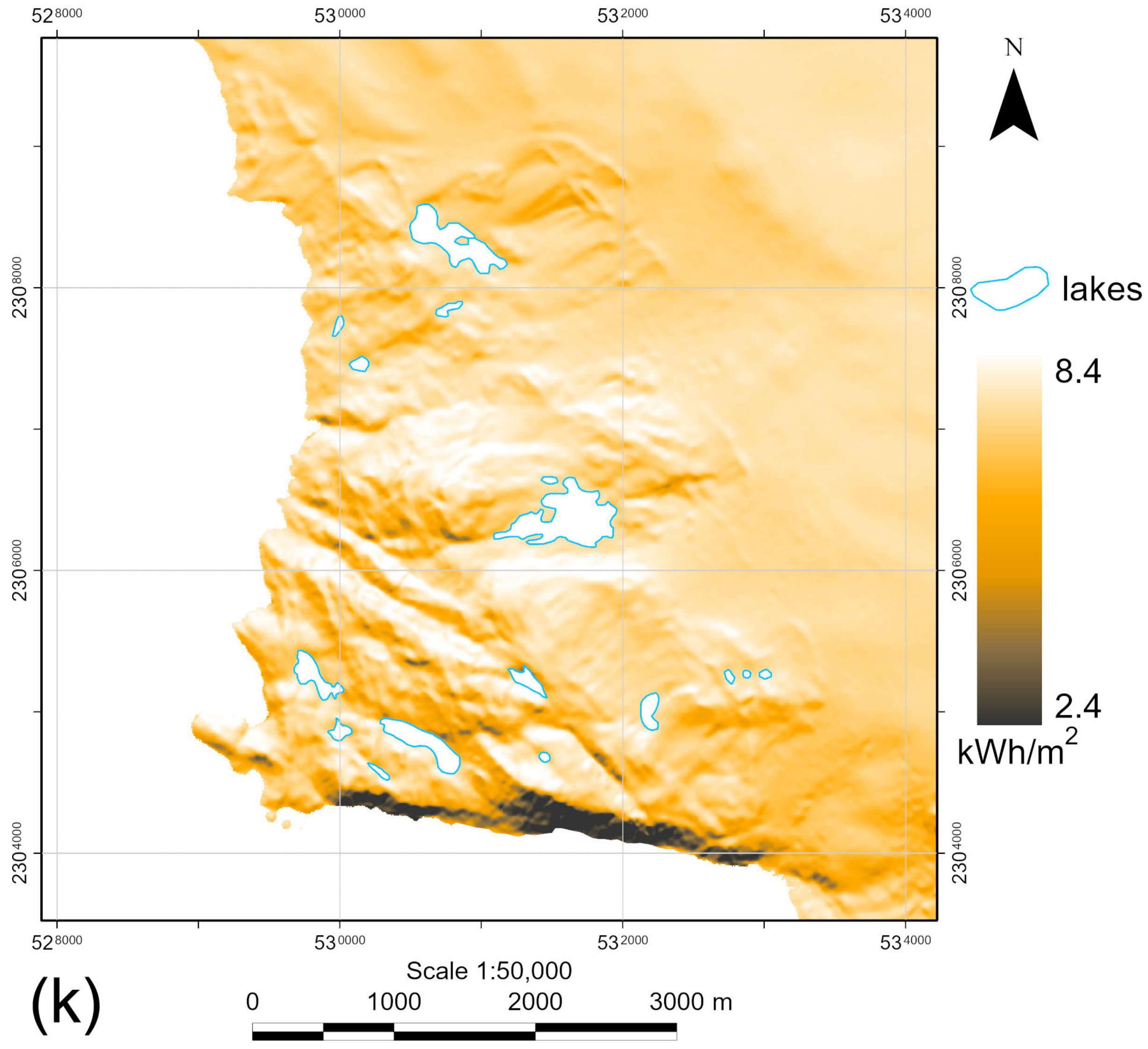

**Fig. 5, cont'd** Breidvågnipa: (k) Total insolation.

*(Continued)*

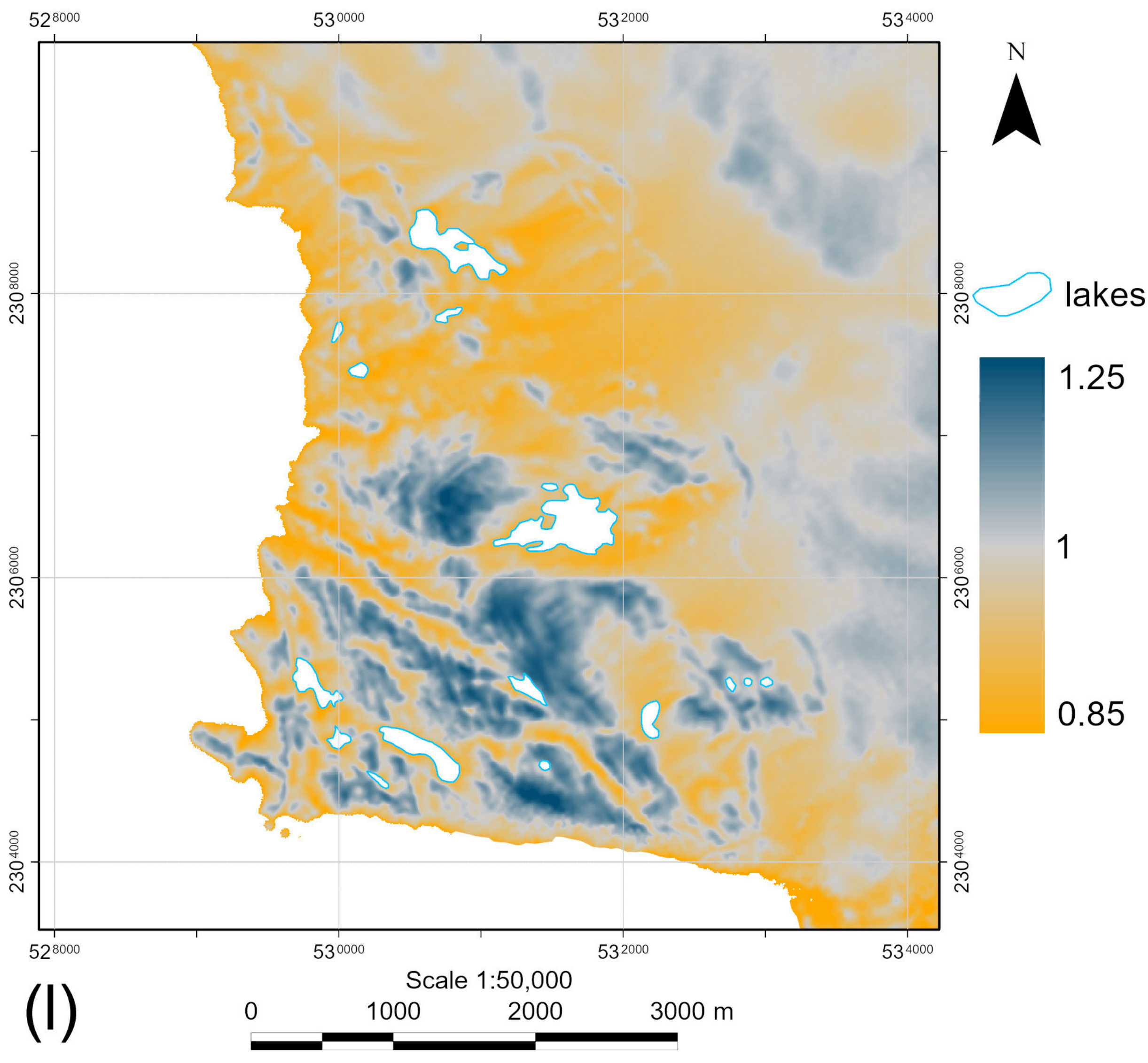

**Fig. 5, cont'd** Breidvågnipa: (l) Wind exposition index.

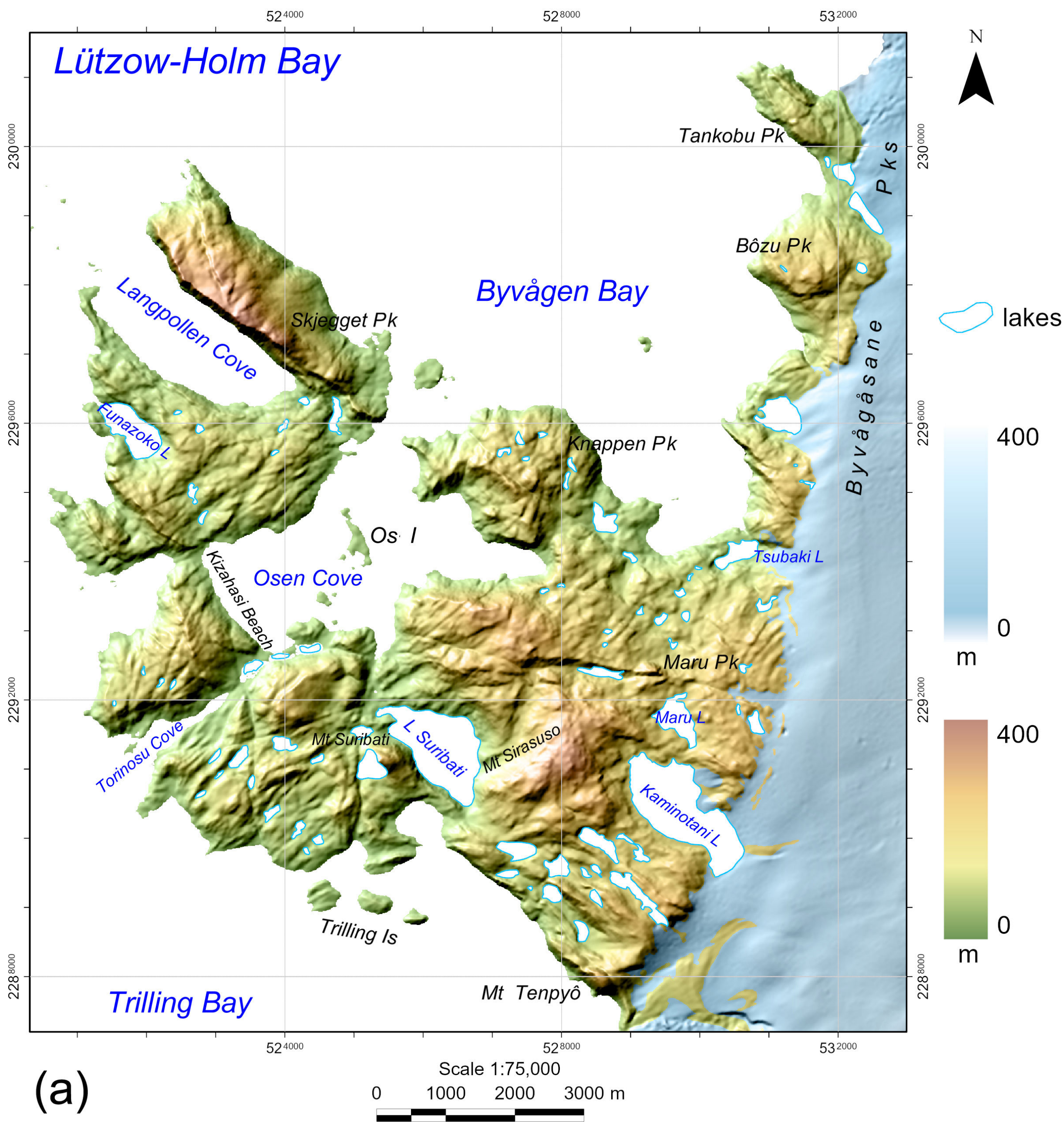

**Fig. 6** Skarvsnes Foreland: (a) Elevation.

*(Continued)*

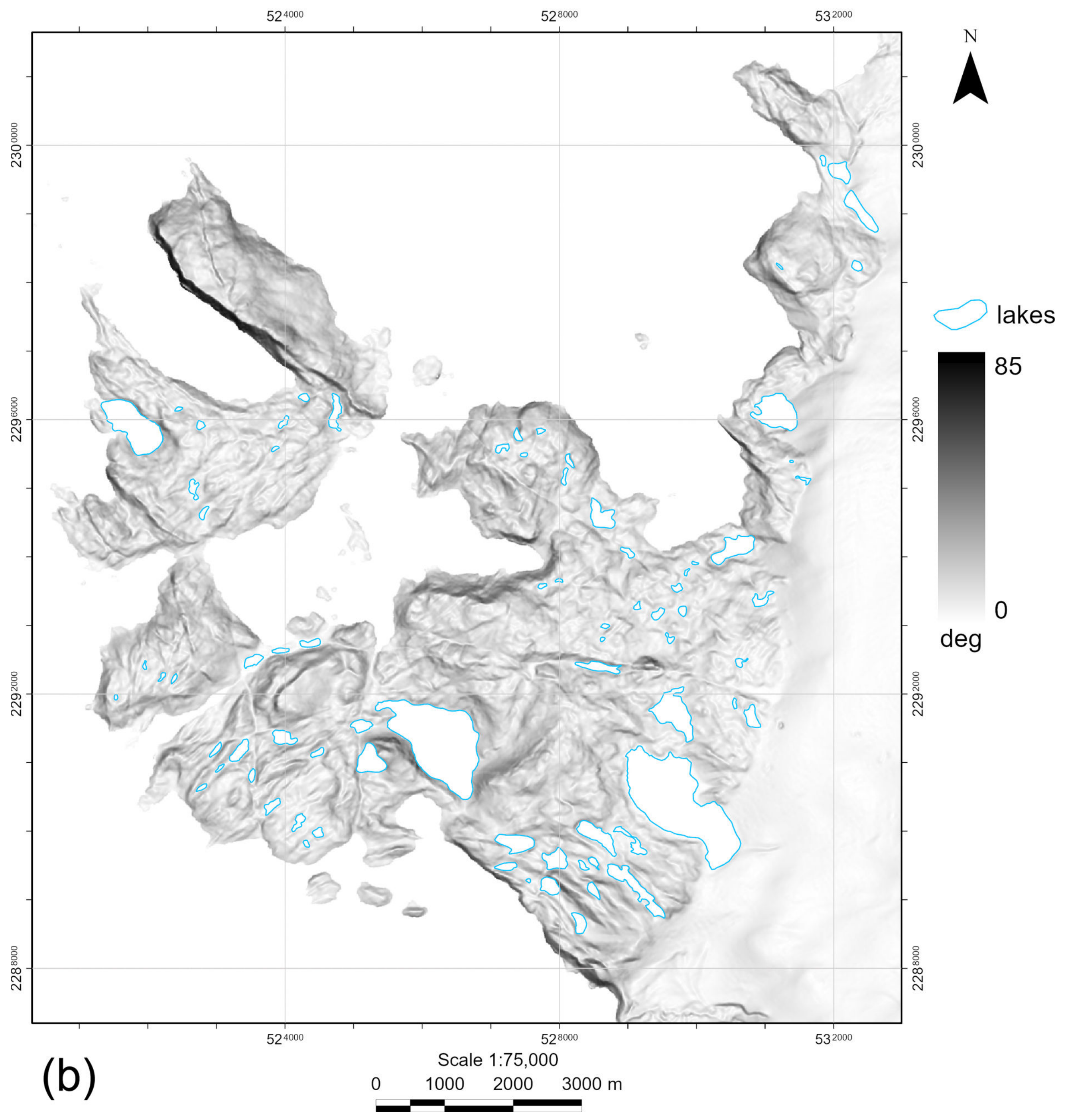

**Fig. 6, cont'd** Skarvsnes Foreland: (b) Slope.

*(Continued)*

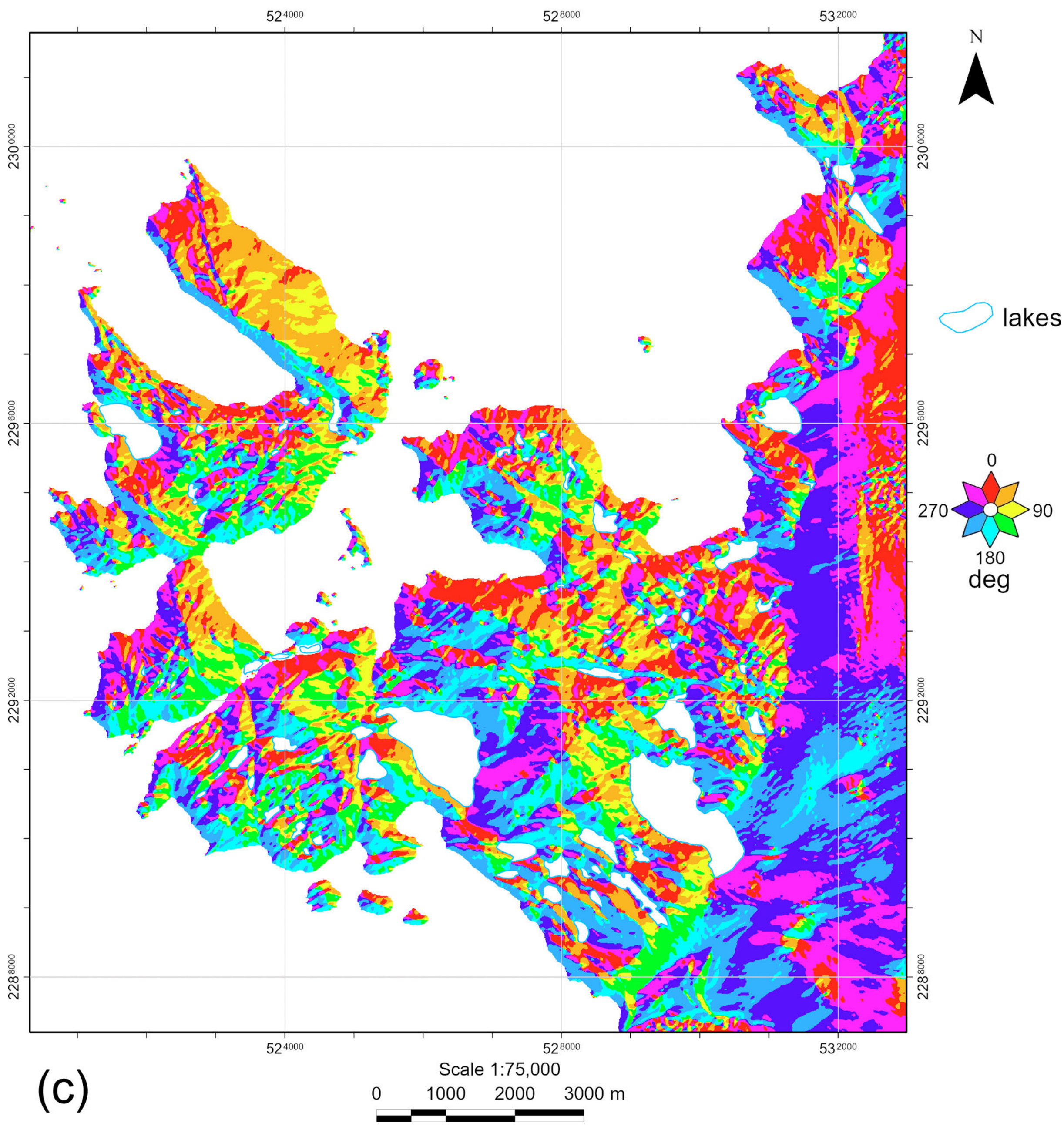

**Fig. 6, cont'd** Skarvsnes Foreland: (c) Aspect.

*(Continued)*

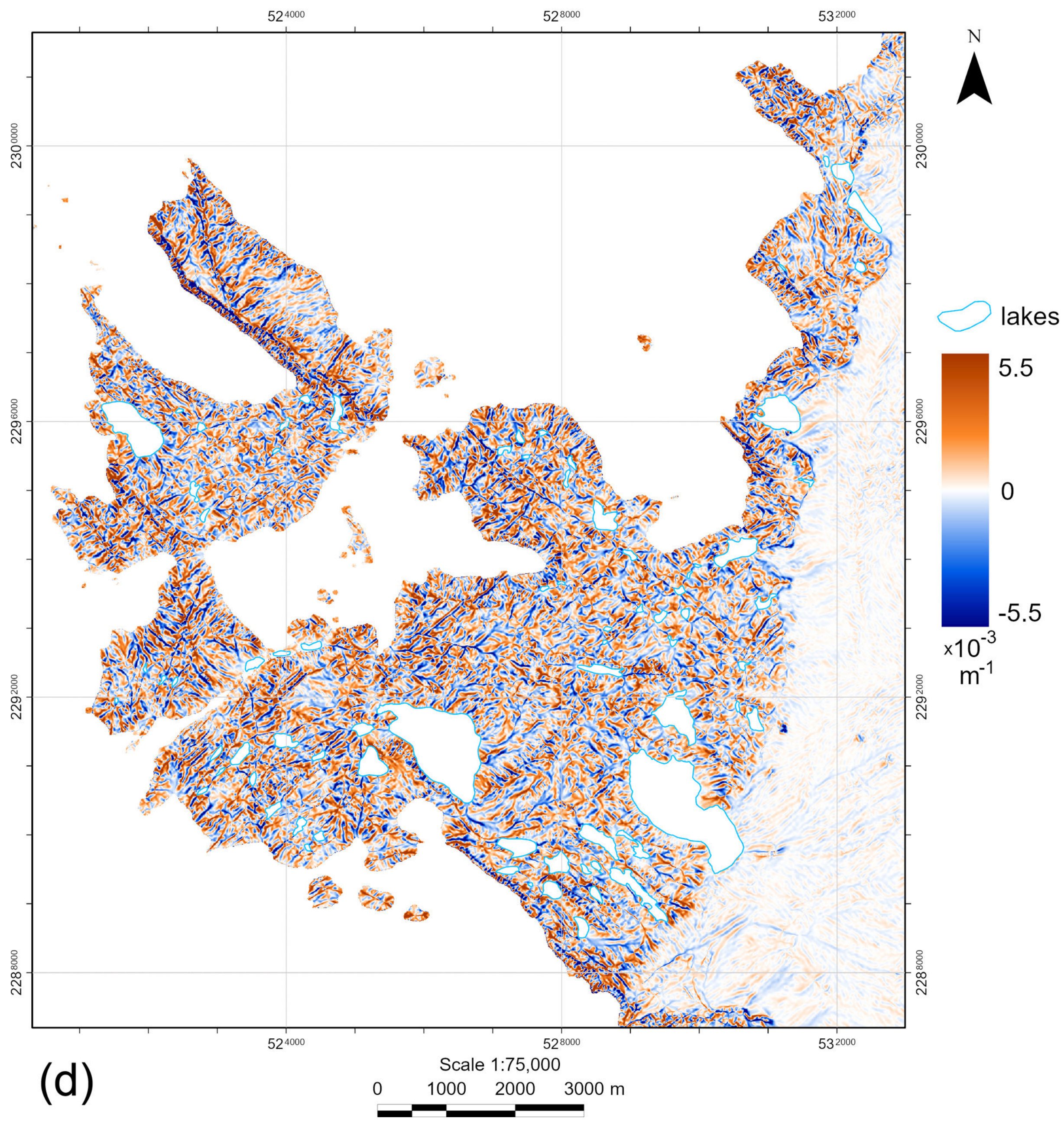

**Fig. 6, cont'd** Skarvsnes Foreland: (d) Horizontal curvature.

*(Continued)*

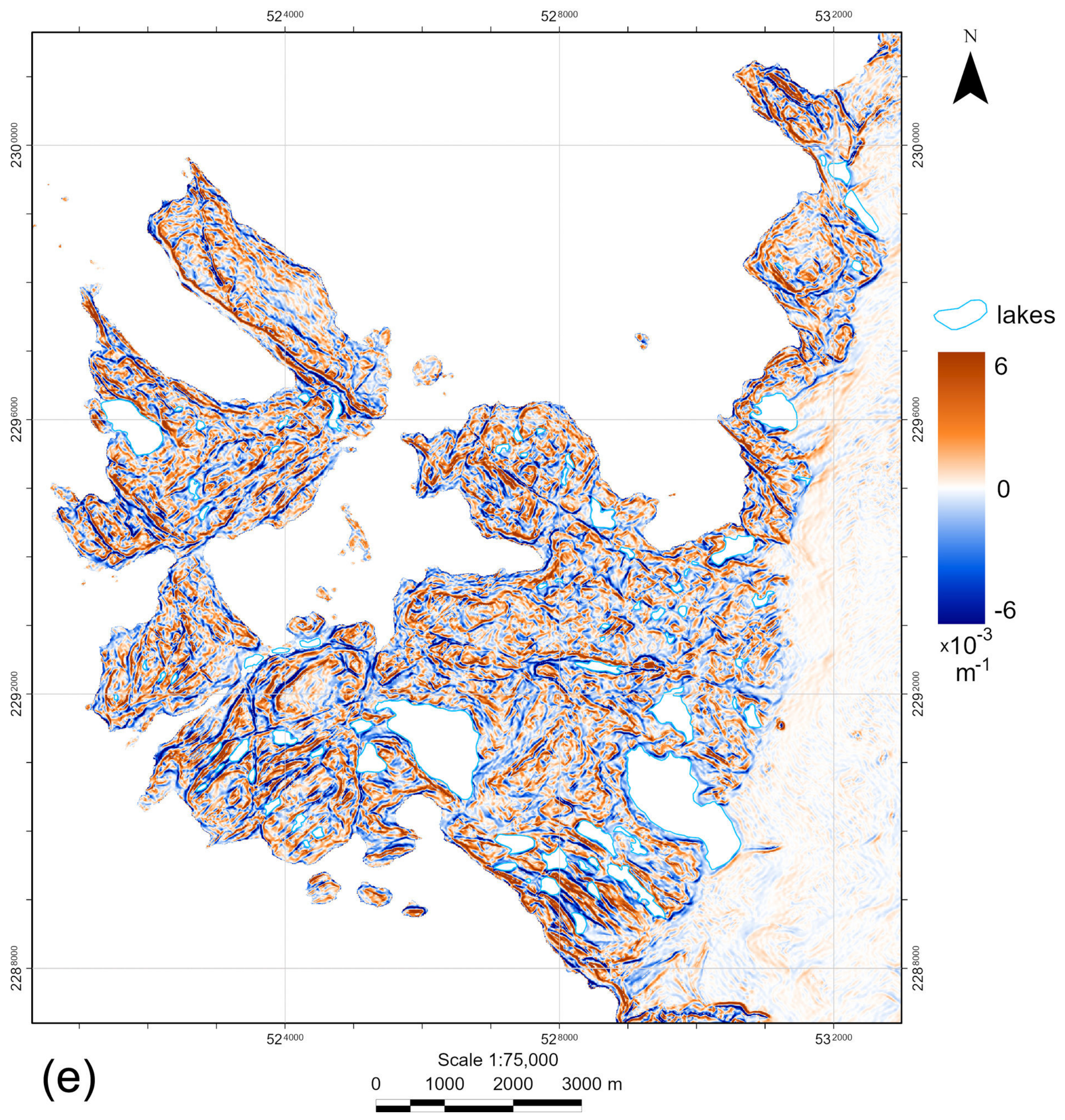

**Fig. 6, cont'd** Skarvsnes Foreland: (e) Vertical curvature.

*(Continued)*

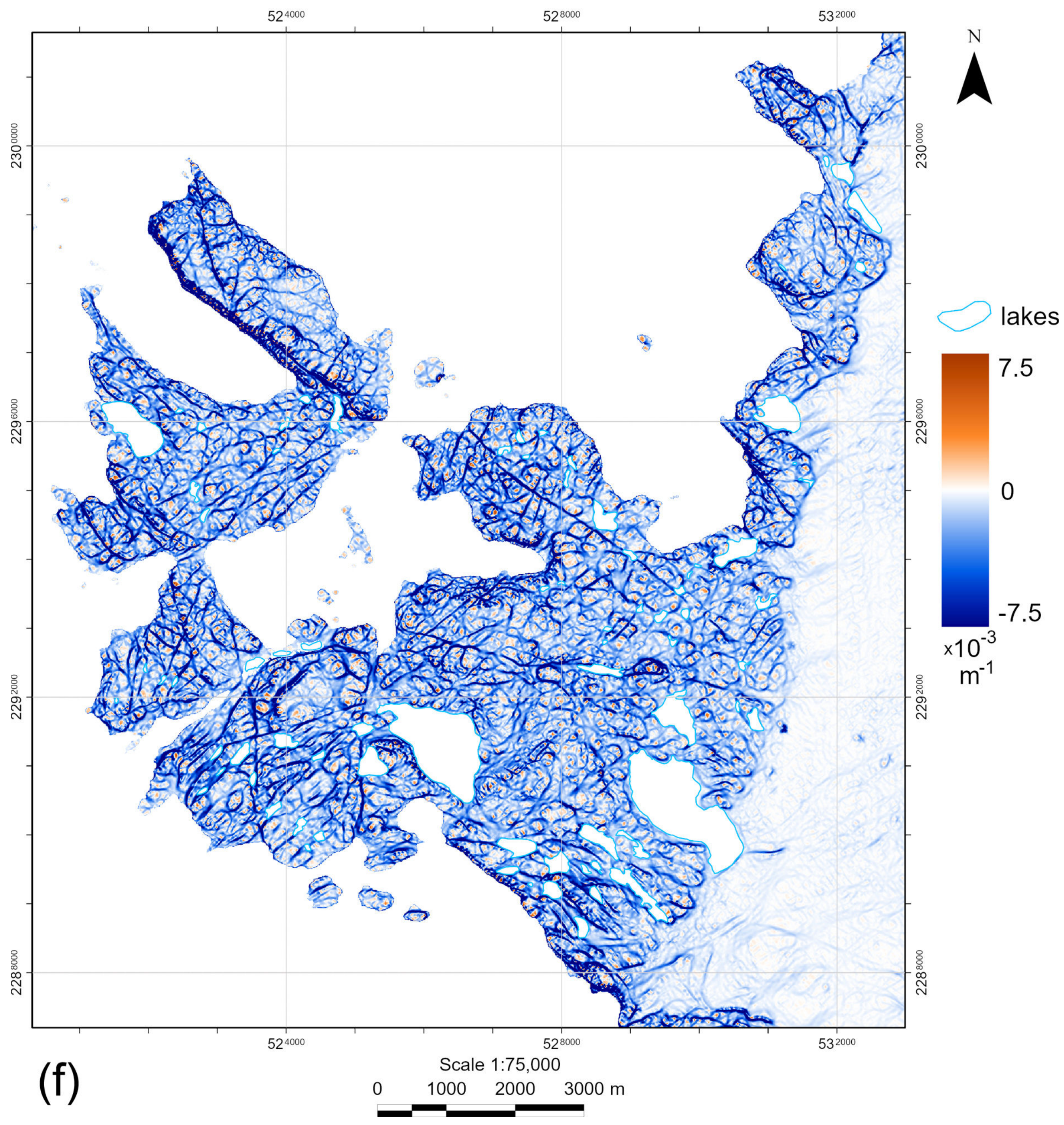

**Fig. 6, cont'd** Skarvsnes Foreland: (f) Minimal curvature.

*(Continued)*

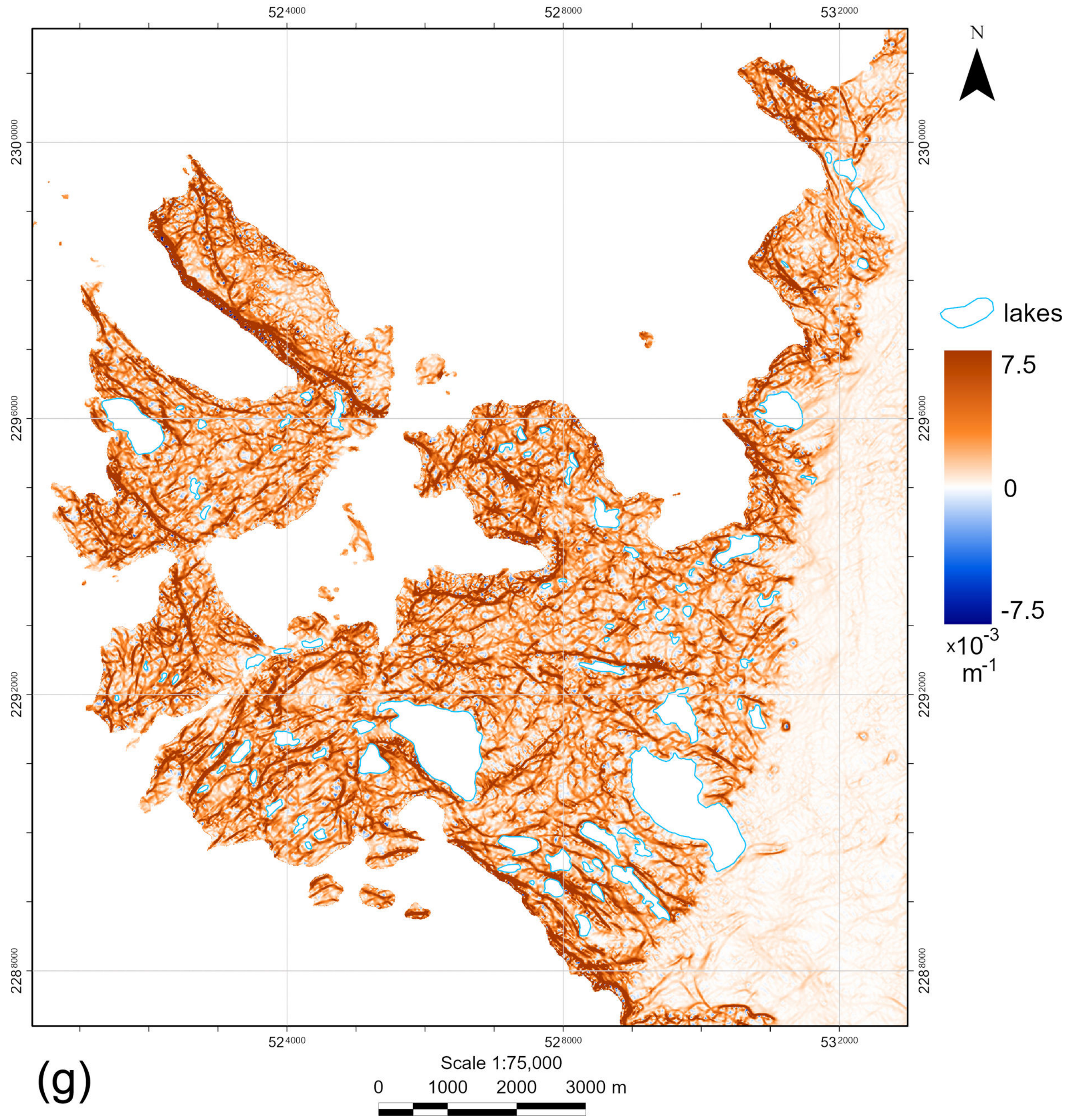

**Fig. 6, cont'd** Skarvsnes Foreland: (g) Maximal curvature.

*(Continued)*

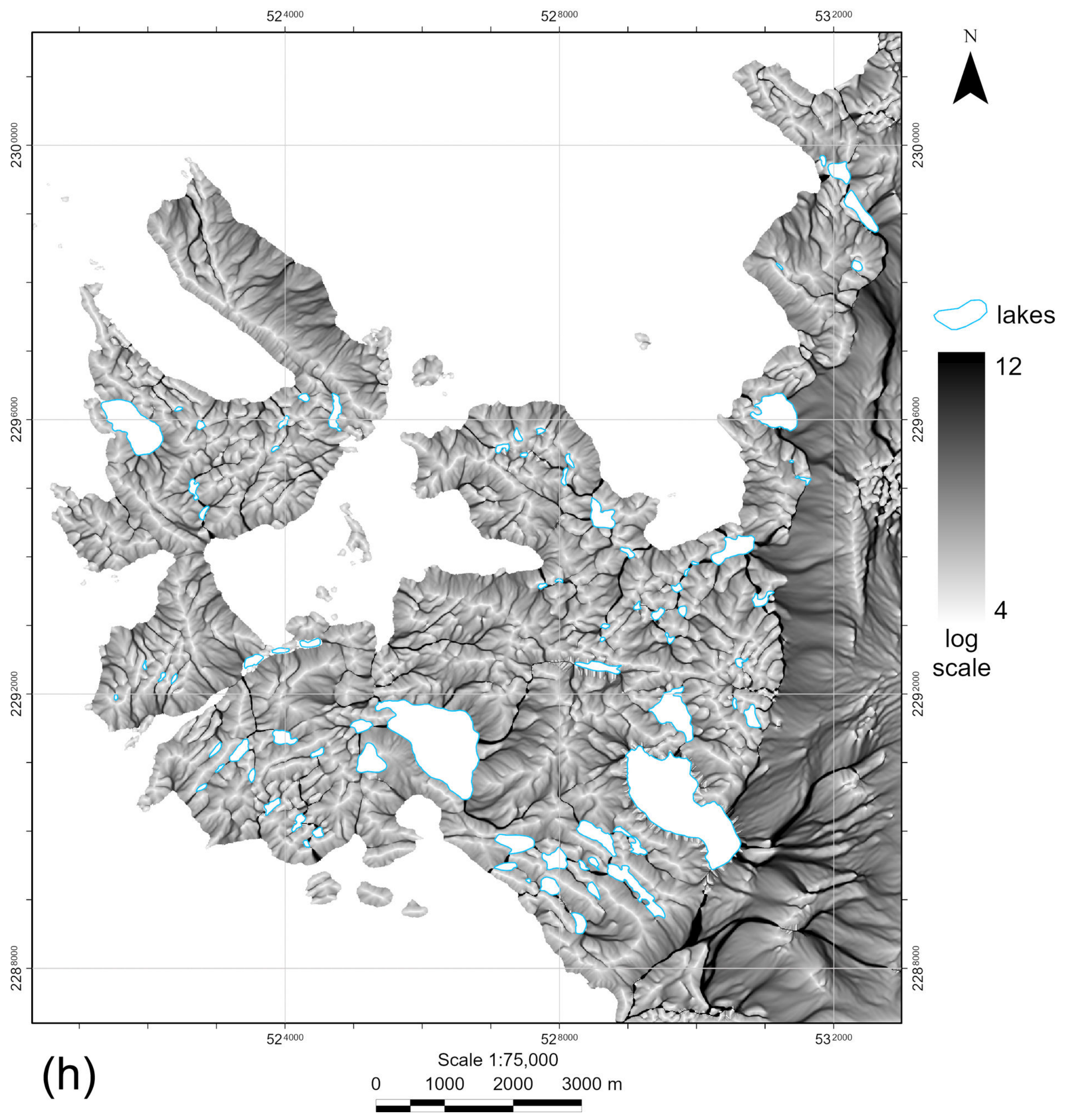

**Fig. 6, cont'd** Skarvsnes Foreland: (h) Catchment area.

*(Continued)*

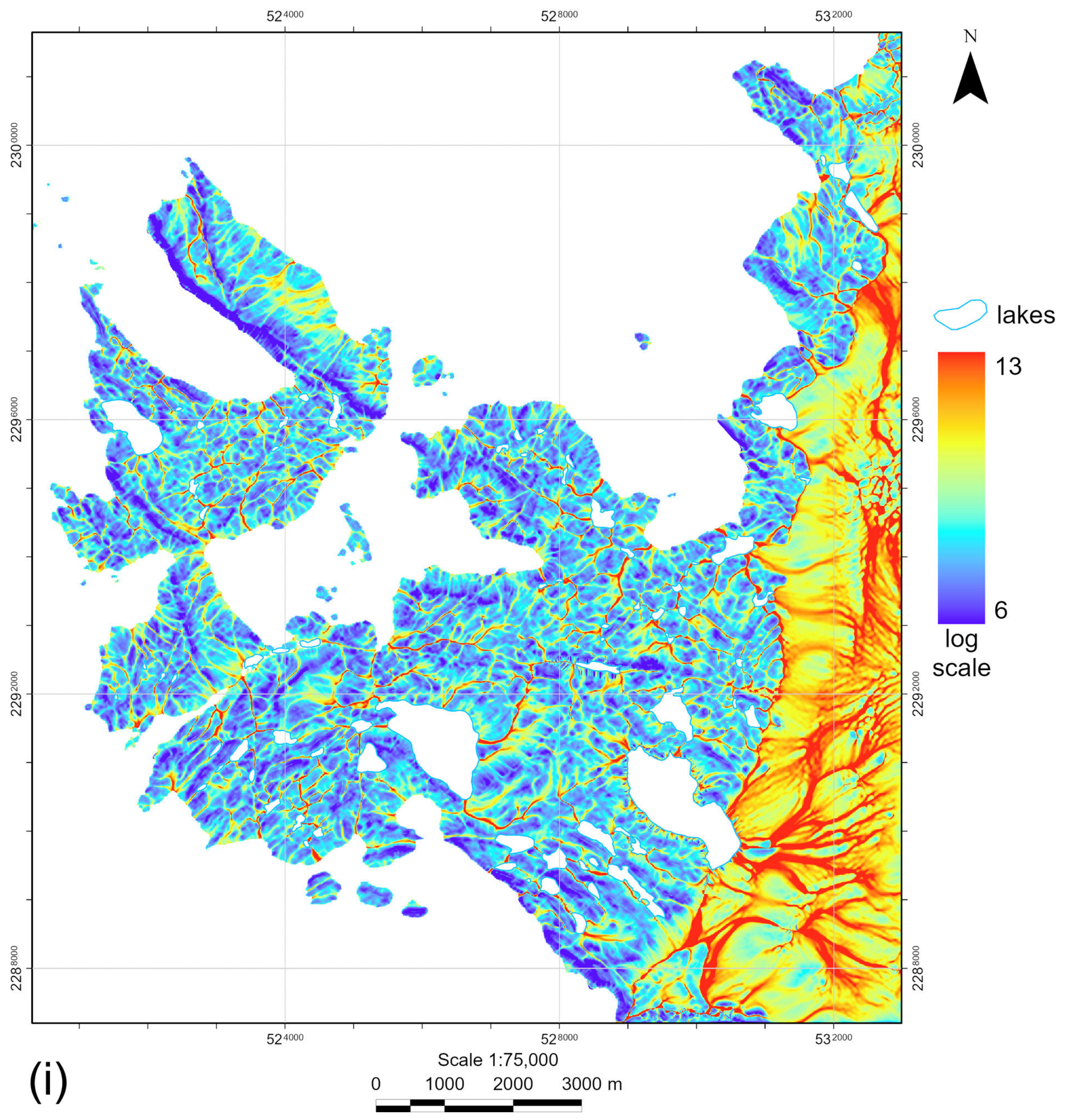

**Fig. 6, cont'd** Skarvsnes Foreland: (i) Topographic wetness index.

*(Continued)*

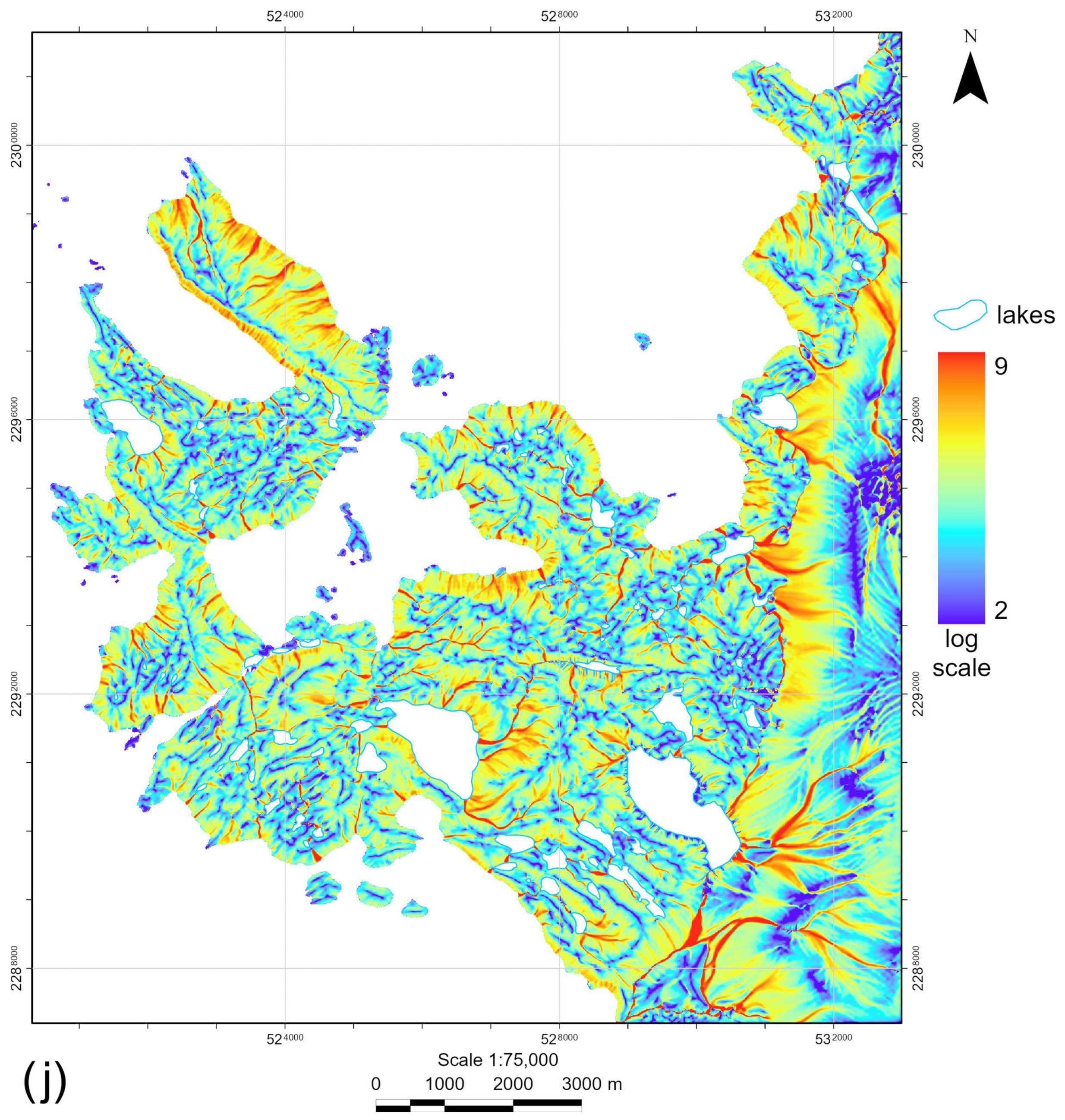

**Fig. 6, cont'd** Skarvsnes Foreland: ( j) Stream power index.

*(Continued)*

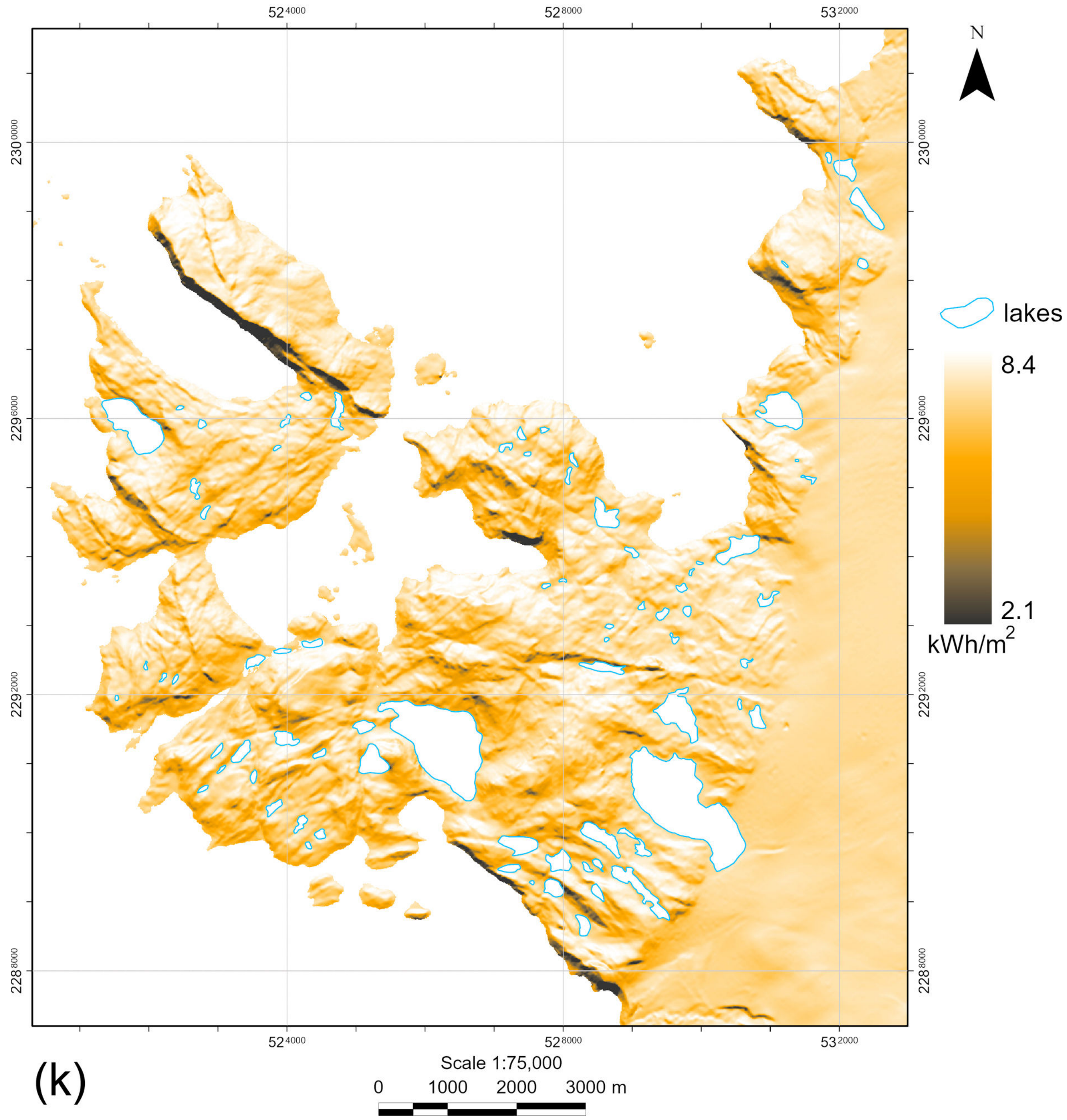

**Fig. 6, cont'd** Skarvsnes Foreland: (k) Total insolation.

*(Continued)*

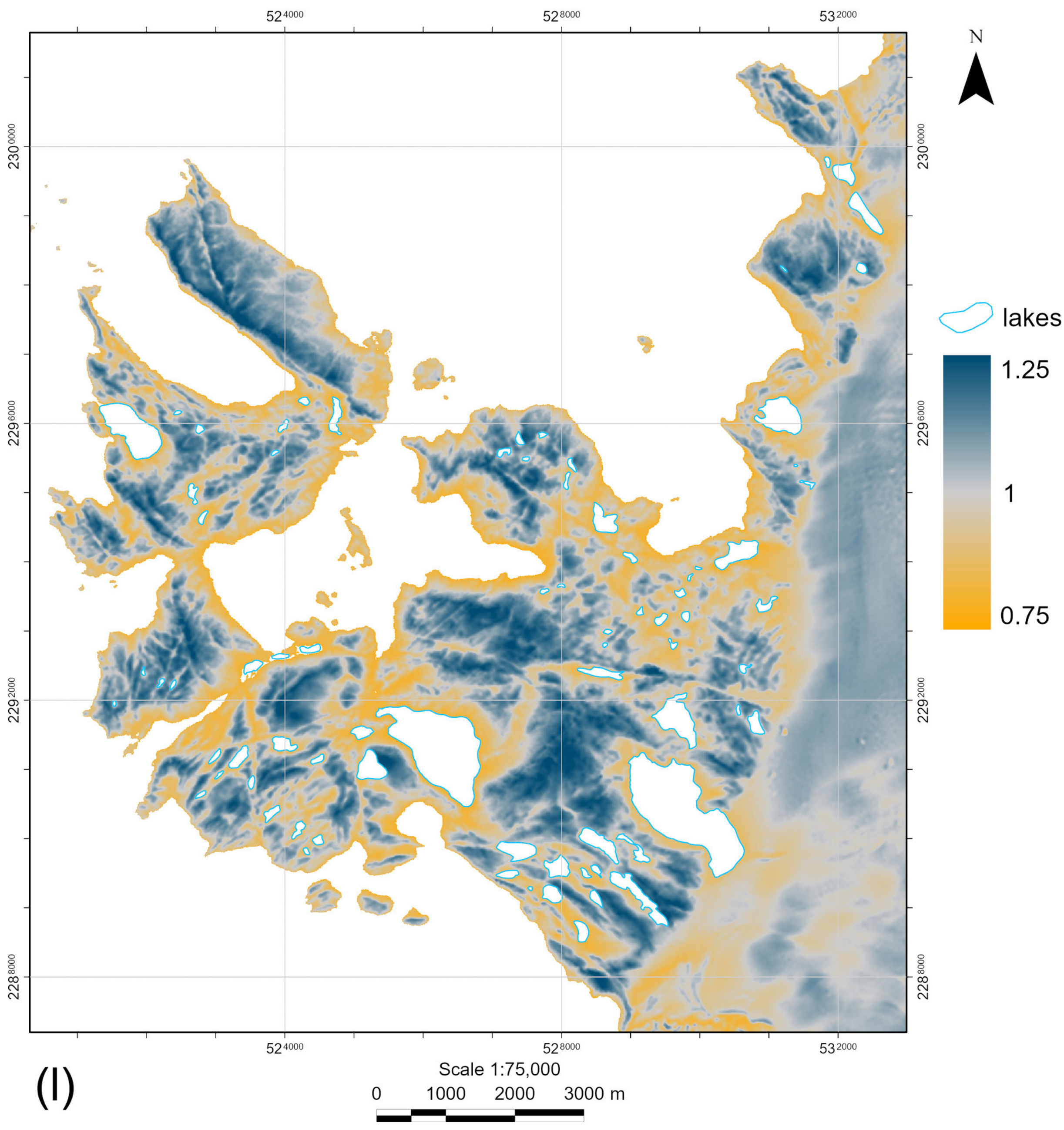

**Fig. 6, cont'd** Skarvsnes Foreland: (l) Wind exposition index.

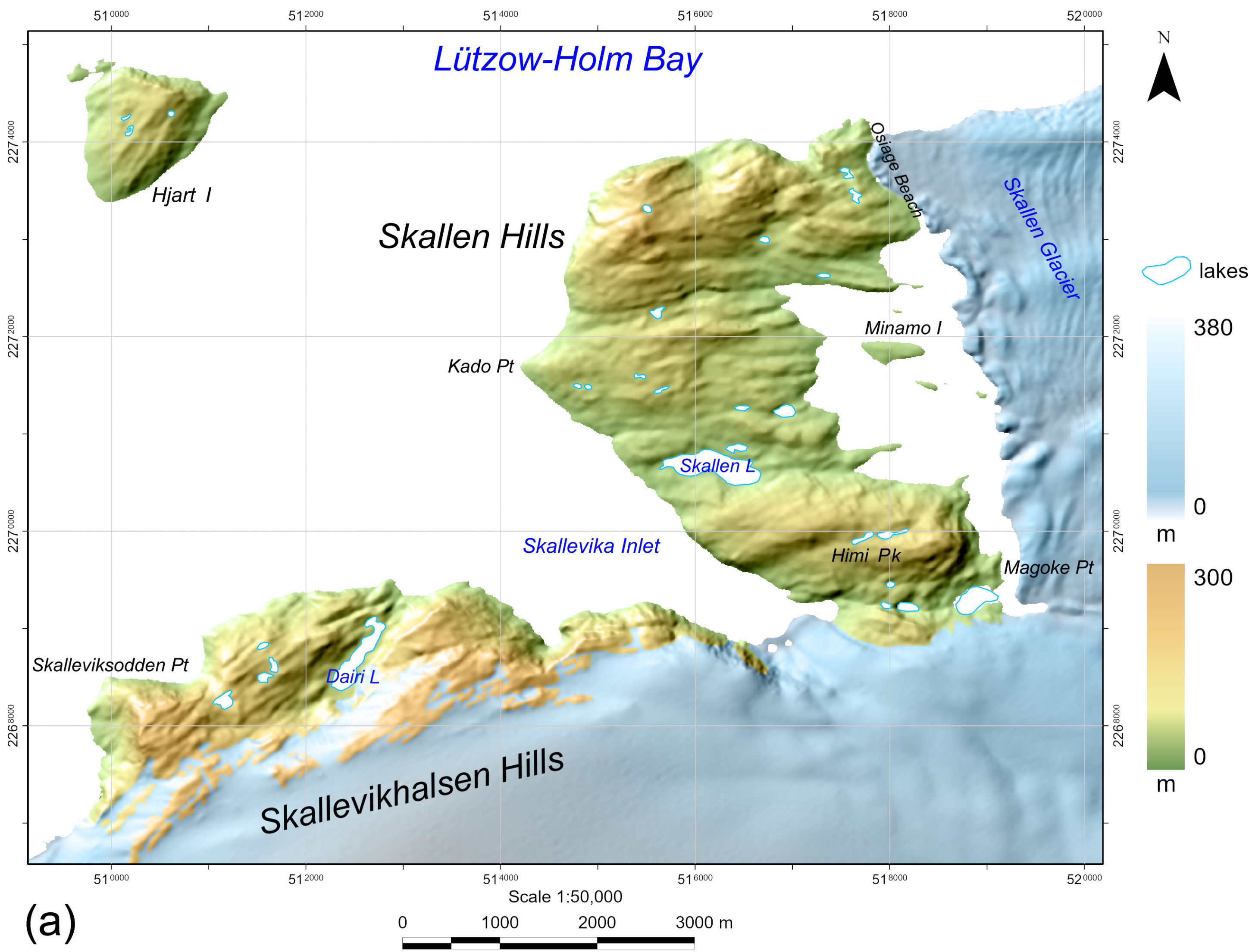

**Fig. 7** Skallen Hills and Skallevikhalsen Hills: (a) Elevation.

*(Continued)*

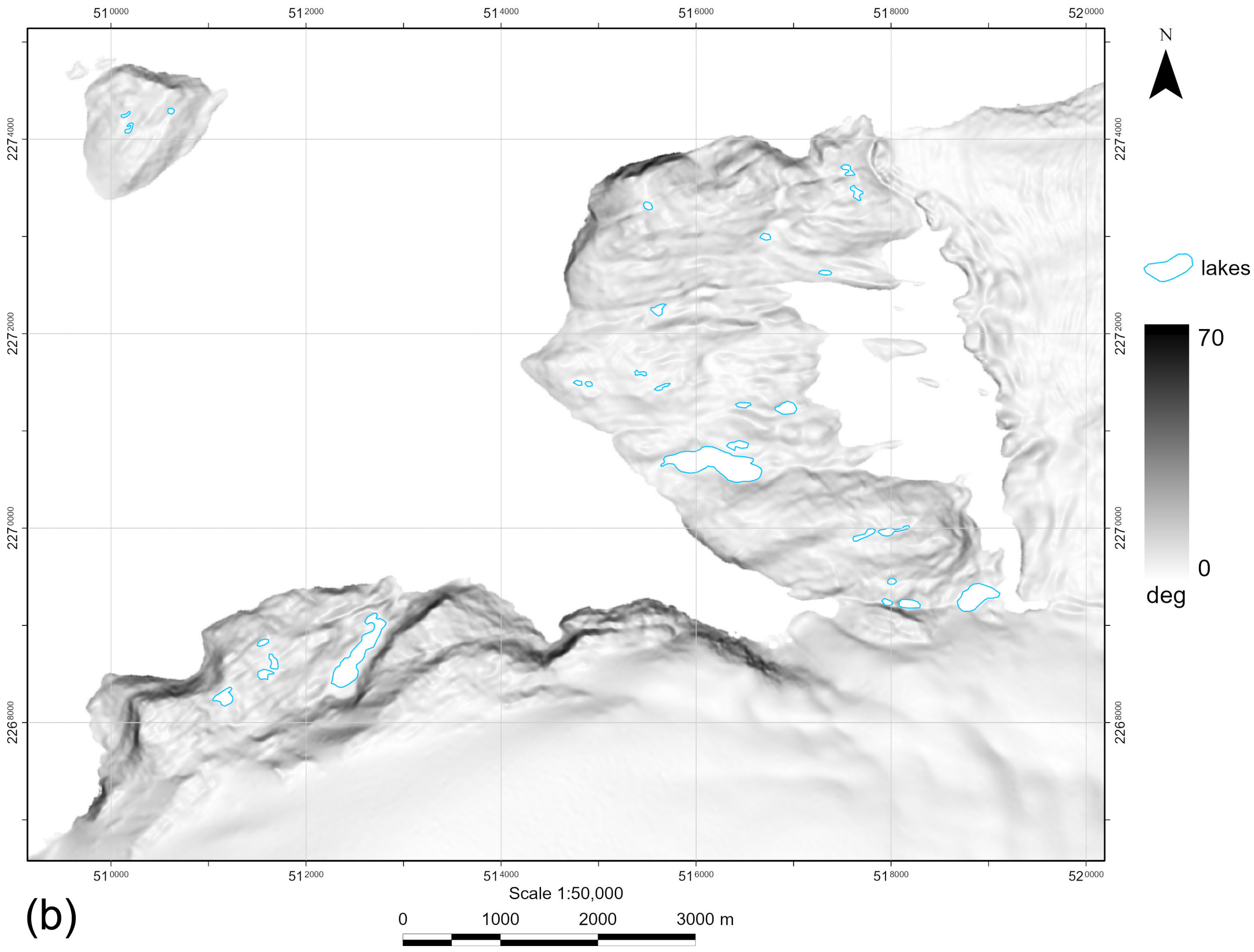

**Fig. 7** Skallen Hills and Skallevikhalsen Hills: (b) Slope.

*(Continued)*

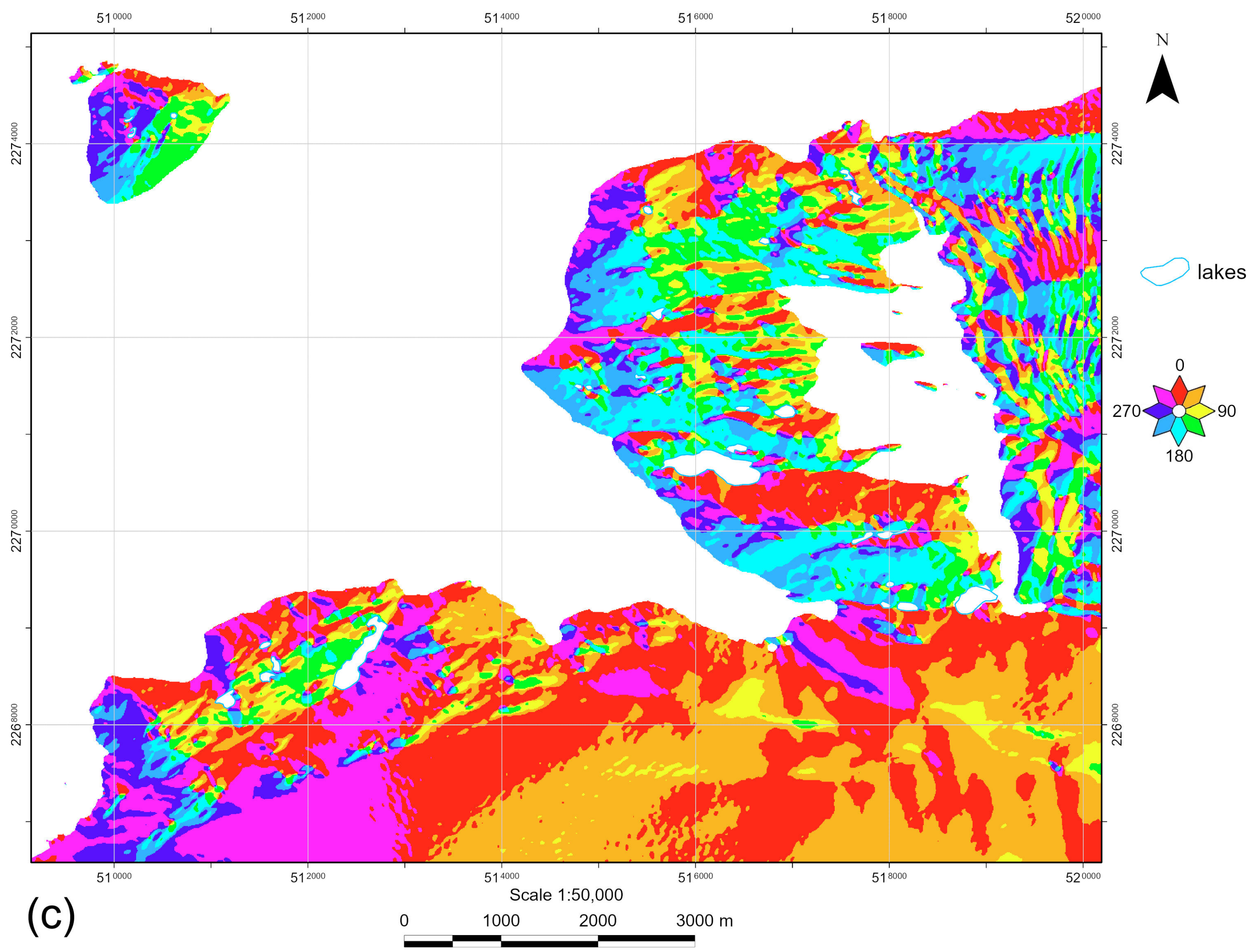

**Fig. 7, cont'd** Skallen Hills and Skallevikhalsen Hills: (c) Aspect.

*(Continued)*

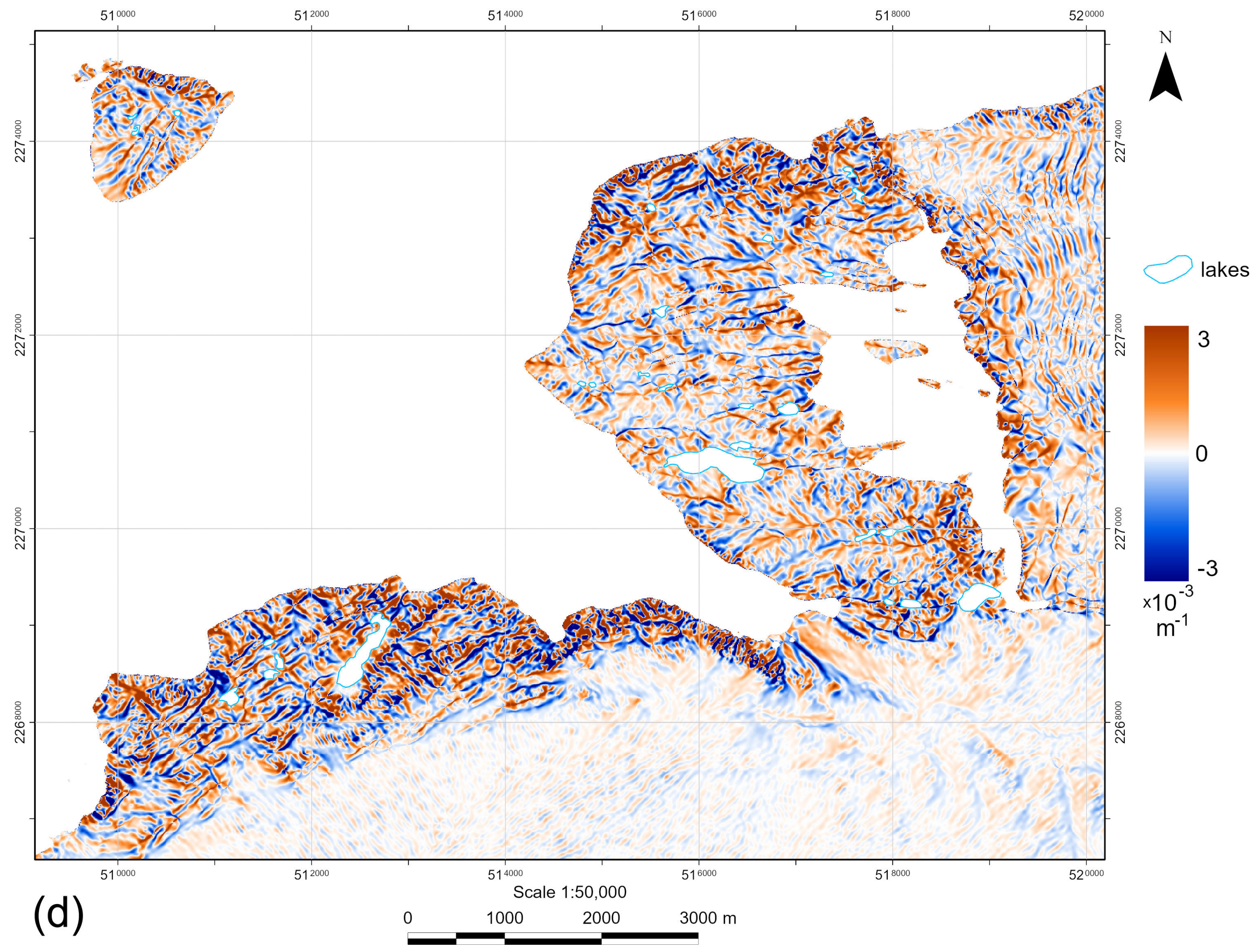

**Fig. 7, cont'd** Skallen Hills and Skallevikhalsen Hills: (d) Horizontal curvature.

*(Continued)*

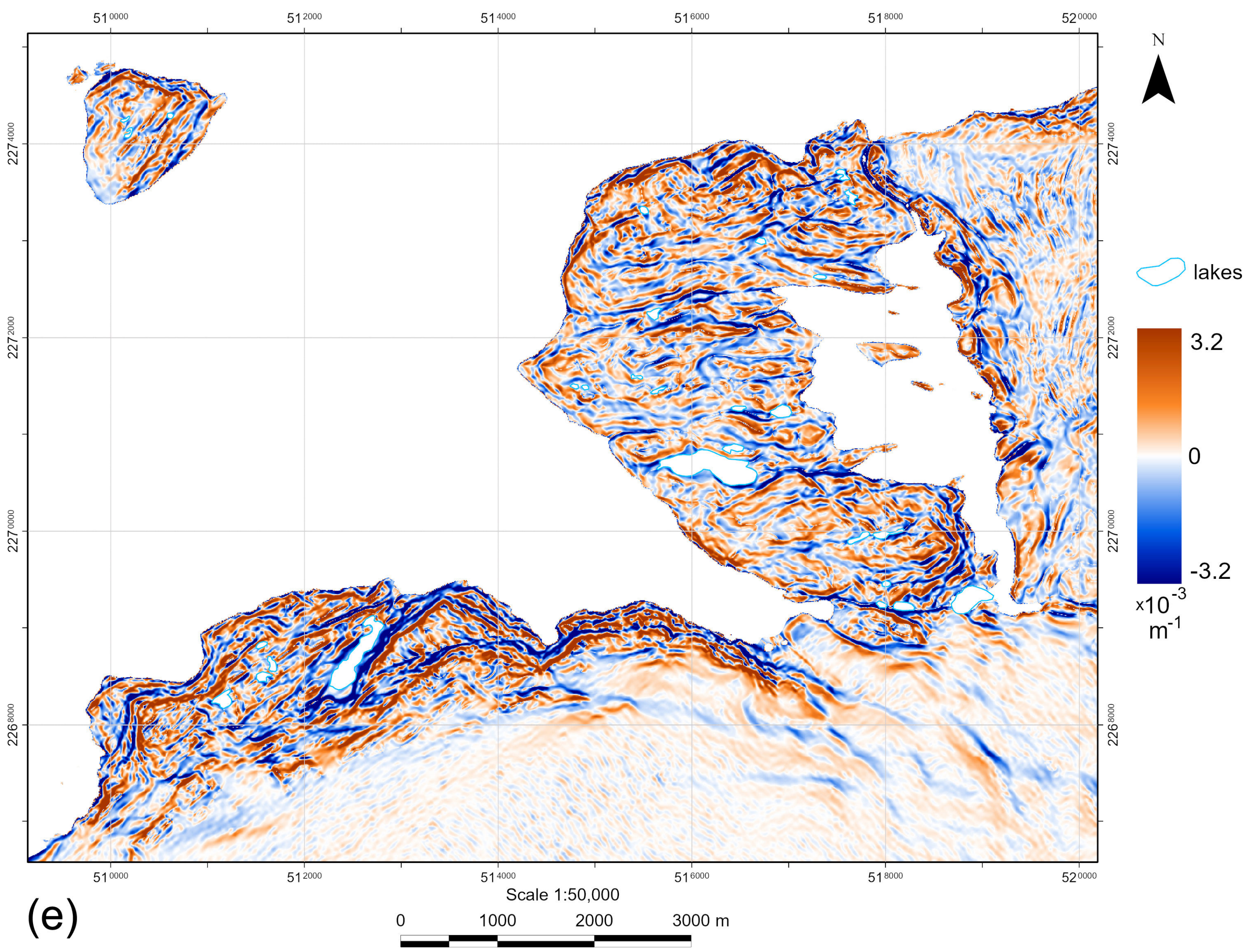

**Fig. 7, cont'd** Skallen Hills and Skallevikhalsen Hills: (e) Vertical curvature.

*(Continued)*

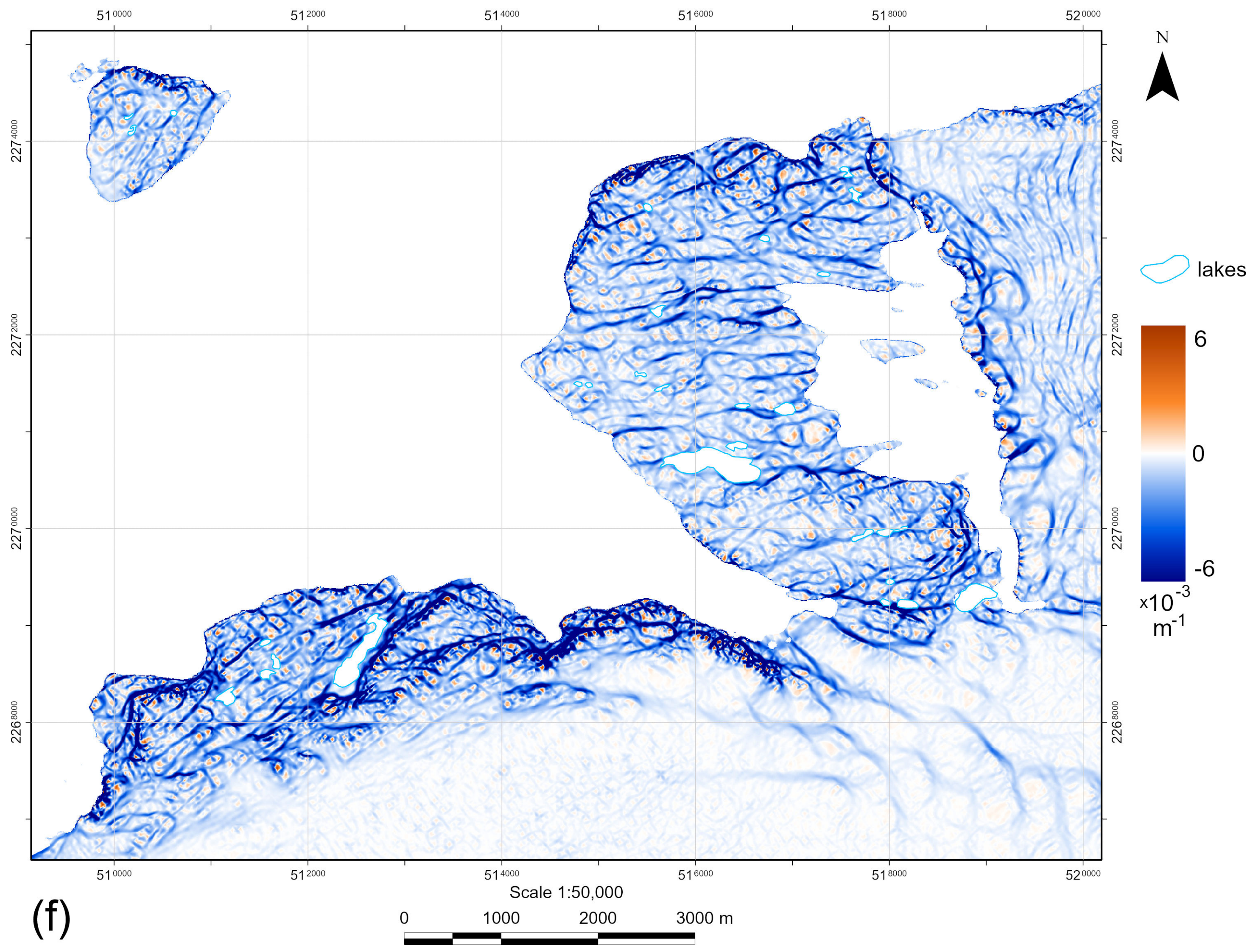

**Fig. 7, cont'd** Skallen Hills and Skallevikhalsen Hills: (f) Minimal curvature.

*(Continued)*

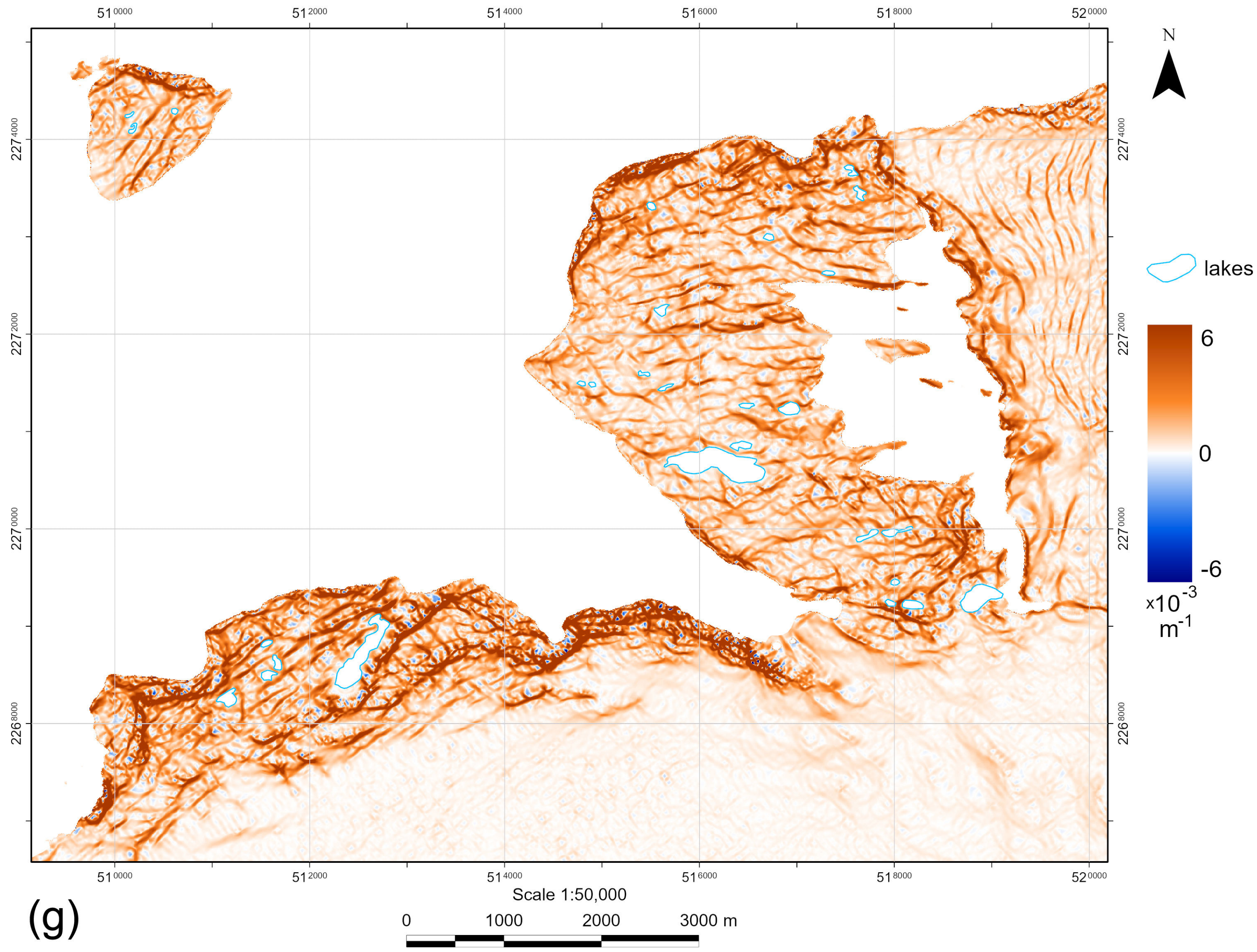

**Fig. 7, cont'd** Skallen Hills and Skallevikhalsen Hills: (g) Maximal curvature.

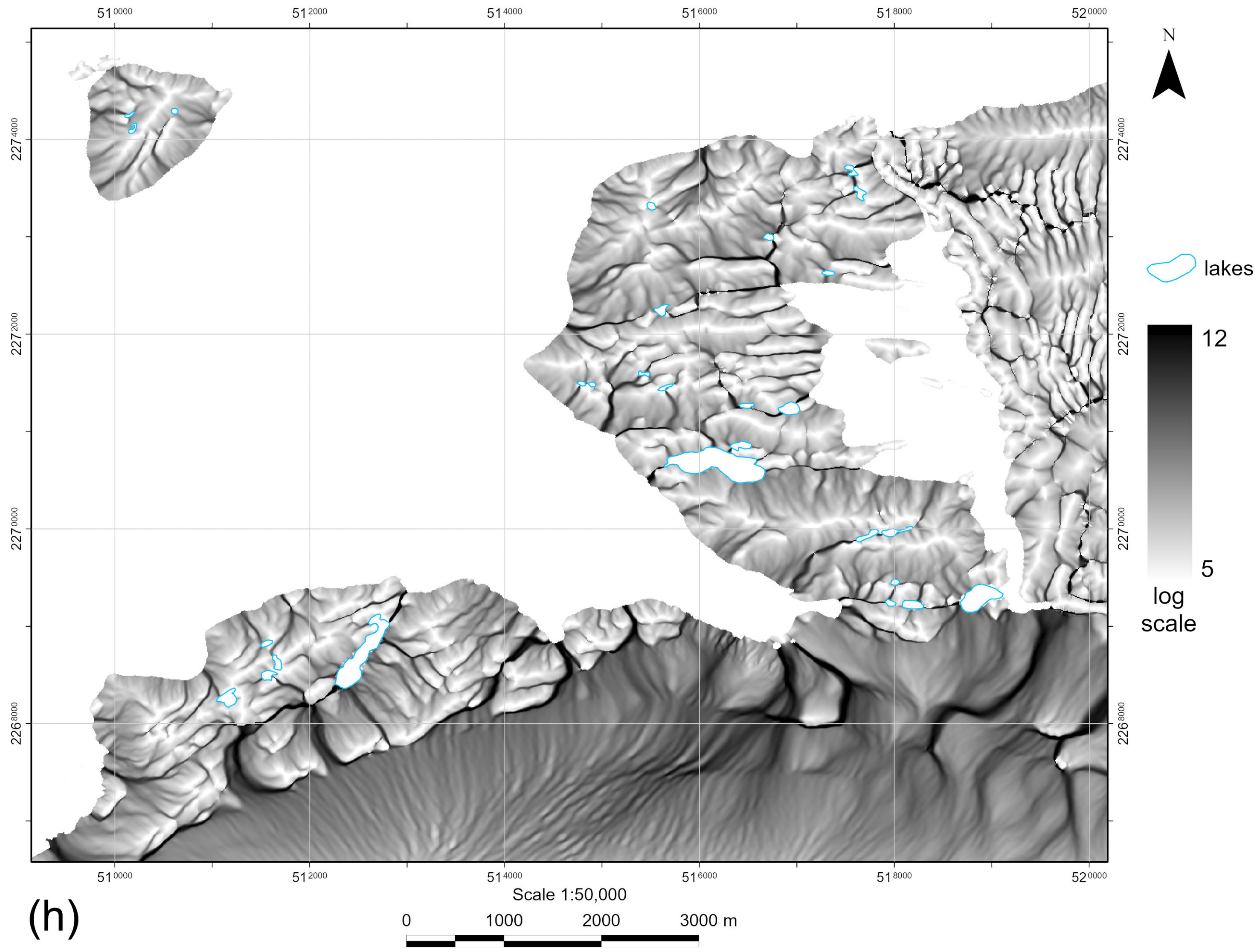

**Fig. 7, cont'd** Skallen Hills and Skallevikhalsen Hills: (h) Catchment area.

*(Continued)*

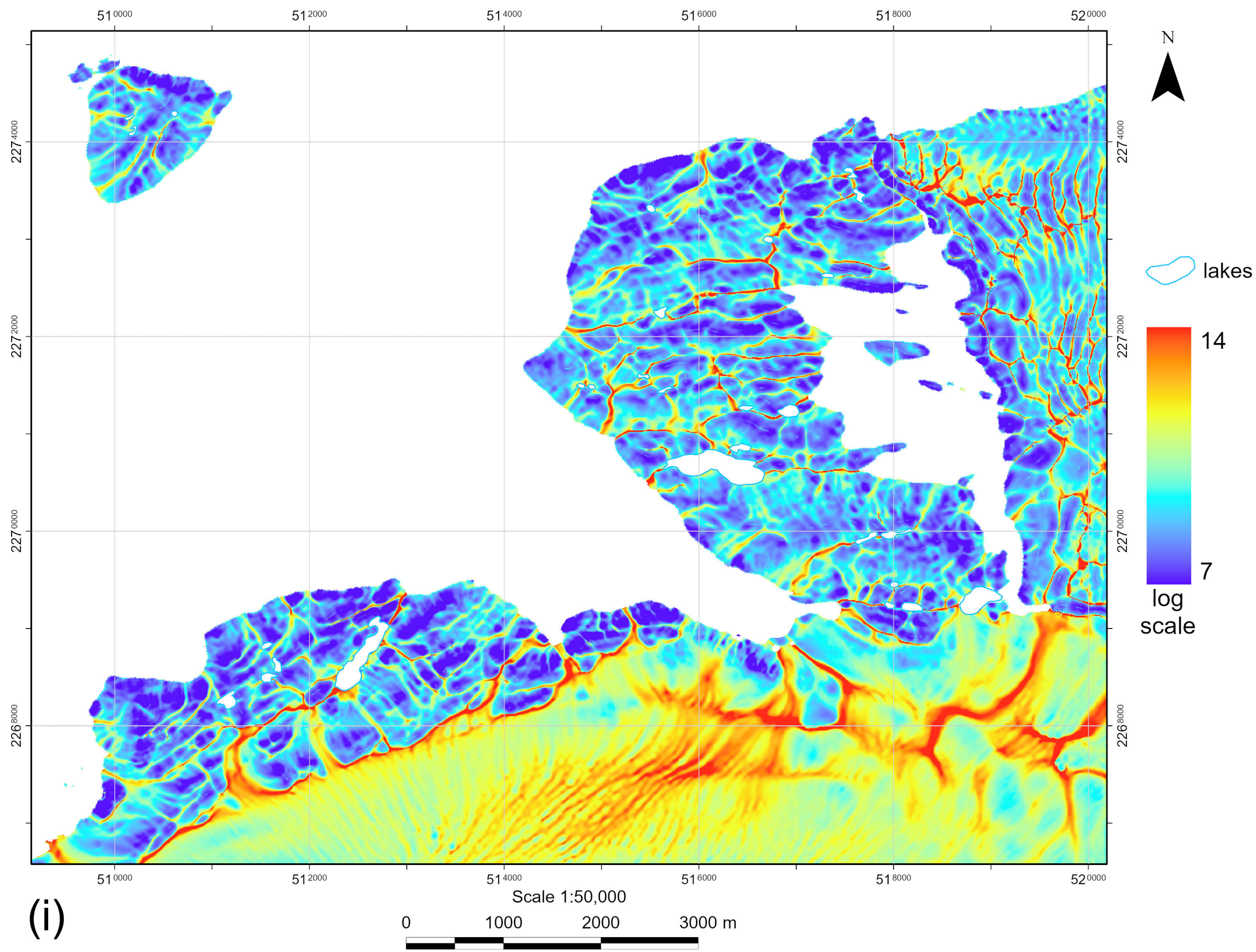

**Fig. 7, cont'd** Skallen Hills and Skallevikhalsen Hills: (i) Topographic wetness index.

*(Continued)*

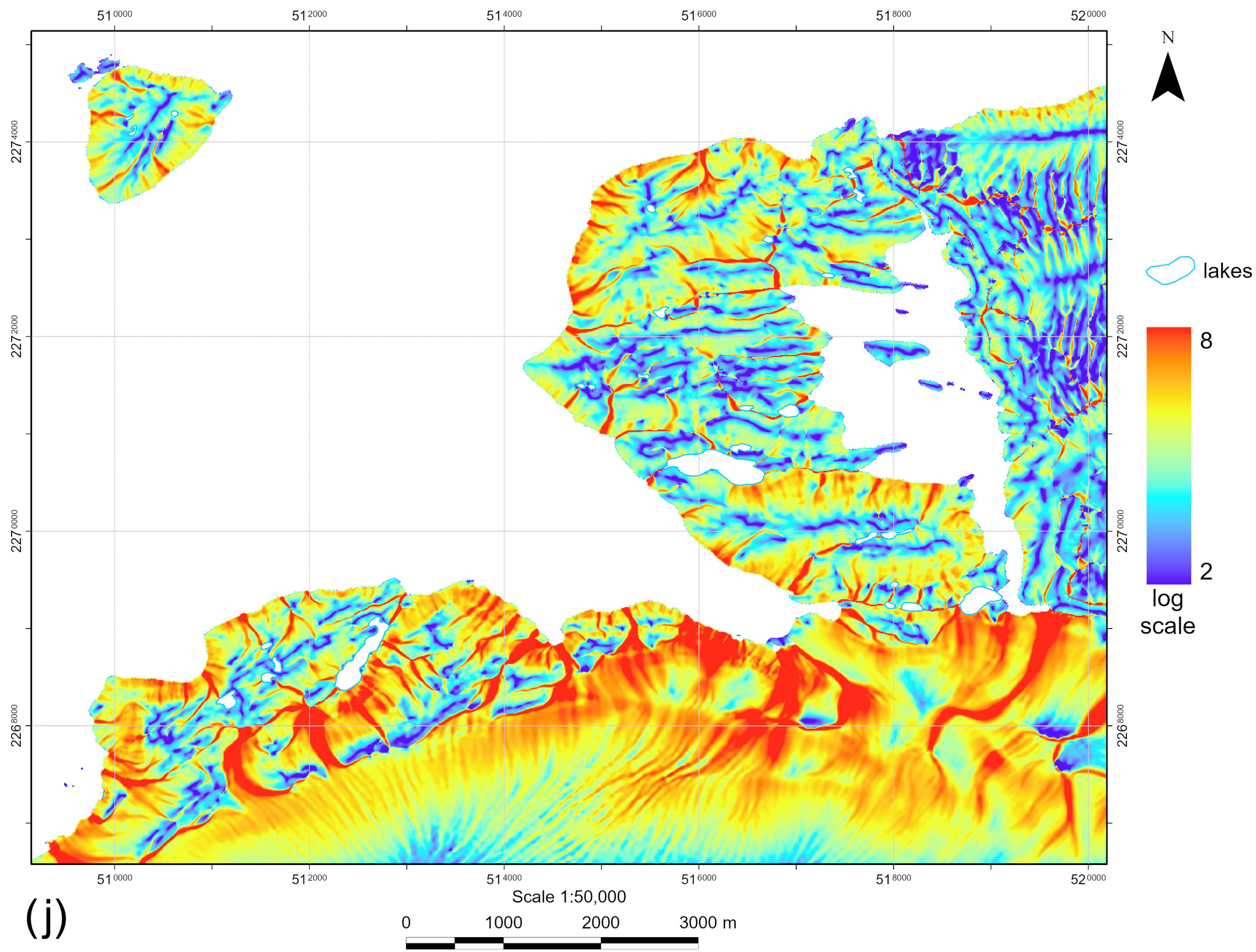

**Fig. 7, cont'd** Skallen Hills and Skallevikhalsen Hills: ( j) Stream power index.

*(Continued)*

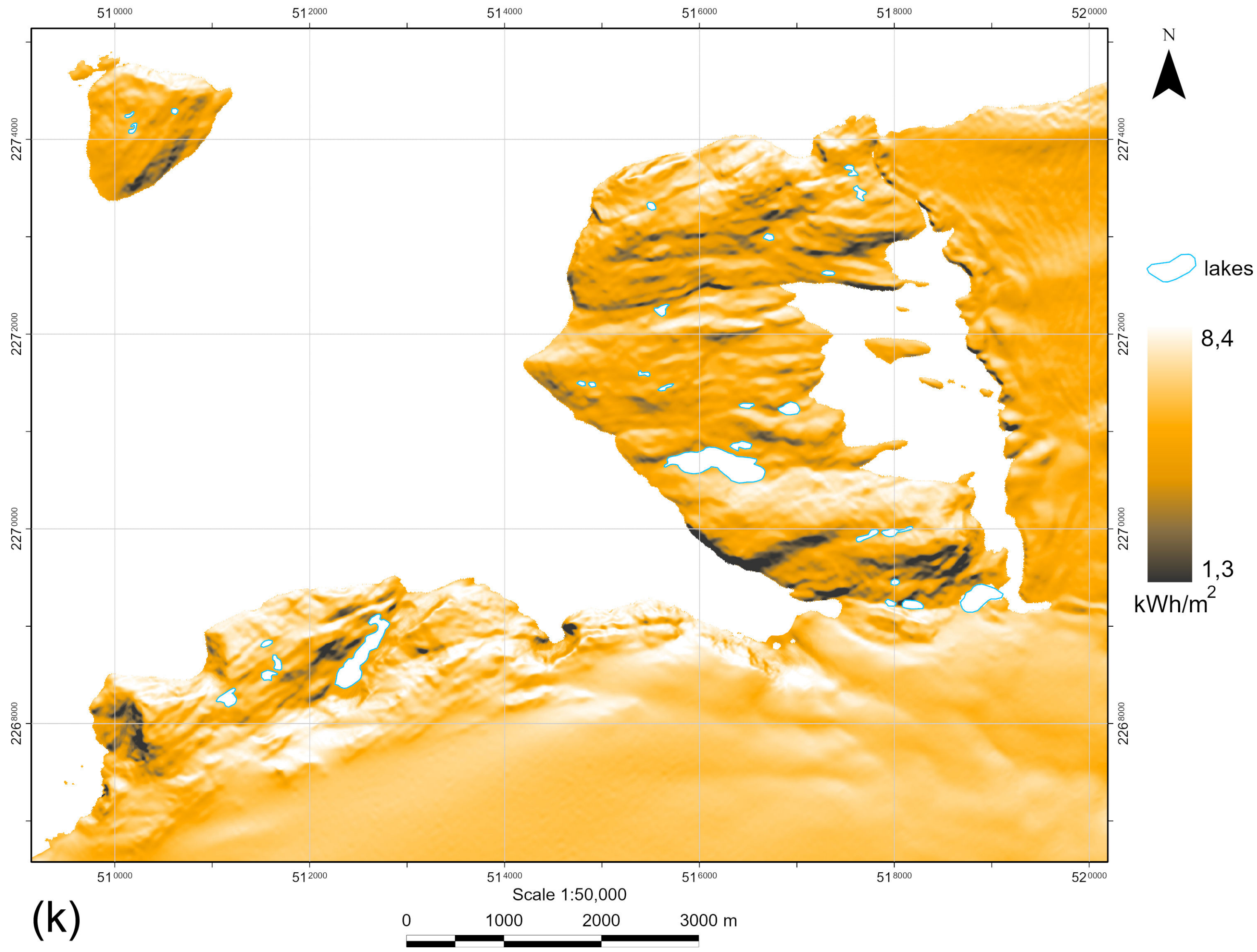

**Fig. 7, cont'd** Skallen Hills and Skallevikhalsen Hills: (k) Total insolation.

*(Continued)*

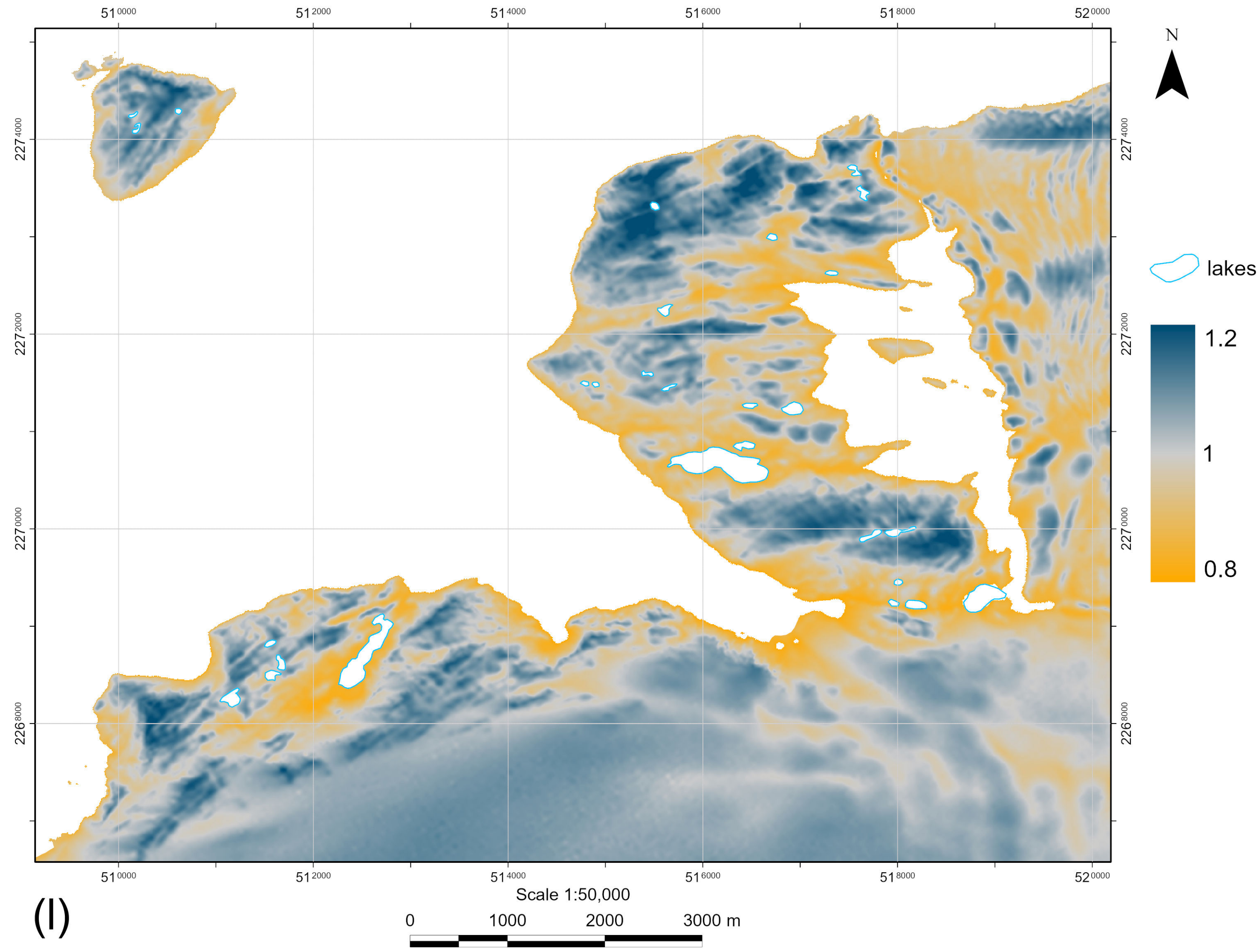

**Fig. 7, cont'd** Skallen Hills and Skallevikhalsen Hills: (l) Wind exposition index.

It is known that topography controls the thermal, wind, and hydrological regimes of slopes, influencing therefore the distribution and properties of soils and vegetation (Huggett and Cheesman, 2002; Florinsky, 2025a).

The thermal regime of slopes depends in part on the incidence of solar rays to the land surface, so it depends on both *G* and *A*. *TIns* directly considers this incidence and better describes the thermal regime (Böhner and Antonić, 2009). Information on the differentiation of slopes by insolation level is critical to predict the spatial distribution of primitive soils and lower plants in ice-free areas. To refine such a prediction, one can use *TWI* digital models describing topographic prerequisites of water migration and accumulation.

A further refinement of such a prediction can be done with *WEx* data. *WEx* digital models are utilized to identify areas affected by and protected from the wind impact (Böhner and Antonić, 2009; Florinsky, 2025a, chap. 2). In periglacial Antarctic landscapes, one of the main meteorological factors determining the microclimate is katabatic wind. Thus, *WEx* maps may be of great importance for modeling the distribution of primitive soil and lower plants, predicting the differentiation of snow accumulation in ice-free areas, and determining the optimal location of buildings and infrastructure at polar stations.

$k_h$ maps display the distribution of convergence and divergence areas of surface flows. Geomorphologically, these are spurs of valleys and crests, respectively. The combination of convergence and divergence areas creates an image of the flow structures (Florinsky, 2025a, chap. 2). $k_v$ maps show the distribution of relative deceleration and acceleration areas of surface flows. Geomorphologically, these maps represent cliffs, scarps, terrace edges, and other similar landforms or their elements with sharp bends in the slope profile (Florinsky, 2025a, chap. 2). In this regard, $k_h$ and $k_v$ digital models may be useful in geomorphological and hydrological studies of the ice-free areas.

Combination of $k_h$ and $k_v$ digital models allows revealing relative accumulation zones of surface flows (Florinsky, 2025a, chap. 2). These zones, marked by both $k_h < 0$ and $k_v < 0$, coincide with the fault intersection sites and are characterized by increased rock fragmentation and permeability. Within these zones, one can observe an interaction and exchange between two types of substance flows: (a) lateral, gravity-driven substance flows moved along the land surface and in the near-surface layer, such as water, dissolved and suspended substances, and (b) vertical, upward substance flows, such as fluids, groundwater of different mineralization and temperature (Florinsky, 2025a, chap. 15). Maps of accumulation zones may be useful for geochemical studies in the ice-free areas.

$k_{max}$ and $k_{min}$ maps are informative in terms of structural geology because they clearly display elongated linear landforms (Florinsky, 2017, 2025a). In Antarctica, such lineaments can be interpreted as a reflection of the local fault and fracture network, which topographic manifestation has been amplified by erosional, exaration, and nival processes (Florinsky, 2023b; Florinsky and Zharnova, 2025). Thus, researchers are able to use $k_{max}$ and $k_{min}$ models for compiling lineament maps and comparing them with other geological data.

*CA* maps can be used to identify the fine flow structure of drainage basins and then to incorporate this information into geochemical and hydrological analysis. *TWI* digital models can be applied for prediction of the ground moisture content in the ice-free areas as well as for forecasting the spatial distribution of snow puddles on adjacent glaciers in summer. *SPI* data can be useful for prediction of slope erosion in the ice-free areas as well as erosion of snow cover and ice by meltwater flows on adjacent glaciers in summer.

## 5 Conclusions

We performed geomorphometric modeling of the key ice-free areas of the Sôya Coast and, for the first time, created series of morphometric models and maps of the most scientifically important morphometric variables for the Flatvaer (Ongul) Islands, Langhovde

Hills, Breidvågnipa, Skarvsnes Foreland, Skallen Hills, and Skallevikhalsen Hills.

The obtained maps rigorously, quantitatively, and reproducibly describe the ice-free topography of the Sôya Coast. New morphometric data can be useful in further geological, geomorphological, glaciological, hydrological, and ecological studies of this Antarctic region.

The study was performed within the framework of the project for creating a physical geographical thematic scientific reference geomorphometric atlas of ice-free areas of Antarctica (Florinsky, 2024, 2025b).

# Appendix A

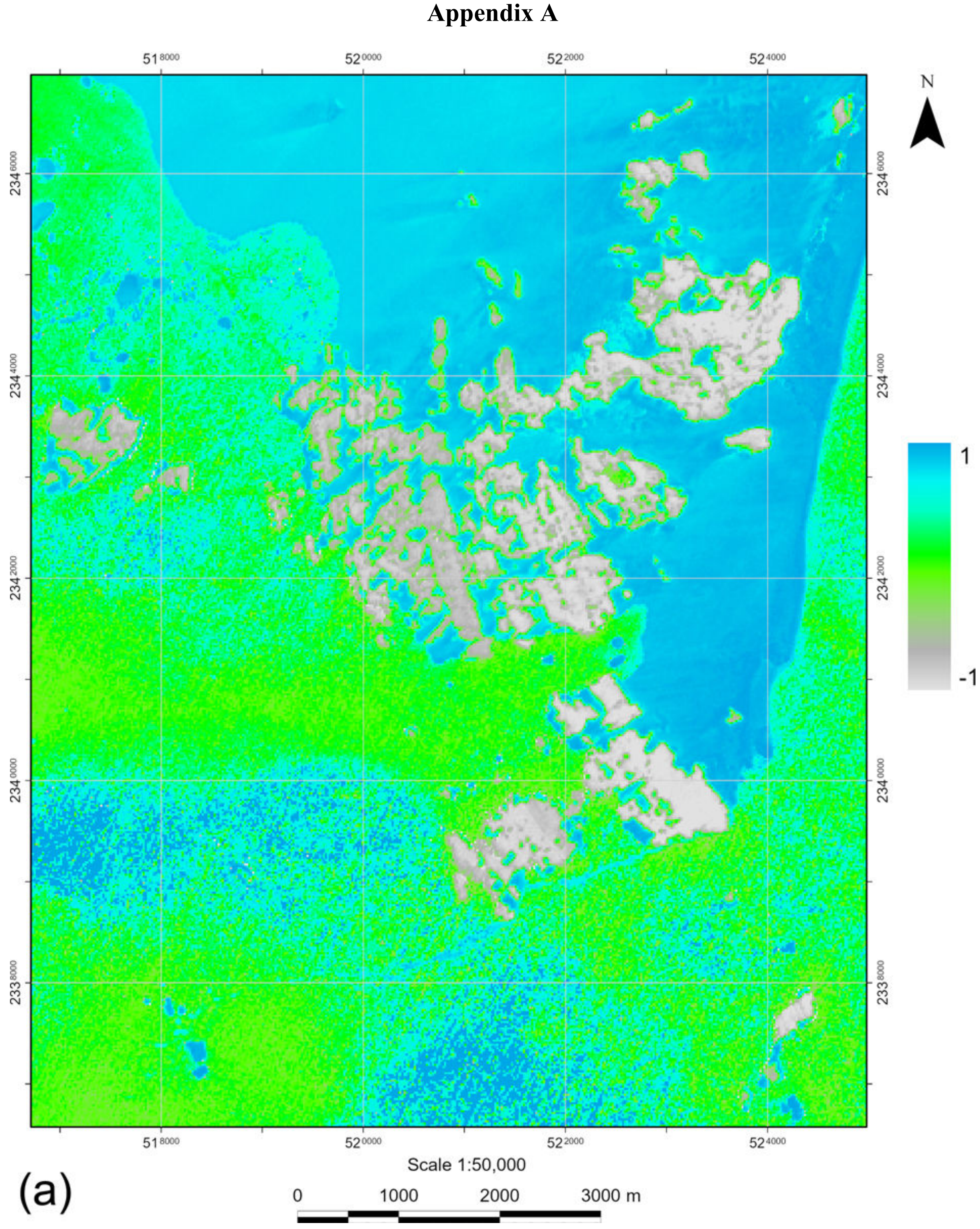

**Fig. A1** MNDWI maps derived from Sentinel-2A MSI data: (a) Flatvaer (Ongul) Islands.

*(Continued)*

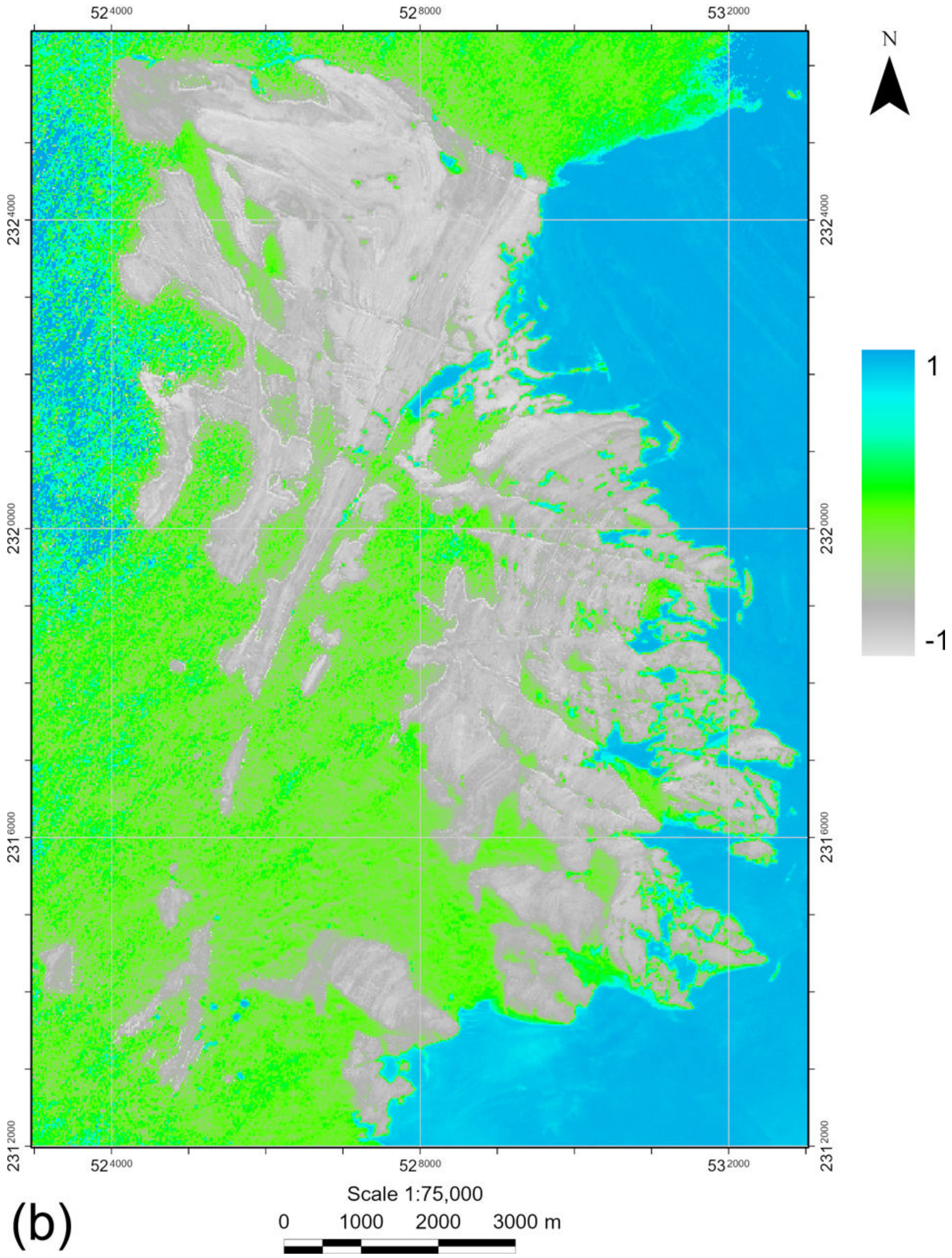

**Fig. A1, cont'd** MNDWI maps derived from Sentinel-2A MSI data: (b) Langhovde Hills.

*(Continued)*

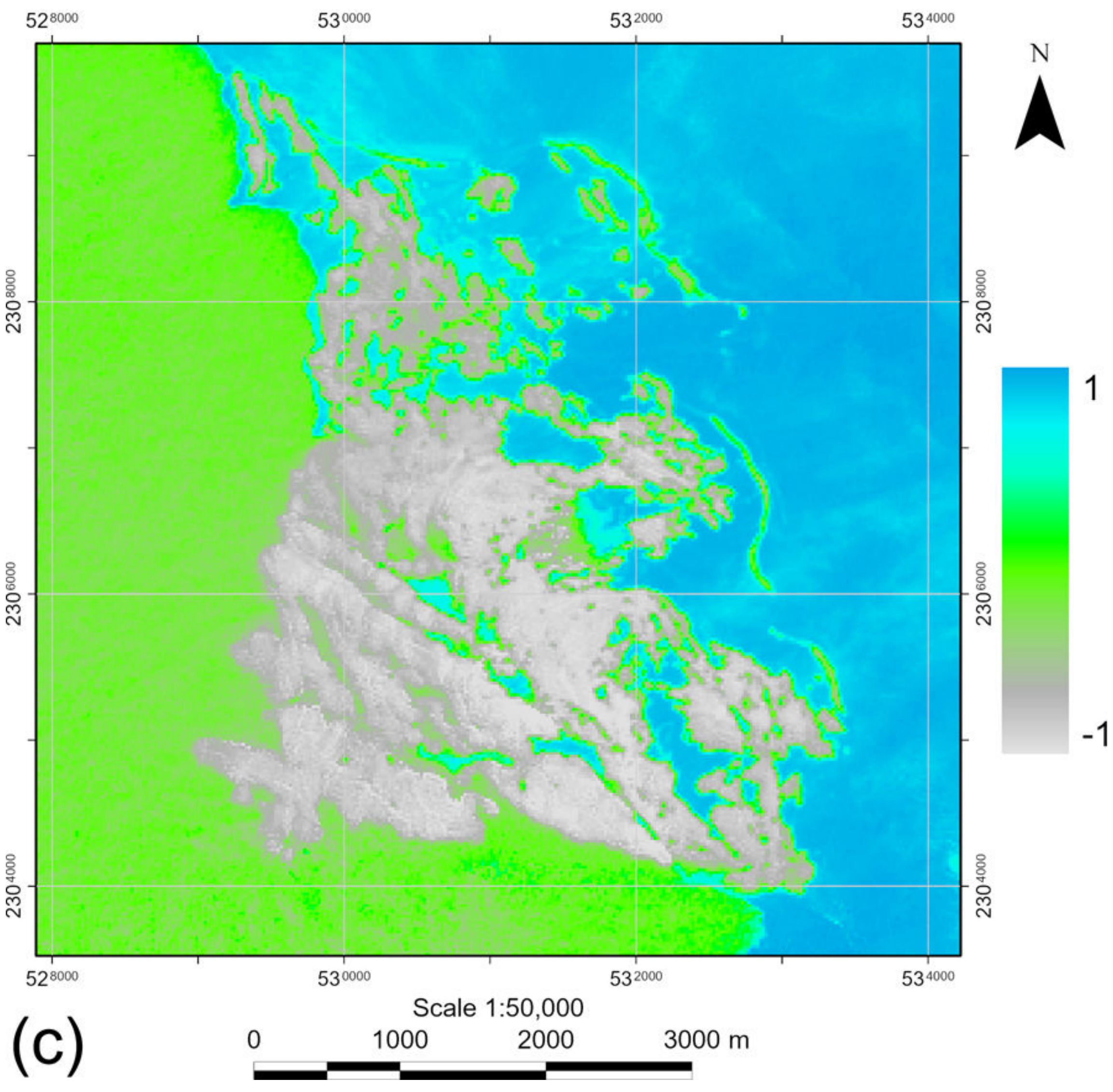

**Fig. A1, cont'd** MNDWI maps derived from Sentinel-2A MSI data: (c) Breidvågnipa.

*(Continued)*

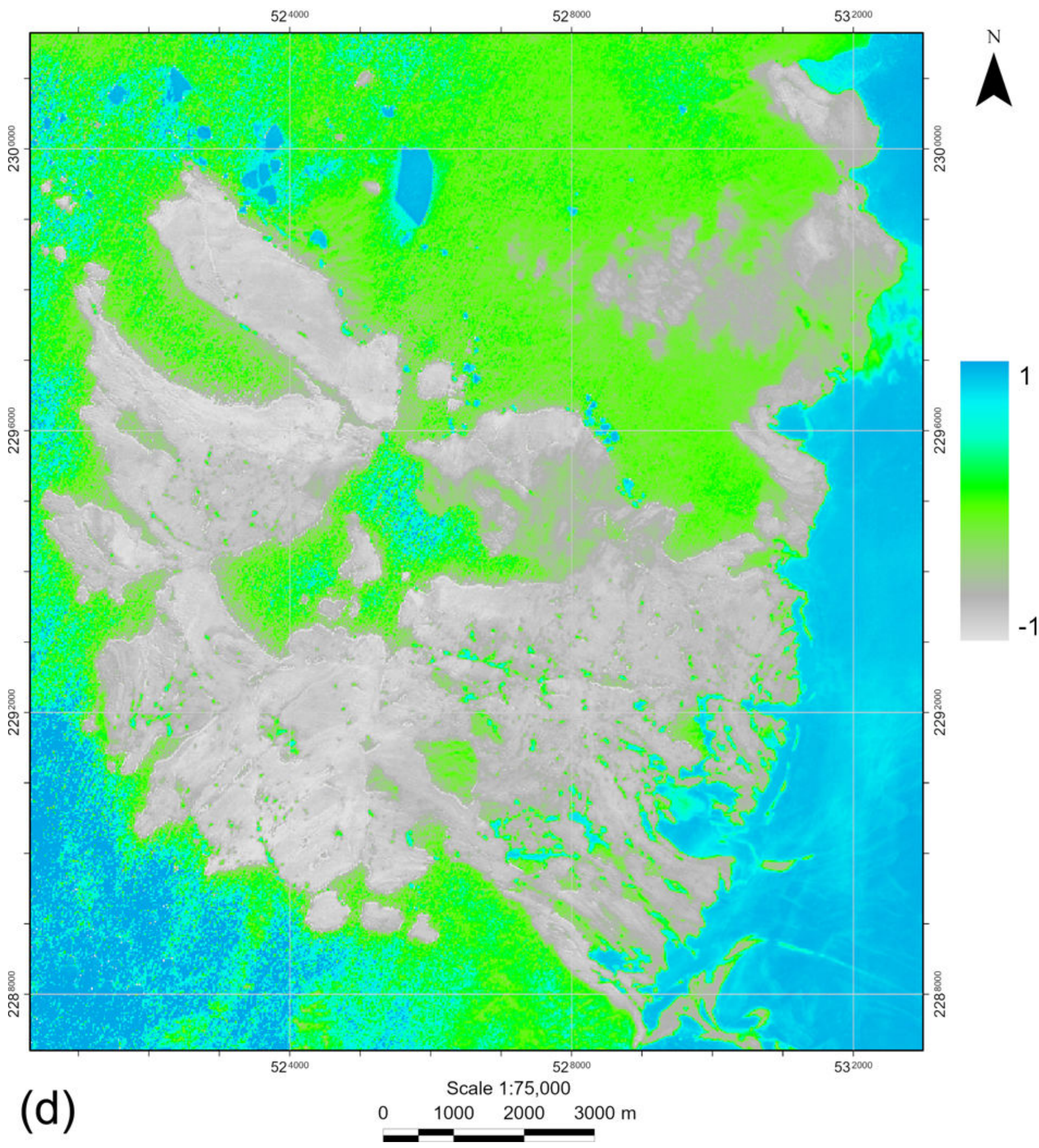

**Fig. A1, cont'd** MNDWI maps derived from Sentinel-2A MSI data: (d) Skarvsnes Foreland.

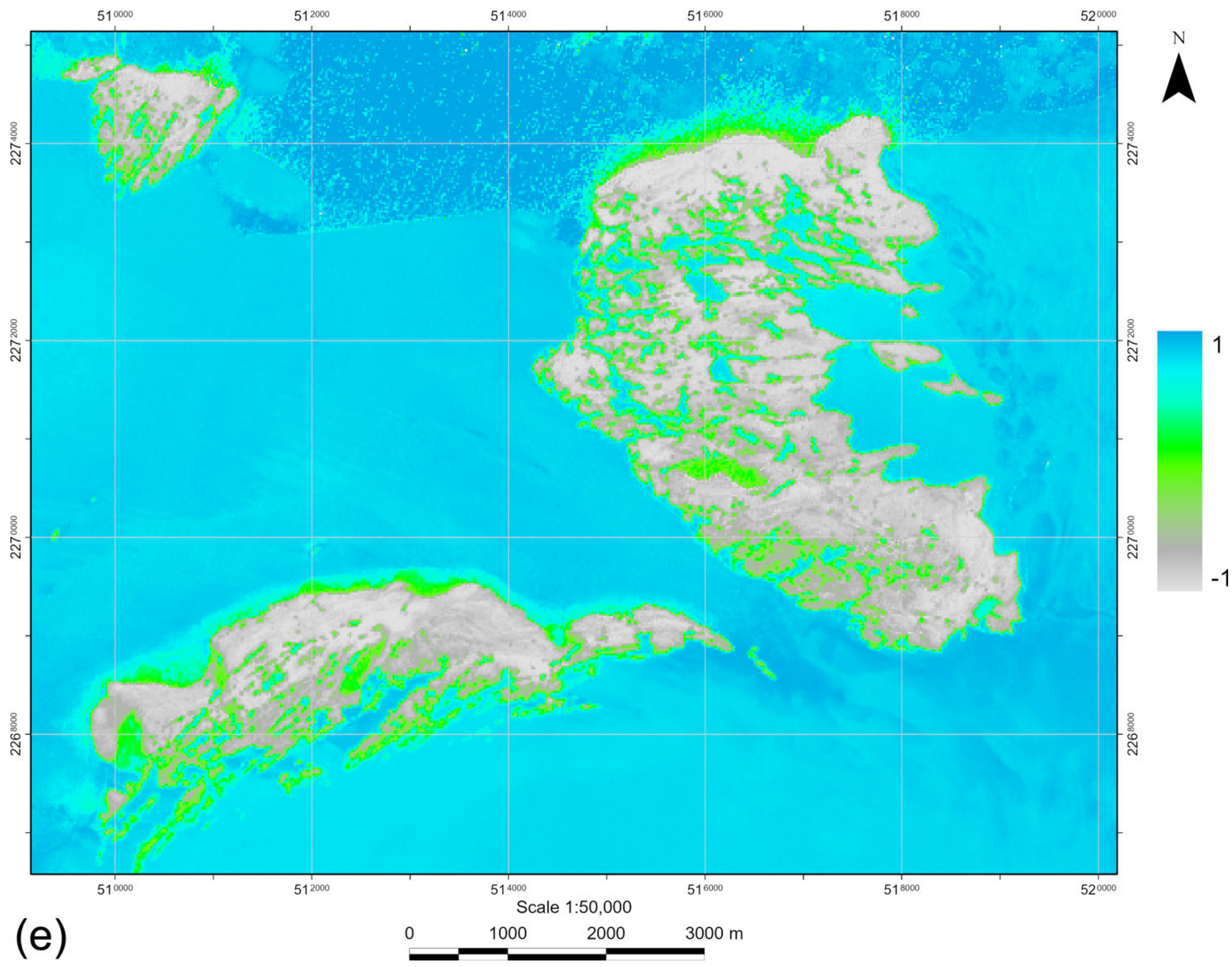

**Fig. A1, cont'd** MNDWI maps derived from Sentinel-2A MSI data: (e) Skallen Hills and Skallevikhalsen Hills.

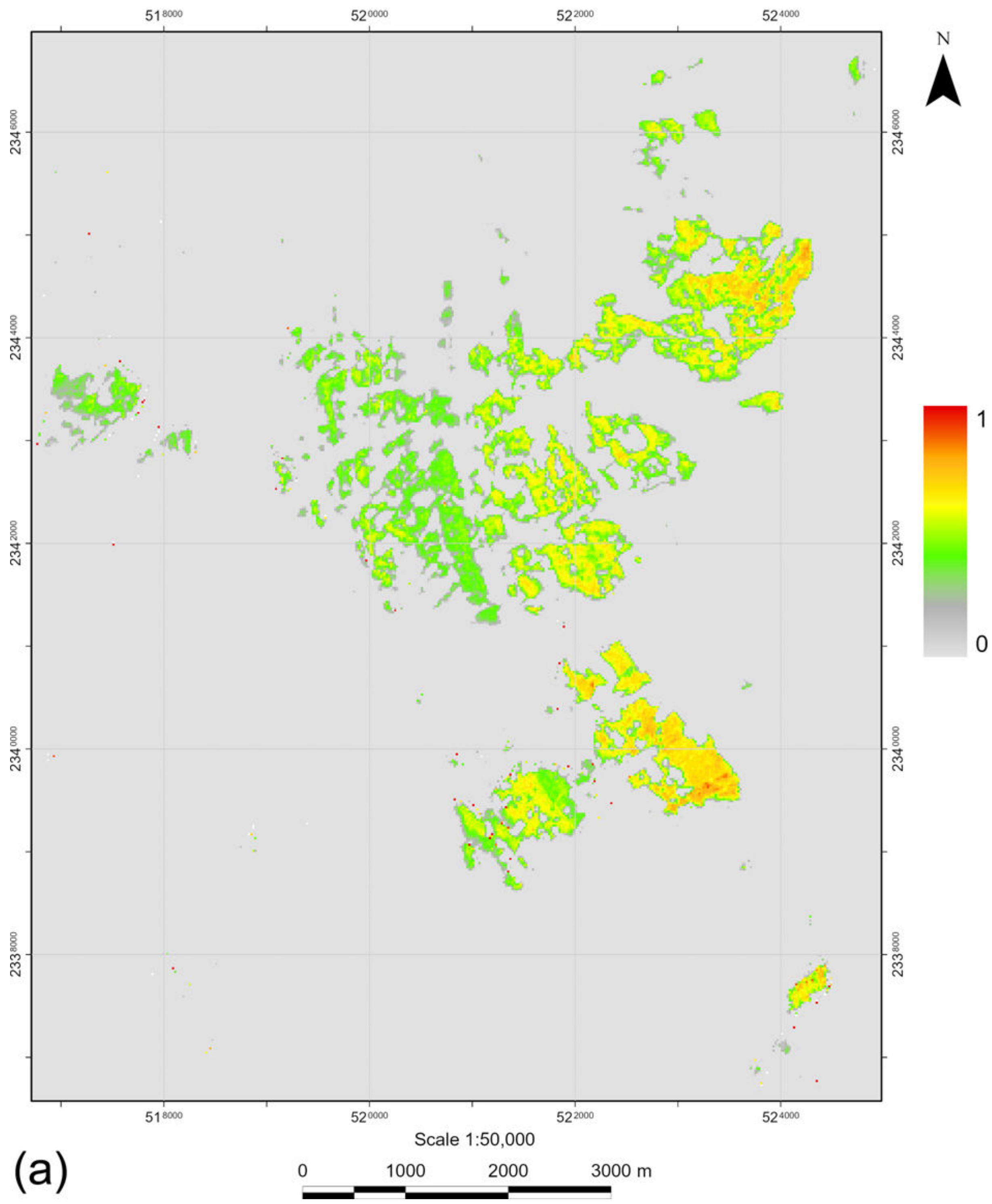

**Fig. A2** FMR maps derived from Sentinel-2A MSI data: (a) Flatvaer (Ongul) Islands.

*(Continued)*

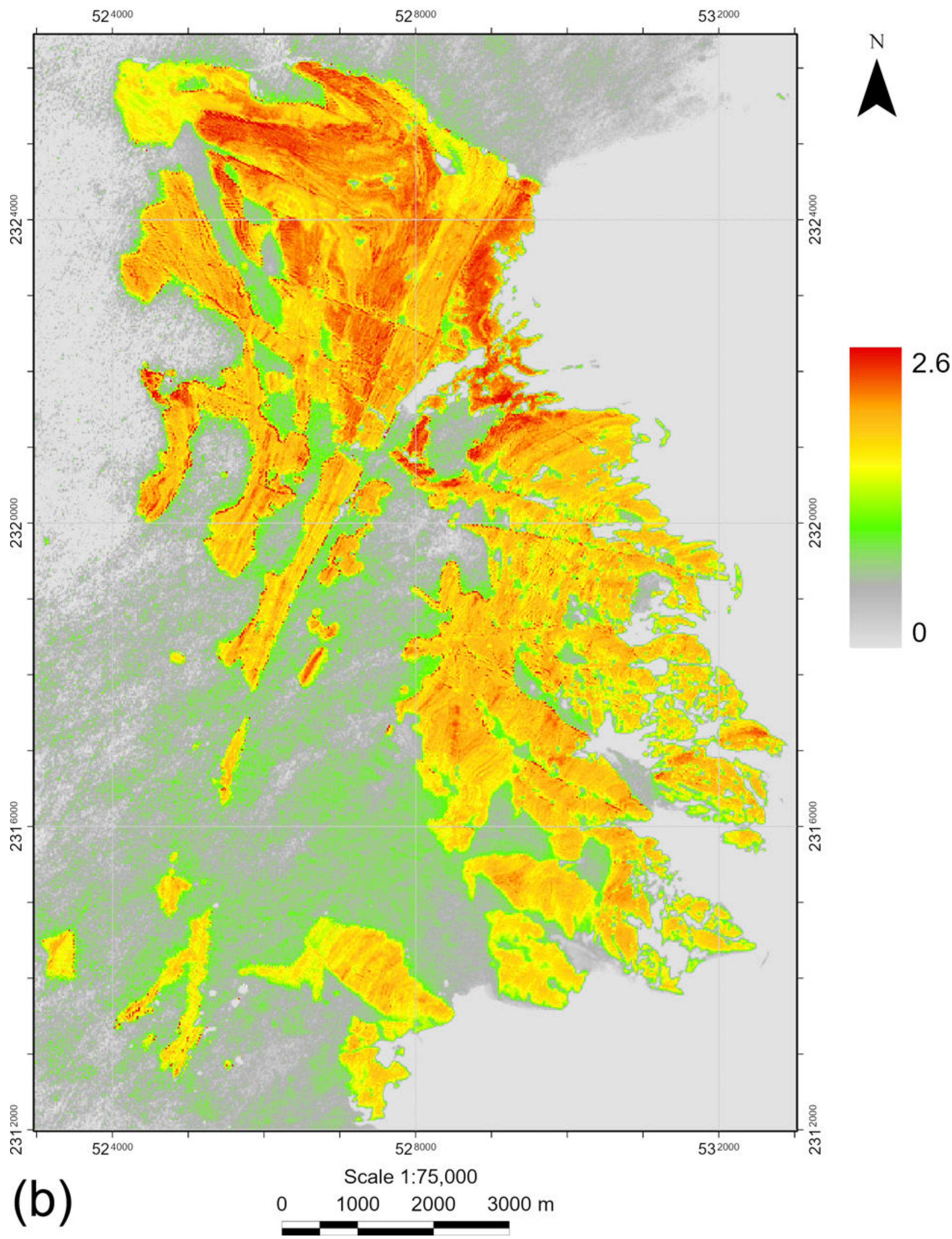

**Fig. A2, cont'd** FMR maps derived from Sentinel-2A MSI data: (b) Langhovde Hills.

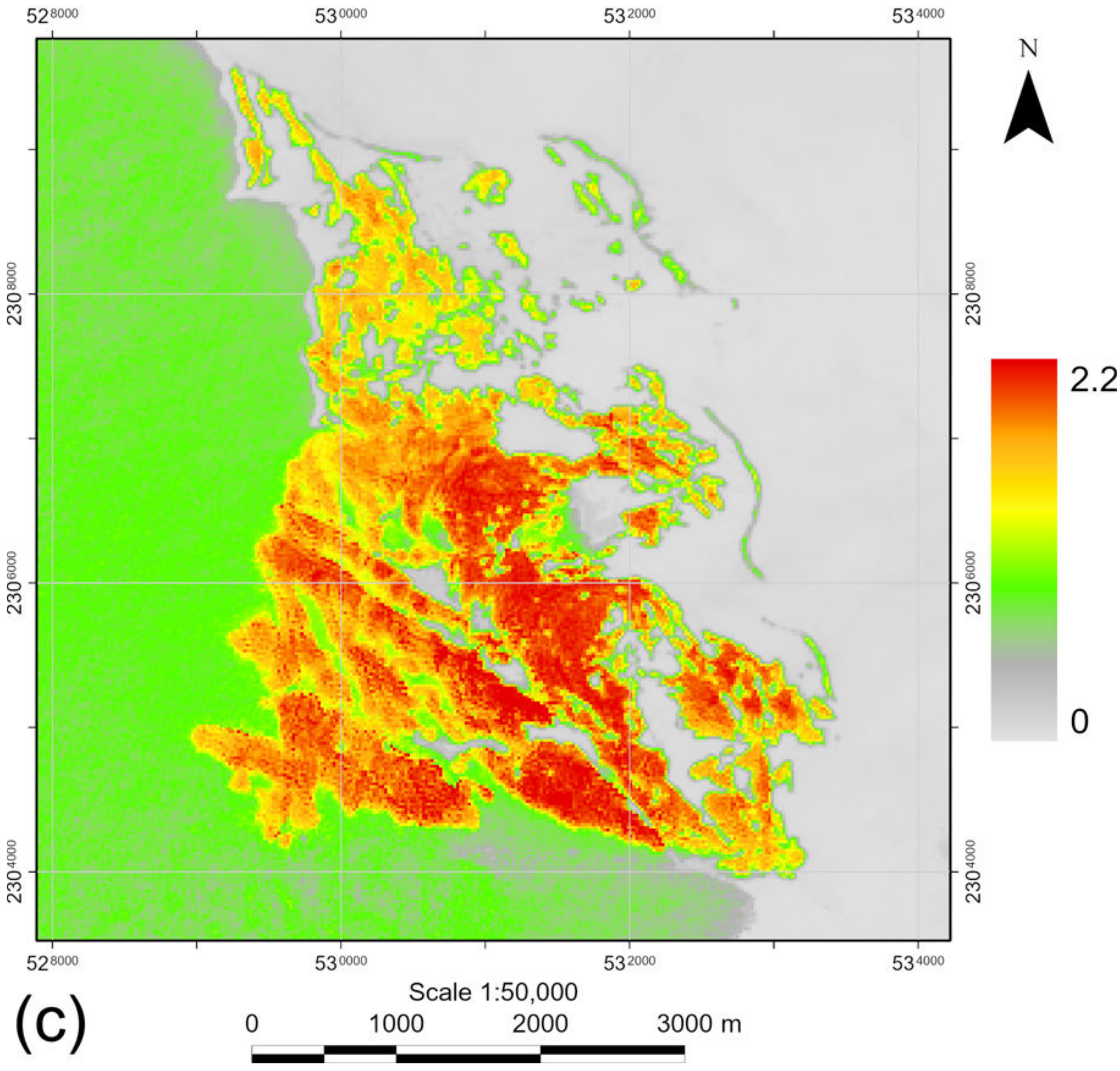

**Fig. A2, cont'd** FMR maps derived from Sentinel-2A MSI data: (c) Breidvågnipa.

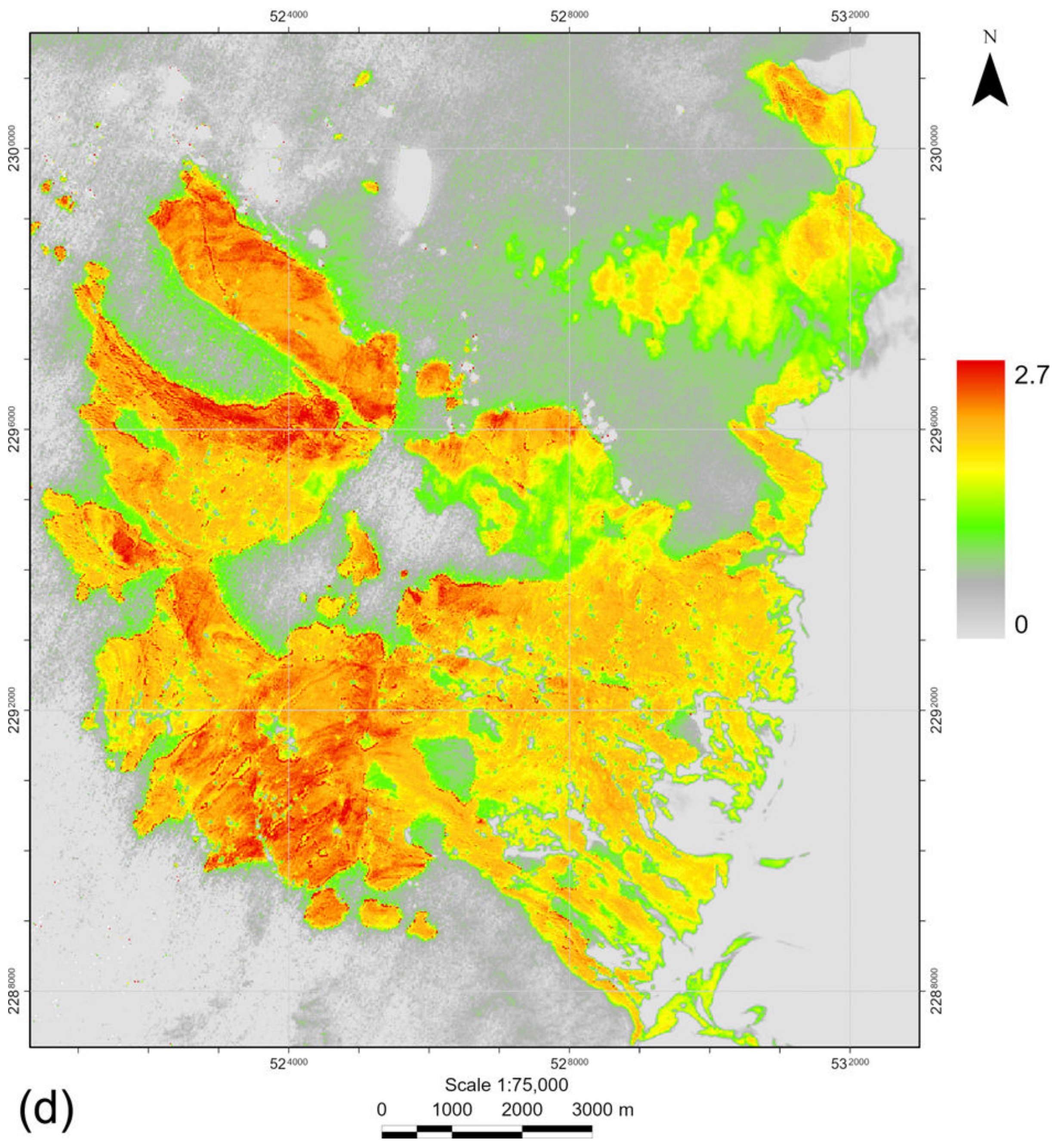

**Fig. A2, cont'd** FMR maps derived from Sentinel-2A MSI data: (d) Skarvsnes Foreland.

*(Continued)*

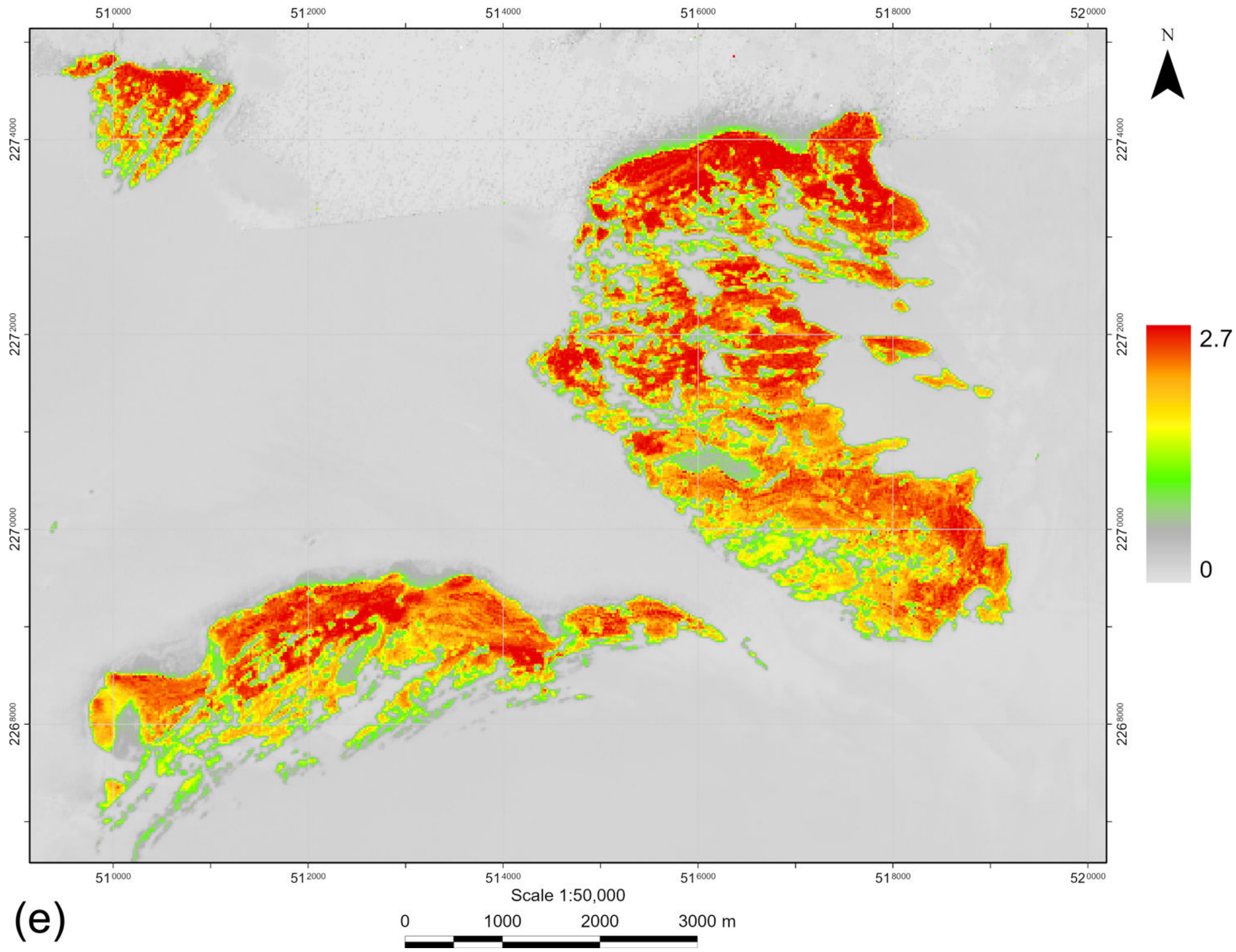

**Fig. A2, cont'd** FMR maps derived from Sentinel-2A MSI data: (e) Skallen Hills and Skallevikhalsen Hills.

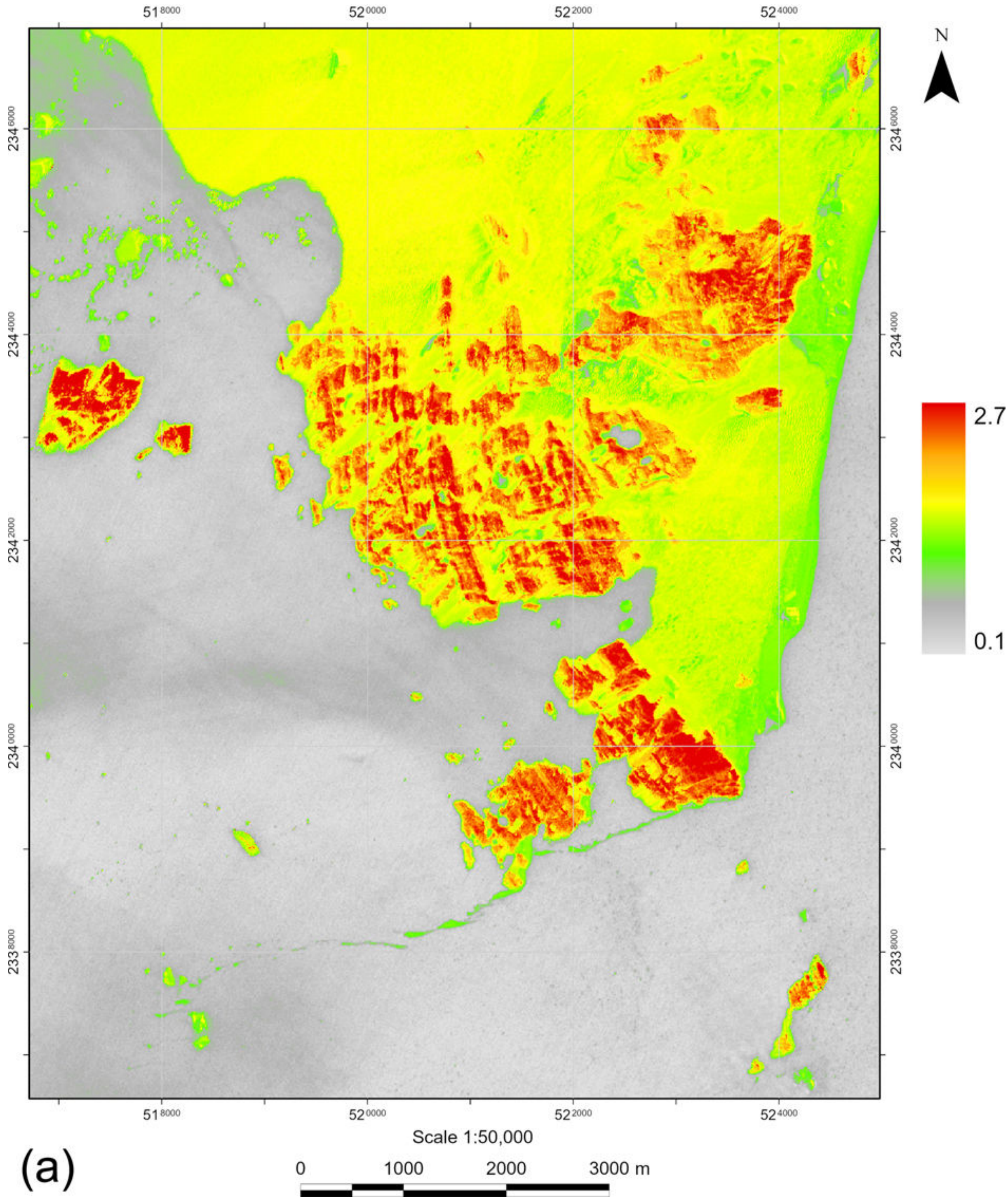

**Fig. A3** IOR maps derived from Sentinel-2A MSI data: (a) Flatvaer (Ongul) Islands.

*(Continued)*

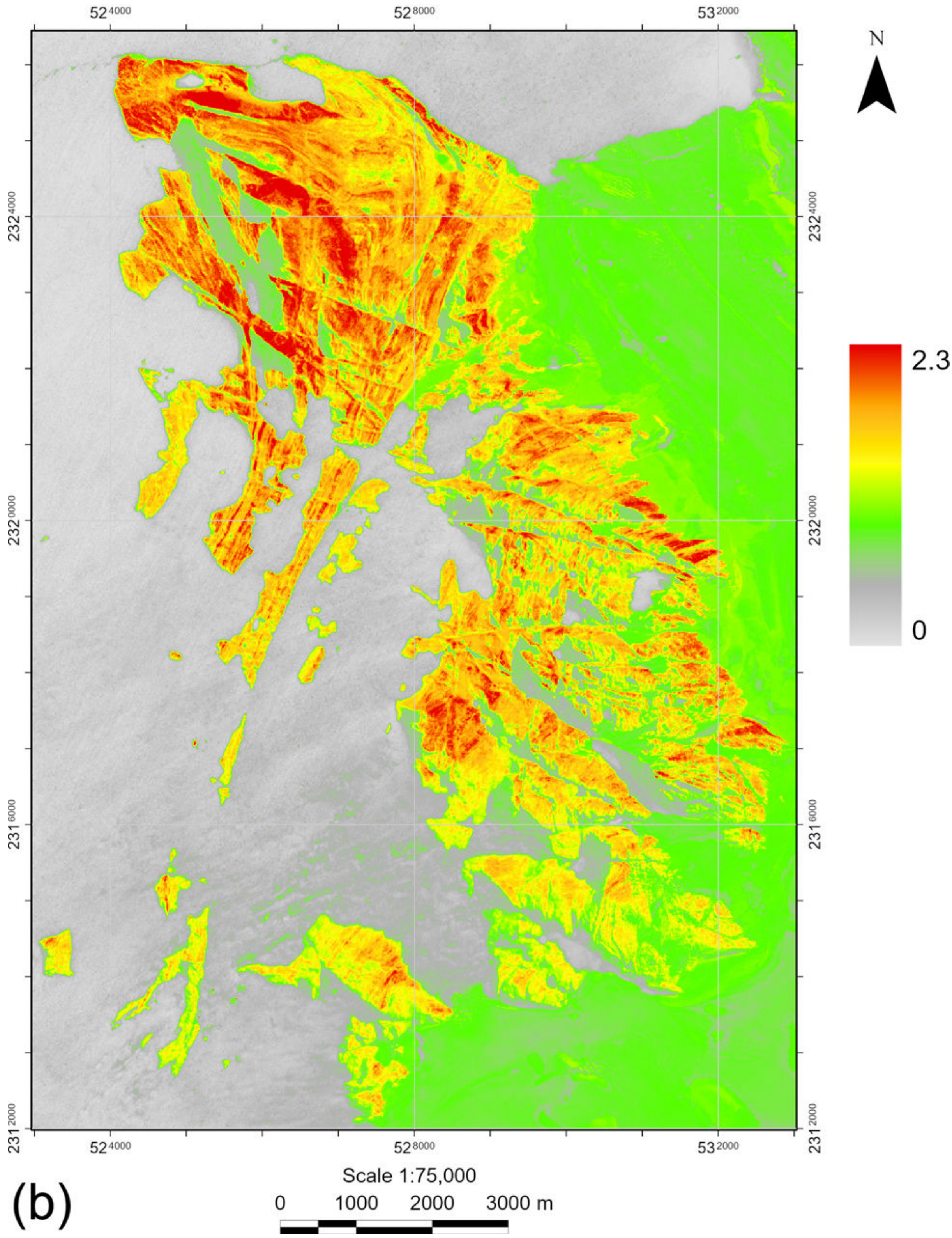

**Fig. A3, cont'd** IOR maps derived from Sentinel-2A MSI data: (b) Langhovde Hills.

*(Continued)*

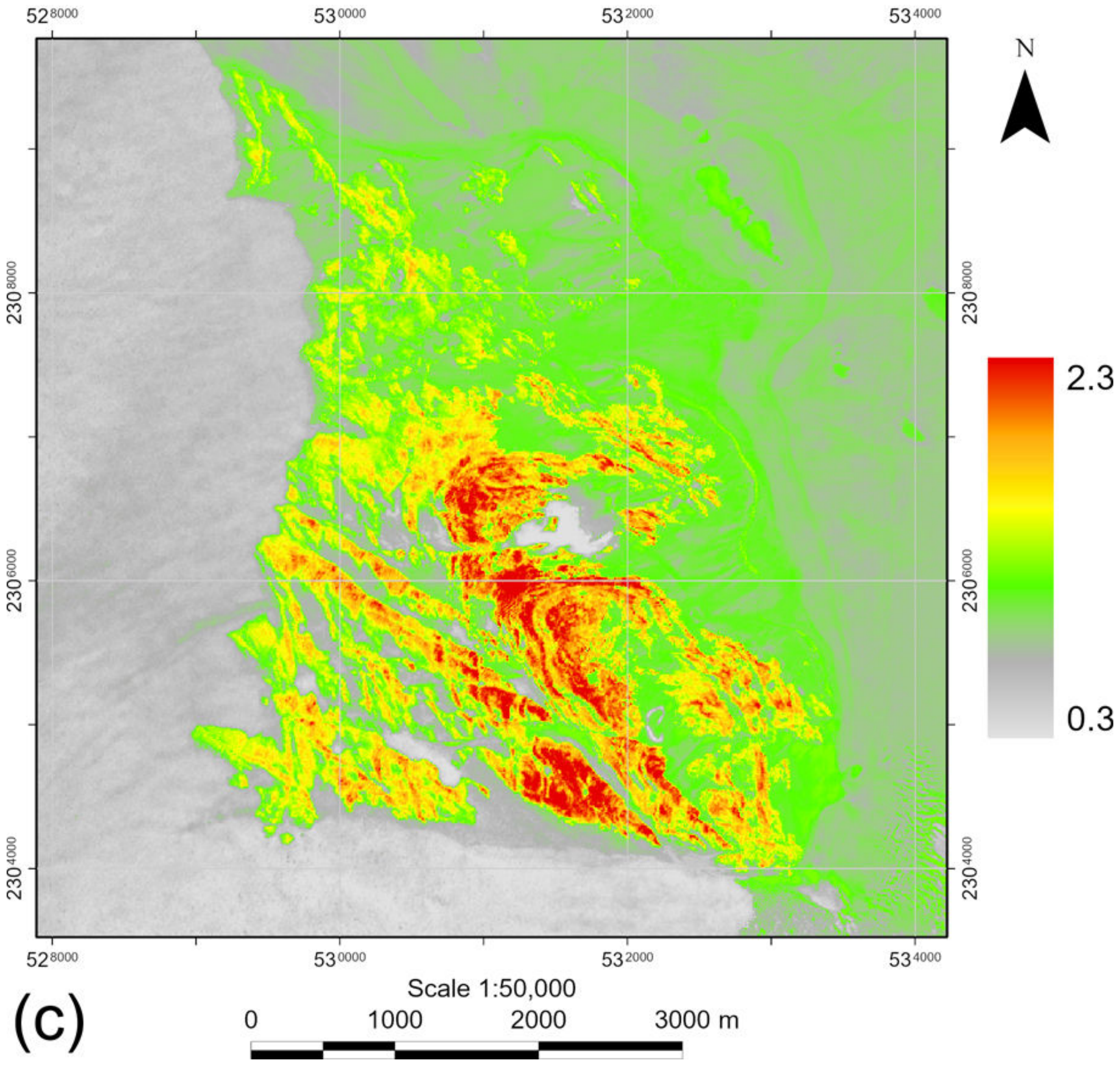

**Fig. A3, cont'd** IOR maps derived from Sentinel-2A MSI data: (c) Breidvågnipa.

*(Continued)*

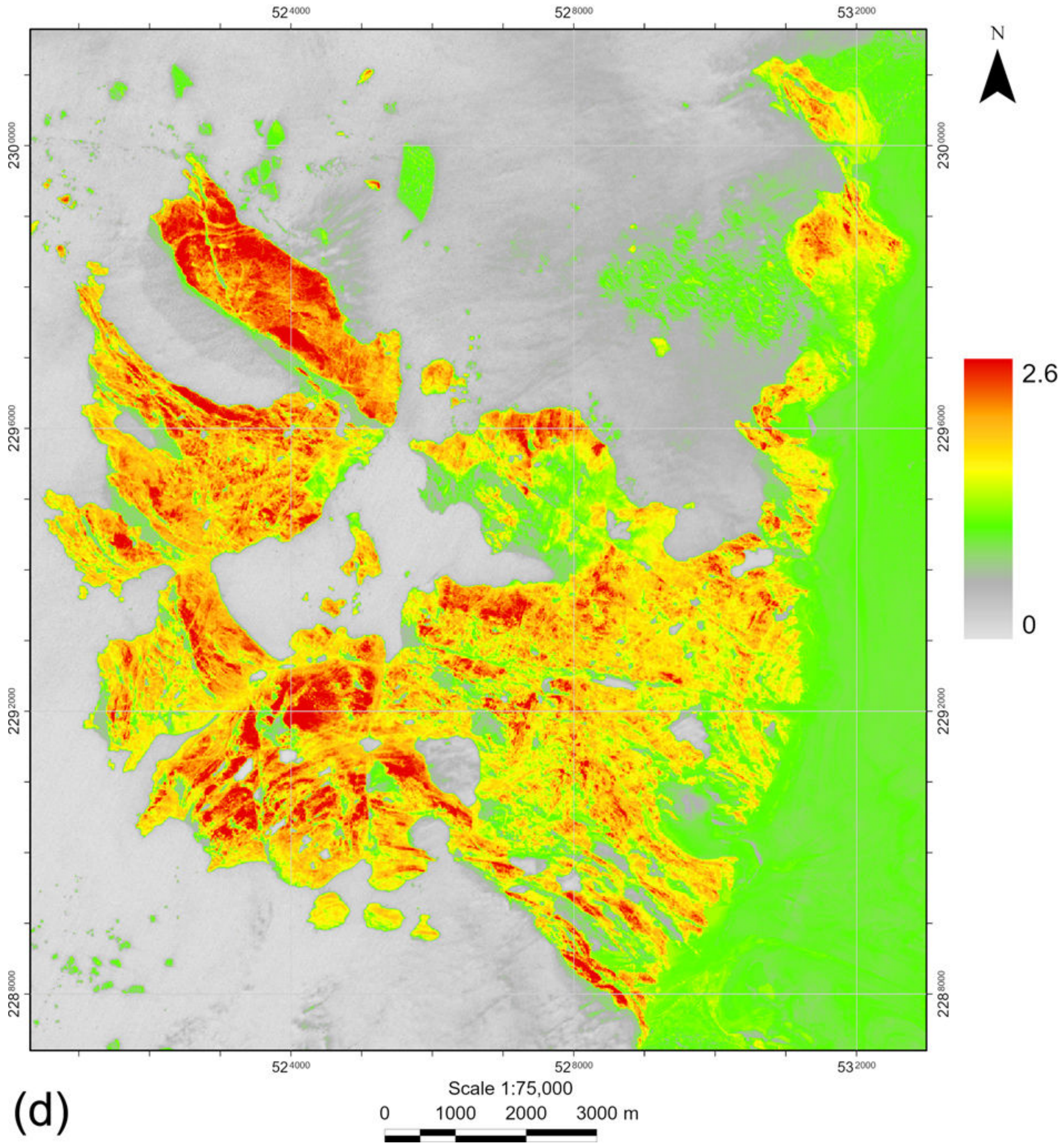

**Fig. A3, cont'd** IOR maps derived from Sentinel-2A MSI data: (d) Skarvsnes Foreland.

*(Continued)*

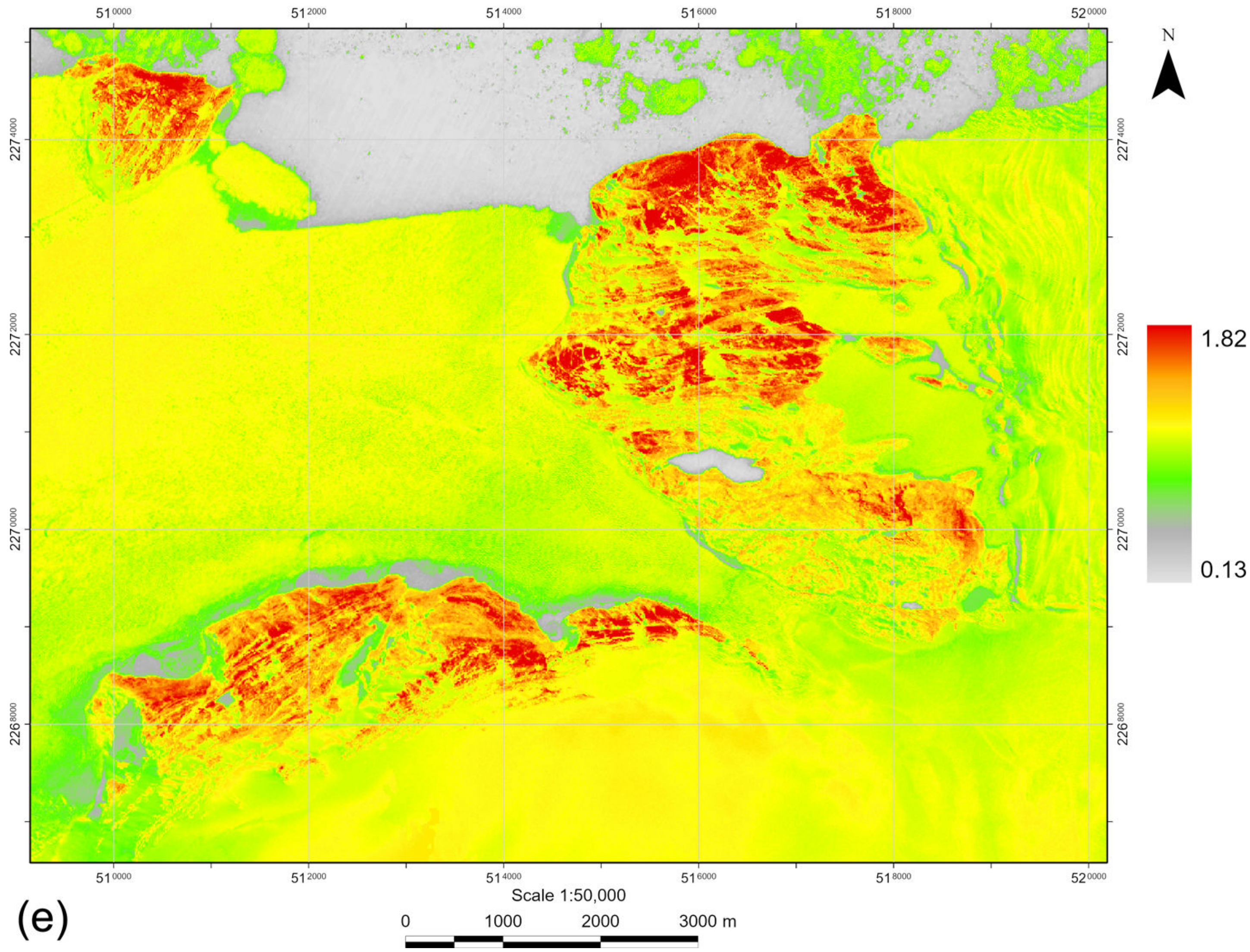

**Fig. A3, cont'd** IOR maps derived from Sentinel-2A MSI data: (e) Skallen Hills and Skallevikhalsen Hills.